\documentclass{SciPost}

\hypersetup{
    colorlinks,
    linkcolor={red!50!black},
    citecolor={blue!50!black},
    urlcolor={blue!80!black}
}

\usepackage[bitstream-charter]{mathdesign}
\DeclareSymbolFont{usualmathcal}{OMS}{cmsy}{m}{n}
\DeclareSymbolFontAlphabet{\mathcal}{usualmathcal}

\fancypagestyle{SPstyle}{
\fancyhf{}
\lhead{\colorbox{scipostblue}{\bf \color{white} ~SciPost Physics }}
\rhead{{\bf \color{scipostdeepblue} ~Submission }}

\fancyfoot[C]{\textbf{\thepage}}
}

\usepackage[utf8]{inputenc}
\usepackage[T1]{fontenc}
\usepackage{ifthen}
\usepackage{graphicx}
\usepackage{times}
\usepackage{babel}
\usepackage[dvipsnames]{xcolor}
\usepackage{bm}
\usepackage[normalem]{ulem}
\usepackage{float}
\usepackage{placeins}
\usepackage{chngcntr}
\usepackage{enumitem}
\usepackage{multirow}

\usepackage{comment}

\newcommand{\figwidth}{0.9\columnwidth}
\newcommand{\hfigwidth}{0.45\columnwidth}

\newcommand{\ifis}[2]{
\ifthenelse{\equal{#1}{}}{}{#2}
}

\def\be{\begin{equation}}
\def\ee{\end{equation}}
\def\bea{\begin{eqnarray}}
\def\eea{\end{eqnarray}}
\def\bsu{\begin{subequations}}
\def\esu{\end{subequations}}
\def\bi{\begin{itemize}}
\def\ei{\end{itemize}}

\newcommand{\op}[1]{\widehat{#1}}
\newcommand{\dagop}[1]{\widehat{#1}^{\dagger}}
\newcommand{\bo}[1]{{\mathbf{\bm{#1}}}}
\newcommand{\mc}[1]{{\mathcal{#1}}}
\newcommand{\wt}[1]{{\widetilde{#1}}}
\newcommand{\wb}[1]{{\overline{#1}}}

\newcommand{\nonu}{\nonumber}
\newcommand{\etal}{~\textsl{et al.}}

\newcommand{\ifi}[1]{\left\{\begin{array}{c@{\ \text{if} \ }l}#1\end{array} \right.}
\newcommand{\bra}[1]{\langle#1\vert}
\newcommand{\ket}[1]{\vert#1\rangle}

\newlength{\templength}

\newcommand{\leads}{\ensuremath{\quad\Longrightarrow\quad}}

\newcommand{\eqn}[1]{(\ref{#1})}

\newcommand{\eq}[2]{\begin{equation}\label{#1}#2\end{equation}}
\newcommand{\eqs}[2]{\begin{subequations}\label{#1}\begin{eqnarray}#2\end{eqnarray}\end{subequations}}
\newcommand{\eqa}[2]{\begin{eqnarray}\label{#1}#2\end{eqnarray}}
\renewcommand{\tanh}{{\rm tanh\,}}
\newcommand{\condi}[2]{{\begin{array}{c}{#1}\\[-0.5em]{\scriptstyle{#2}}\end{array}}}

\usepackage{lmodern}

\setlist[itemize]{nosep}
\setlist[enumerate]{nosep,nolistsep}

\newcommand{\pin}[1]{{\begin{color}[rgb]{1.0,0,0.667}{#1}\end{color}}}		
\newcommand{\org}[1]{{\begin{color}[rgb]{1.0,0.5,0}{#1}\end{color}}}		
		
\newcommand{\REM}[1]{\ifthenelse{0=1}{#1}{}}

\newcommand{\ve}{\varepsilon}

\begin{document}

\pagestyle{SPstyle}

\begin{center}{\Large \textbf{\color{scipostdeepblue}{
Lee-Huang-Yang dynamics 
emergent from a direct Wigner representation\\
}}}\end{center}

\begin{center}\textbf{
King Lun Ng\textsuperscript{1$\star$},
Maciej B. Kruk\textsuperscript{1} and
Piotr Deuar\textsuperscript{1$\dagger$}
}\end{center}

\begin{center}
{\bf 1} Institute of Physics, Polish Academy of Sciences, Aleja Lotnik\'ow
32/46, 02-668 Warsaw, Poland
\\[\baselineskip]
$\star$ \href{mailto:email1}{\small klng@ifpan.edu.pl}\,,\quad
$\dagger$ \href{mailto:email2}{\small deuar@ifpan.edu.pl}
\end{center}

\section*{\color{scipostdeepblue}{Abstract}}
\textbf{\boldmath{%
We demonstrate how the beyond-mean-field Lee-Huang-Yang (LHY) corrections and its related physics can be naturally incorporated into the representation of an ultracold Bose gas using the truncated Wigner approach without invoking effective energy terms or local density approximation. By generating a Bogoliubov ground-state representation with appropriately tailored bare interaction strength $g_{0}$ and condensate density $n_{0}$, the expected initial energy and densities are obtained while retaining access to quantum effects beyond the reach of the extended Gross-Pitaevskii equation (EGPE) formulation. This approach enables the study of correlations, coherence decay, single realisations, and the onset of quantum fluctuation effects with growing interaction strength. Numerical demonstrations for a weakly interacting single-component Bose gas show that observables deviate significantly from both the plain GPE and the EGPE incorporating LHY corrections. In regimes of strong interaction, many of the interference effects predicted by the GPE and EGPE are suppressed, and the EGPE offers no improvement over the plain GPE compared to the full Wigner model. In the weakly interaction limit, the EGPE appears accurate but resolving its deviation from mean-field results requires extensive ensemble averaging.
%
}}

\vspace{\baselineskip}

\noindent\textcolor{white!90!black}{%
\fbox{\parbox{0.975\linewidth}{%
\textcolor{white!40!black}{\begin{tabular}{lr}%
  \begin{minipage}{0.6\textwidth}%
    {\small Copyright attribution to authors. \newline
    This work is a submission to SciPost Physics. \newline
    License information to appear upon publication. \newline
    Publication information to appear upon publication.}
  \end{minipage} & \begin{minipage}{0.4\textwidth}
    {\small Received Date \newline Accepted Date \newline Published Date}%
  \end{minipage}
\end{tabular}}
}}
}


\vspace{10pt}
\noindent\rule{\textwidth}{1pt}
\tableofcontents
\noindent\rule{\textwidth}{1pt}
\vspace{10pt}







\section{Introduction}
\label{INTRO}
The Lee-Huang-Yang (LHY) correction to the Bogoliubov ground-state energy \cite{Lee57a,Lee57,Huang57} has been the focus of much study in recent years in the context of ultracold gases \cite{Anderson1995,Davis1995,Bradley1995}, particularly in the study of quantum droplets in Bose-Bose mixtures \cite{Petrov15,Petrov16} both experimentally \cite{Cabrera18,Cheiney18,Ferioli19,Semeghini18,DErrico19,Guo2021} and theoretically ( \!\!\!\cite{Astrakharchik18,Zin18,Zin22,Zin22b,Zin21,Zin21b,Tylutki20,Cikojevic19,Cikojevic21,Spada24} and many more), LHY fluids \cite{Skov21} and supersolids \cite{Leonard2017}.
Its growing importance is due to its involvement in the expanding range of currently studied phenomena in which beyond-mean-field effects play a decisive role, such as \cite{Yogurt22,Abdullaev2024}.

In most treatments, LHY corrections are introduced via an effective density-dependent mean-field energy term obtained from the thermodynamic limit of a uniform segment of the system. 
This additional term is incorporated into the Gross-Pitaevskii equation (GPE) under a local density approximation (LDA) to give an extended GPE (EGPE). 
While this approach has been remarkably successful in predicting the appearance of quantum droplets and some of their properties, it has clear limitations. For example, it neglects two-body correlations and quantum depletion effects in the zero temperature state, and it is not always justified if the density varies on short spatial or temporal scales. Moreover, the degree to which LHY-induced effects preserve the coherence assumed in the EGPE formulation remains uncertain. In low-density regions of a cloud, incoherent fluctuations can be locally more dominant than in high-density regions, further challenging the LDA-based picture.

Here our aim is to develop a more first-principles approach in which the LHY-mediated physics arises naturally, without invoking the local density and mean-field assumptions. This could in principle allow one to access the accuracy of the EGPE from various perspectives, and to treat highly nonuniform or nonequilibrium scenarios, including the breaking of the perfect coherence of the GPE and EGPE, the appearance of condensate fragmentation, as well as droplets at finite temperature and in single experimental realisations. On an underlying level, LHY physics -- and phenomena such as quantum droplets -- do not arise from a coherent effective LHY term, but instead emerge from the addition of many local and short-lived fluctuations. Therefore, it is conceptually valuable to represent these effects in detail and to study how this plays out.

In this context, we cannot omit to mention the recent work of Spada\etal \cite{Spada24}, who used a path integral Monte Carlo approach to generate a quantum droplet in two dimensions from first principles without relying on any LDA or EGPE. This work demonstrated the feasibility of such a line of inquiry and was able to provide substantial non-mean-field information on droplet properties. Earlier Monte Carlo investigations of uniform and three-dimensional (3D) droplet systems going beyond LHY include \cite{Spada23,Cikojevic18,Cikojevic19,Cikojevic21}. Our angle of attack here is different and complementary, we aim to develop an approach to representing LHY and droplet physics numerically that is applicable to dynamics and scales more easily to large systems than Monte Carlo methods (though the 16384 particles achieved by \cite{Spada24} is impressive). 

We employ the truncated Wigner approximation (TWA) \cite{Wigner1932} to represent the state of the system. While the TWA is a well-established method, we will see that an accurate description of the LHY energy contribution exceeds the standard TWA procedure. This is because we must deal in a numerically explicit way with the energy divergences that are normally somewhat opaquely renormalised only ``on paper'' (analytic) \cite{Lee57,Lee57a,Lifshitz1998,Huang1963,Weiss16}. 
As it turns out, carrying this out successfully also sheds light on how the renormalisation can be seen from an operational perspective. 
As the topic is less trivial than it may appear, in this work, we restrict ourselves to the single-component, repulsive, contact-interacting Bose gas. This is the fundamental building block for later extensions to attractive interacting gases or quantum droplets.

Although direct study of the quantum droplets in two-component Bose-Bose mixtures is not our present focus (we remain within the single-component gas), a direct representation of the bosonic field without introducing LHY terms by fiat has significant potential applications to the droplet case. 
For example, in quantum droplets the usual LHY term in the EGPE is not very well justified when external potentials sharper than a healing length are present, when jolts faster than the healing time take place, or in low-density regions in the tails and inside vortices. Knowing how to capture LHY physics without an explicit LHY term would, for instance, allow one to inspect how exactly the instability inherent in droplet excitations that was highlighted from the outset \cite{Petrov15} is ``cured'' on a microscopic scale, or to investigate the conditions under which the cure becomes ineffective. At nonzero temperature, it is not reasonable to assume the condensate profile remains smooth and amenable to the LDA, so a direct representation of the field is broader in applicability than the nonzero temperature modifications previously made to the Bose-Bose EGPE \cite{Ota20,Boudjemaa18a,Boudjemaa21,Guebli21,Boudjemaa17,Aybar19}. %
This is especially relevant when it comes to studying issues of droplet fragmentation or interactions between droplets. 

This is a long and careful paper, as required to properly explain and demonstrate the considerations and behaviour of LHY physics in the single realisation framework developed here.
It is organised as follows:
Sec.~\ref{BOG} presents the known starting framework and notation, and recalls the Bogoliubov description of a uniform interacting Bose gas in terms of the bare interaction, standard mean-field and LHY energies, as well as how a Bogoliubov ground state is implemented as a TWA ensemble for subsequent dynamical evolution.
Sec.~\ref{OBS} presents observable calculations using the bare interaction (including LHY energy) that explicitly account for a finite computational lattice and box size, and discusses the divergence problems that arise.  
Sec.~\ref{MATCH} then explains the proposed 
mechanism that matches a \emph{numerical} representation in terms of a modified bare interaction to the dressed mean-field, as opposed to the traditional ``on paper'' renormalisation (Sec.~\ref{g0steps}). An implementation of this procedure within the TWA is described in Sec.~\ref{cutoff}, which provides a numerical inclusion of the LHY physics into the system. 
Additional technical details on relevant small side issues are provided in the appendices.
Finally, in Sec.~\ref{DYN} we present numerical examples of nonequilibrium dynamics and, in particular, examine how well the EGPE deals with effects beyond the standard mean-field. 

We will go into the above topics in some detail in order to establish a well described baseline on which future studies of other systems—such as attractive gases, droplets, or supersolids—can be easily built.

\section{Basics of the Bogoliubov and TWA descriptions}
\label{BOG}

\noindent The implementation procedure we have in mind is to:
\begin{itemize}
\item calculate Bogoliubov quantities; 
\item sample the resulting field configurations using the Wigner representation; 
\item then calculate observables from the Wigner samples, either immediately or after some period of time evolution.
\end{itemize}
To grasp how to implement LHY physics properly in a numerical implementation of the Wigner representation, we need first to understand how the LHY physics appears in terms of the bare interaction and to inspect the associated divergence problems.

\subsection{Hamiltonian}

The standard Hamiltonian for an interacting one-component Bose gas in $d$-dimensional free space with coordinates $\bo{x}$ is written in 2nd quantisation as
\begin{equation} \label{H1D}
\hat{H} = \int\!d^d\bo{x}\,\dagop{\Psi}(\bo{x})\left[
-\frac{\nabla^2}{2} + \int\!d^d\bo{x}'\,\dagop{\Psi}(\bo{x}')\frac{U(\bo{r})}{2}\op{\Psi}(\bo{x}')
\right]\op{\Psi}(\bo{x}),
\end{equation}
where $\hbar=m=1$ has been assumed. Here $\op{\Psi}(\bo{x})$ is the usual Bose field operator, and the inter-particle potential $U$ depends only on $\bo{r}=\bo{x}-\bo{x}'$. 

In anticipation of implementation on a numerical lattice, let us immediately convert to a discretised space with volume element $\Delta v=(\Delta x)^d$ per lattice point and box length $L$ (volume $V=L^d$). The Hamiltonian may then be written approximately as the sum of kinetic and interaction terms, $\op{H}\approx\op{H}_{\rm kin}+\op{H}_{\rm int}$, and the interaction energy is 
\eq{Hint}{
\op{H}_{\rm int} = \frac{(\Delta v)^2}{2}\sum_{\bo{x},\bo{x}'} \dagop{\Psi}(\bo{x})\dagop{\Psi}(\bo{x}')U(\bo{r})\op{\Psi}(\bo{x}')\op{\Psi}(\bo{x}),
}
while the kinetic energy is conveniently represented in momentum $\rm{k}$-space using the Fourier transformed field
\eq{Psik}{
\op{\Psi}(\bo{k}) = \frac{\Delta v}{(2\pi)^{d/2}}\sum_{\bo{x}}\, e^{-i\bo{k}\cdot\bo{x}}\,\op{\Psi}(\bo{x}).
}
The $\rm{k}$-space has $d$ dimensions spanned by a lattice of values with spacing $\Delta k=2\pi/L$ and maximum values $|\bo{k}|\le k_{\rm max}=\pi/\Delta x$.
This choice of normalisations ensures that the $\rm{k}$-space field is also a density, and the total number of particles is
\eq{Nop}{
\op{N} = \Delta v\sum_{\bo{x}} \dagop{\Psi}(\bo{x})\op{\Psi}(\bo{x}) = (\Delta k)^d\sum_{\bo{k}} \dagop{\Psi}(\bo{k})\op{\Psi}(\bo{k}).
}
We are approximating a continuum field on a numerical lattice, not a true lattice system, so Fourier transforms and a parabolic spectrum are the appropriate discretisation, rather than a finite difference approach. Writing $k^2=|\bo{k}|^2$, the kinetic energy becomes 
\eq{Hkin}{
\op{H}_{\rm kin} =(\Delta k)^d\sum_{\bo{k}} \frac{k^2}{2}\dagop{\Psi}(\bo{k})\op{\Psi}(\bo{k}).
}
For the purpose of understanding renormalisation and regularisation within the Bogoliubov approximation, it is in fact convenient to consider the $\rm{k}$-space representation of the interaction term: 
\eq{Hintk}{
\op{H}_{\rm int} = \frac{(\Delta k)^{2d}}{2V}\sum_{\bo{k},\bo{k}',\bo{k}''}g_{\bo{k}}\, \dagop{\Psi}(\bo{k}')\dagop{\Psi}(\bo{k}'')\op{\Psi}(\bo{k}''+\bo{k})\op{\Psi}(\bo{k}'-\bo{k}),
}
where the $\rm{k}$-space interaction strength is
\eq{gk}{
g_{\bo{k}} = \Delta v\sum_{\bo{r}} e^{-i\bo{k}\cdot\bo{r}}\, U(\bo{r}),
}
with reverse transform
\eq{gx}{
U(\bo{r}) = \frac{1}{V}\sum_{\bo{k}}e^{i\bo{k}\cdot\bo{r}}g_{\bo{k}} 
\approx \frac{1}{(2\pi)^d}\int d^d\bo{k} \,e^{i\bo{k}\cdot\bo{r}}\,g_{\bo{k}}.
}
In the uniform system, the Bogoliubov modes and frequencies will turn out to depend only on $g_{\bo{k}}$.
The typical procedure then approximates the true potential by a contact delta-interacting substitute,
\eq{deltag}{
U(\bo{r}) \rightarrow g_0\delta(\bo{r}) \approx \ifi{g_0/\Delta v, & \bo{r}=0, \\0, & \bo{r}\neq 0,}
}
which implicitly assumes that the range of $U(\bo{r})$ is smaller than $\Delta x$, 
and one has $g_{\bo{k}}=g_0$ for all $\bo{k}$. 


\subsection{Bogoliubov description}
\label{BOGDES}
Following the Bogoliubov method \cite{Bogoliubov47}, the field operator is split into a linear combination of a condensate and a fluctuation component:
\eq{BOGpsi}{
\op{\Psi}(\bo{x})=\phi(\bo{x})\,\op{a}_{0}+\delta\op{\Psi}(\bo{x}),
}
with the normalisation 
$\Delta v\sum_{\bo{x}}|\phi(\bo{x})|^2 = 1$,
so that the condensate occupation is $N_0=\langle\dagop{a}_0\op{a}_0\rangle$.
We work in a formulation
in which the fluctuations (except for shot noise in the condensate) have support 
only in modes orthogonal to the condensate, following approaches in Refs. \cite{Castin98,Leggett01,Sinatra02}. We also adopt the convention of a minus on the $v^*$ term as per \cite{Steel98,Isella06,FINESS-Book-Ruostekoski,Weiss04}. The fluctuation field is then expanded as:
\eq{dpsimj}{
\delta\op{\Psi}(\bo{x}) = \sum_{j>0}\left[ u_j(\bo{x}) \op{b}_j - v_j^*(\bo{x}) \dagop{b}_j\right], 
}
with the quantum number $j=1,2,\dots$ counting the excited single-particle states.

Two assumptions are made. The first is that $\delta\op{\Psi}$ is small compared to $\phi\,\op{a}_0$, allowing one to neglect all cubic and quartic terms in $\delta\op{\Psi}$ in the Hamiltonian. A second less obvious assumption is that $\phi$ is static, so that one can also omit all the $\mc{O}(\delta \op{\Psi})$ terms, because they vanish 
when $\phi(\bo{x})$ is a solution of the Gross-Pitaevskii equation. 
This leads to
\eq{Hbogbog}{
\op{H}_{\rm Bog} = E_0 + \sum_{j>0} \ve_j\dagop{b}_{j}\op{b}_{j},
}	
which is an ideal gas of quasiparticles $\op{b}_j$ above the Bogoliubov ground state $\ket{B}$, defined by the properties $\op{b}_j\ket{B} = 0$ and $\op{a}_0\ket{B} = \sqrt{N_0}\ket{B}$.
The c-number contribution $E_0$ is closely related to the LHY energy.
The inverse transformation of the quasiparticle operator is
\eq{bkakpx}{
\op{b}_j = \sum_{\bo{x}} \Delta v \left[u^*_j(\bo{x})\delta\op{\Psi}(\bo{x}) + v^*_j(\bo{x})\delta\dagop{\Psi}(\bo{x})\right],
}
and the discretised orthogonality conditions \cite{Castin98}:
\eqs{uu-vvx}{
\sum_{\bo{x}}\Delta v \left[u^*_j(\bo{x})u_{j'}(\bo{x})-v^*_j(\bo{x})v_{j'}(\bo{x})\right]&=&\delta_{jj'},\\
\sum_{\bo{x}} \Delta v \left[v_j(\bo{x})u_{j'}(\bo{x})-u_j(\bo{x})v_{j'}(\bo{x})\right]&=&0.
}

In the particular case of a uniform gas in a box with periodic conditions, \eqn{bkakpx} can be cast into the familiar plane wave form
\eq{dpsim}{
\delta\op{\Psi}(\bo{x}) = \frac{1}{\sqrt{V}}\sum_{\bo{k}\neq 0}\left[ U_{\bo{k}}e^{i\bo{k}\cdot\bo{x}} \op{b}_{\bo{k}} - V^*_{\bo{k}}e^{-i\bo{k}\cdot\bo{x}} \dagop{b}_{\bo{k}}\right], 
}
with a sum over all nonzero $\bo{k}$ values that appear on the numerical lattice. The mode functions are 
\begin{equation} \label{uv_plane}
u_j(\bo{x})=\frac{U_\bo{k}}{\sqrt{V}}e^{i\bo{k}_j\cdot\bo{x}};\qquad v_j(\bo{x})=\frac{V_{\bo{k}}}{\sqrt{V}}e^{i\bo{k}_j\cdot\bo{x}},
\end{equation}
with $j$ labelling the allowed momentum vectors $\bo{k}_j$ on the lattice.
The $\rm{k}$-space annihilation operator, $\op{a}_{\bo{k}} = \delta\op{\Psi}(\bo{k})(\Delta k)^{d/2}$, is given by
\eq{akp}{
\op{a}_{\bo{k}} = \frac{\Delta v}{\sqrt{V}}\sum_{\bo{x}}\ e^{-i\bo{k}\cdot\bo{x}}\,\delta\op{\Psi}(\bo{x})
=U_{\bo{k}}\op{b}_{\bo{k}}-V^*_{-\bo{k}}\dagop{b}_{-\bo{k}},
}
and the identities become
\eq{uu-vv}{
U_{\bo{k}}U^*_{\bo{k}'}-V_{\bo{k}}V^*_{\bo{k}'}=\delta_{\bo{k},\bo{k}'}\ ,\qquad
U_{\bo{k}}V_{\bo{k}'}-V_{\bo{k}}U_{\bo{k}'}=0\ ,\qquad
\op{b}_{\bo{k}} = U^*_{\bo{k}}\op{a}_{\bo{k}} + V^*_{\bo{k}}\dagop{a}_{-\bo{k}},
}
where $U_{\bo{k}}=U_{\bo{k}}^*=U^*_{-{\bo{k}}}$ and $V_{\bo{k}}=V_{\bo{k}}^*=V^*_{-{\bo{k}}}$.
The dimensionless Bogoliubov coefficients 
follow the simple analytical expressions:
\begin{equation}\label{BOG_uv}
U_{\bo{k}} = \frac{\varepsilon_{\bo{k}}+E_{\bo{k}}}{2\sqrt{\varepsilon_{\bo{k}}E_{\bo{k}}}},\qquad 
V_{\bo{k}} = \frac{\varepsilon_{\bo{k}}-E_{\bo{k}}}{2\sqrt{\varepsilon_{\bo{k}}E_{\bo{k}}}},
\end{equation}
where the free-particle energy is 
\eq{Ek=}{
E_{\bo{k}}=\frac{k^{2}}{2},
}
and the excitation energy 
is
\eq{epsilon}{
\varepsilon_{\bo{k}}=\sqrt{E_{\bo{k}}(E_{\bo{k}}+2g_{\bo{k}}n_0)}
= \frac{k}{2}\sqrt{k^2+4g_{\bo{k}}n_0} \ge E_{\bo{k}}.
}
Here the condensate density is $n_0=N_0|\phi|^{2}$.

In this work we will restrict ourselves to the uniform system with periodic boundary conditions when constructing the Wigner representation. In this uniform system, the natural length unit is the healing length 
\eq{heal=gen}{
\xi = \frac{\hbar}{\sqrt{mgn}},
}
with corresponding chemical potential $\mu=gn$.
For definiteness, in Secs.~\ref{BOG} and~\ref{OBS} we define units using the bare interaction and condensate density,
\eq{heal=gen0}{
\xi_0 = \frac{\hbar}{\sqrt{mg_0n_0}},\qquad\qquad \mu_0=g_0n_0,
}
while in Secs.~\ref{MATCH} and~\ref{DYN}, we move to using preset quantities $g_{\rm set}$ and $n_{\rm set}$, satisfying $g_{\rm set}n_{\rm set}=1$, for which $g_0$ and $n_0$ are varied and optimised to achieve matching.
%
%
Under the rescaling
\eqs{utrans}{
x\to\frac{x}{\xi_0},\
k\to k\xi_0,\
\ve\to\frac{\ve}{\mu_0},\ 
n\to n\xi_0^d,\ 
\op{\Psi}\to\op{\Psi}\xi_0^{d/2},\ 
v_j\to v_j\xi_0^{d/2},\\ 
\qquad\text{and}\qquad t\to\frac{\mu_0 t}{\hbar},\
k_{\rm max}\to k_{\rm max}\,\xi_0, \quad\quad\quad\quad\quad\quad\quad
}
where $t$ is the time and $k_{\rm max}$ is the momentum cutoff in the numerical box, the excitation spectrum becomes
\eq{epsilonu}{
\varepsilon_{\bo{k}}= \frac{k}{2}\sqrt{k^2+4g_{\bo{k}}/g_0} \condi{\leads}{g_{\bo{k}}=g_0} \frac{k}{2}\sqrt{k^2+4}.
}
The latter concerns the contact interacting gas. 
This is the only case considered hereafter.



\subsection{Mean-field energy}
For a uniform short-range interacting gas with coupling constant $g$, the zero-temperature 
``mean-field'' energy due to contact interactions is considered to be
\eq{Emf}{
E_{\rm mf} = \frac{g}{2} nN = \frac{g V (n_0+\delta n)^2}{2},
}
given the quantum depletion density $\delta n=(\Delta v/V)\sum_{\bo{x}}\langle\delta\dagop{\Psi}(\bo{x})\delta\op{\Psi}(\bo{x})\rangle$, and the 
total density is $n=n_0+\delta n$. 
%
\REM{
\org{
If our numerical lattice were to properly resolve $U(\bo{r})$ then the expectation of the interaction energy \eqn{Hint} can be written $\langle\op{H}_{\rm int}\rangle = 
\frac{V}{2}\,\Delta v\sum_{\bo{r}}\, U(\bo{r}) G_2(\bo{r})$ with the two body correlation $G_2(\bo{r})=\langle \dagop{\Psi}(\bo{x})\dagop{\Psi}(\bo{x}+\bo{r})\op{\Psi}(\bo{x}+\bo{r})\op{\Psi}(\bo{x})\rangle$, which does not depend on position $\bo{x}$ in a uniform system.} The mean-field assumption of no explicit two-body correlations \org{then sets $G_2(\bo{r})=n^2$, and one would obtain:}
\eq{Emfres}{
E_{\rm res}^{\rm mf} = \frac{Nn}{2}\sum_{\bo{r}}\Delta v\,U(\bo{r}) = \frac{g_{\bo{k}=0} V (n_0+\delta n)^2}{2},
}
\org{where \eqn{gk} has been used on the right to identify $g=g_{\bo{k}=0}$. Taking this naively, \eqn{deltag} could then imply $g=g_0$}
}

There is some subtlety regarding how the value of $g$ used here corresponds to the underlying interaction potential $U(\bo{r})$. A typical quantum gas exhibits a separation of length scales such that the actual range of the interparticle potential, $r_0$, is far smaller than any other relevant length scales. This makes a simulation with $\Delta x\lesssim r_0$  that would resolve the potential impractical, not to mention that the exact form of the potential is often not very well known. Instead, 
a simplified contact interaction \eqn{deltag} is introduced into the microscopic model.

In this case it is widely known that the simple identification $g=g_0$ should not be used. The purpose of a mean-field energy such as \eqn{Emf} is to incorporate all interaction effects below the resolution of the model into the value of $g$, whereas the effects of the contact potential \eqn{deltag} on the energy turn out to depend on the momentum cutoff, which is chosen arbitrarily by the simulator. These issues will be discussed in Sec.~\ref{OBS}. 

From the experimental side, the value of $g$ is 
obtained phenomenologically through observations of scattering processes or by comparing the behaviour of the cloud to a mean-field evolution with a given $g$. From the modelling side, when going beyond the mean-field description, one needs to choose a ``bare'' interaction strength $g_0$ that 
reproduces the experimental phenomenology described by the ``dressed'' $g$. 
In 
3D, a nontrivial renormalisation procedure is used to relate these quantities, 
as is explained in detail in 
Ref.~\cite{Weiss04}. Basically, in addition to the bare $s$-wave scattering length $a_s$ contained in $g_0=4\pi\hbar^2a_s/m$, one incorporates the 2nd Born correction to the scattering length into the mean-field coupling constant $g$. As per:
\eq{g=g1g2}{
g = g_0+g_1,
}
where the contribution of the 2nd term in the Born expansion is
\eqa{g1Weiss}{
g_1 = -\frac{1}{V}\sum_{\bo{k}\neq0} \frac{g_{\bo{k}}^2}{k^2}.
}
This leads to a decomposition of the mean-field energy as 
\eq{Emf01}{
E_{\rm mf} = E_{\rm mf}^{\rm bare} + E_{\rm mf}^{\rm dress},\qquad\text{where}\qquad
E_{\rm mf}^{\rm bare}=\frac{g_0n^2V}{2},\quad 
E_{\rm mf}^{\rm dress}=\frac{g_1n^2V}{2}.
}
For later use it is helpful to note that the bare mean-field energy $E_{\rm mf}^{\rm bare}$ can be further 
decomposed into a condensate contribution (the most reduced baseline for ground state energy) and a fluctuation contribution: 
\eq{E0mf}{
E_{\rm mf}^{\rm bare} = E_{\rm mf}^0+E_{\rm mf}^{\delta n},\qquad\text{with}\qquad
E_{\rm mf}^{0} = \frac{g_0n_0^2V}{2},\quad
E_{\rm mf}^{\delta n}= \frac{g_0V}{2}\delta n\left(2n_0+\delta n\right).
}

A caveat regarding the explicit use of  $g_1$ is that its ultraviolet (UV) divergence 
in 3D makes it 
suitable for ``on paper'' renormalisation, but less so for numerical implementation. The solution of this practical issue is addressed in Sec.~\ref{MATCH}. 
\subsection{LHY correction}
\label{LHYMF}
The LHY energy term is an additional correction to the mean-field ground-state energy arising from the quantum fluctuations described by Bogoliubov theory, such that the total ground-state energy is 
\eq{Etotgs}{
E_0 = E_{\rm mf}+E_{LHY}^{\infty}.
} 
A naive application of Bogoliubov theory leads to divergent values for this correction. For dilute Bose gases in 3D with hard-sphere interactions, Lee, Huang and Yang employed a corrected pseudopotential for the interaction between particles, leading to a convergent integral for the ground-state energy over all momenta \cite{Lee57a,Lee57,Huang57}. The renormalisation can be formulated in terms of the first and second Born approximations without recourse to specific potentials such as hard spheres \cite{Brueckner57,Weiss04}.
The values of the LHY energy
in three-, two-, and one-dimensional systems (3D, 2D, and 1D, respectively), after renormalisation, are accepted to be \cite{Lee57,Lee57a,Weiss04,Rakshit19b}:
\eqs{LHY}{
\frac{E_{LHY}^{\infty}(g_0,n_0,k_c)}{V} =& \frac{m^{3/2}}{\hbar^3}(g_0n_0)^{5/2}\,\frac{8}{15\pi^2}\qquad\quad\qquad&{\rm(3D)},\label{LHY3d}\\
=& -\frac{m}{\hbar^2}\frac{(g_0n_0)^2}{8\pi}\ \log\left[\frac{\hbar^2k_c^2}{mg_0n_0\sqrt{e}}\right]\quad\ &{\rm(2D)},\qquad\label{LHY2d}\\
=& -\frac{\sqrt{m}}{\hbar}(g_0n_0)^{3/2}\,\frac{2}{3\pi}\qquad\quad\qquad\ &{\rm(1D)}.\label{LHY1d}
}
These expressions are subject to the choice of whether one uses the condensate density $n_0$ or the total density $n$, and the bare coupling $g_0$ or the dressed coupling $g$, in the prefactors. While the formal renormalisation structure of effective field theory strictly distinguishes bare and dressed quantities, different literature sources adopt different conventions in these prefactors. The difference is usually minor because it lies at a higher order of accuracy than Bogoliubov approach, see Appendix~\ref{BAREDRESS} and Figs.~\ref{fig:Ekc}(d-f). We will operationally take preset dressed values $g_{\rm set}$ and $n_{\rm set}$ in Sec.~\ref{MATCH}.

The original treatment of the LHY correction was developed for a homogeneous gas under the assumption that the local density approximation (LDA) is valid. For a non-homogeneous gas which the variation of the density profile occurs on the scale of the healing length, the breaking of the LDA leads to a modification of the LHY energy. Such effects have also been quantified, for example, for quantum droplets in Bose-Bose mixtures\cite{Zin22}.


\subsection{Truncated Wigner representation}
\label{WIG1}

The representation of a Bogoliubov state in equilibrium at temperature $T$ has been derived in several forms \cite{Steel98}, here we use the convention from the formulation of Ruostekoski\etal \cite{Isella06,FINESS-Book-Ruostekoski}, in which
\eq{Rusto-wigj}{
\Psi_{\rm W}(\bo{x}) = \phi_0(\bo{x})\beta_0 + \sum_{j>0} \left[u_j(\bo{x})\beta_j-v_j^*(\bo{x})\beta_j^*\right],
}
where $\beta_j$ are Gaussian complex noises distributed according to 
\eq{betaTj}{
W(\beta_j) \propto \exp\left[-2|\beta_j|^2\tanh\left(\frac{\ve_j}{2k_BT}\right)\right],
}
i.e., a Gaussian distribution with zero mean and standard deviation 
\eq{sigmaT}{
\sigma=\frac{1}{2\sqrt{\tanh\left(\ve_j/(2k_BT)\right)}}.
}
The expression \eqn{Rusto-wigj} can be compared to \eqn{dpsimj} to identify $\beta_j\sim\op{b}_j$.
In the $T\to0$ limit, the distribution becomes
\eq{betaTj0}{
W(\beta_j) =\frac{2}{\pi} \exp\left(-2|\beta_j|^2\right),
}
with standard deviation $\tfrac{1}{2}$ in the real and imaginary parts, so that $\langle|\beta_j|^2\rangle=\tfrac{1}{2}$ for all modes.


In the case of a uniform gas, \eqn{Rusto-wigj} simplifies to the plane-wave form of excited states, with $\ve_j=\ve_{\bo{k}\neq0}$ in all cases. We specifically use
\eq{Rusto-wig}{
\Psi_{\rm W}(\bo{x}) = 
\phi_0(\bo{x})+\frac{\wt{\beta}_0}{\sqrt{V}} + \frac{1}{\sqrt{V}}\sum_{\bo{k}\neq 0} \left[U_{\bo{k}}\beta_{\bo{k}}e^{i\bo{k}\cdot\bo{x}}-V_{\bo{k}}^*\beta_{\bo{k}}^*e^{-i\bo{k}\cdot\bo{x}}\right].
}
Here we make one modification compared to \cite{FINESS-Book-Ruostekoski}, which is to take a coherent condensate
rather than the more involved expression found there, in order to allow for a symmetry-broken initial condition for dynamics in Sec.~\ref{DYN}. That is, the distribution of $\wt{\beta}_0$ is also given by \eqn{betaTj0}. 
The excited-state population is 
\eq{NcRusto}{
N_{\rm ex} = \sum_{\bo{k}\neq0}\left[ \left(|U_{\bo{k}}|^2+|V_{\bo{k}}|^2\right)\left(|\beta_{\bo{k}}|^2-\frac{1}{2}\right) + |V_{\bo{k}}|^2\right].
}

In the TWA, remarkably, evolution follows essentially the standard Gross-Pitaevskii equation (GPE)\cite{Steel98}:
\eq{TWAGPE}{
i\frac{d\Psi_{\rm W}(\bo{x})}{dt} = \left[-\frac{\nabla^2}{2} + V(\bo{x}) + g\left(|\psi_{\rm W}(\bo{x})|^2-\frac{1}{\Delta v}\right)\,\right]\Psi_{\rm W}(\bo{x}).
}
In cases where a restricted subspace of modes $\op{\Psi}(\bo{x})=\sum_i\phi_i(\bo{x})\op{a}_i$ is desired, 
a slight modification to a projected form of the equation is required \cite{Davis01c,Davis01b,Blakie05}. 
This is, for example, appropriate for plane-wave modes $\phi_{\bo{k}}=e^{i\bo{k}\cdot\bo{x}}/\sqrt{V}$ in 3D space, where a spherical UV cutoff $k_c$ is imposed to retain only momenta with $|\bo{k}|\le k_c$. 
For the TWA, such a projection takes the form:
\eq{TWAPGPE}{
i\frac{d\Psi_{\rm W}(\bo{x})}{dt} = \mc{P}\left\{\left[-\frac{\nabla^2}{2} + V(\bo{x}) + g\Big(|\psi_{\rm W}(\bo{x})|^2-\delta_{\mc{P}(\bo{x})}\Big)\,\right]\Psi_{\rm W}(\bo{x})\right\}.
}
Here, $\mc{P}$ is a projector onto the desired subspace, ensuring that the evolution remains within the designated mode space, while
\eq{deltaP}{
\delta_{\mc{P}}(\bo{x}) = \sum_j|\phi_j(\bo{x})|^2 \to\frac{M_c}{M}\,\frac{1}{\Delta v}.
}
The final expression, independent of $\bo{x}$, applies to a plane-wave subspace containing $M_c$ modes, compared to $M=V/\Delta v$ in the full ``square $\rm k$-space lattice'' case. For an unrestricted subspace, $\delta_{\mc{P}}$ takes the familiar value of $1/\Delta v$, as expected from a discretised delta function.

Expectation values of observables in the TWA must be calculated with greater care than in classical fields, namely by evaluating the ensemble mean of the appropriate Weyl symbols \cite{Steel98}. In particular, 
the local density $n_{\rm W}(\bo{x})$, kinetic energy, local two-body correlation, interaction energy for a contact interaction \eqn{deltag}, and normalised pair-correlation are, respectively:
\eqs{TWobs}{
\left\langle\dagop{\Psi}(\bo{x})\op{\Psi}(\bo{x})\right\rangle &=& \left\langle |\Psi_{\rm W}(\bo{x})|^2-\frac{\delta_{\mc{P}}(\bo{x})}{2}\right\rangle=n_{\rm W}(\bo{x}),\label{TWAn}\\
H_{\rm kin} &=& (\Delta k)^d\sum_{\bo{k}} \frac{k^2}{2}\left\langle|\Psi_{\rm W}(\bo{k})|^2-\frac{1}{2(\Delta k)^d}\right\rangle,\qquad\label{TWAHkin}\\
G_2(\bo{x}) &=& \left\langle\dagop{\Psi}(\bo{x})\dagop{\Psi}(\bo{x})\op{\Psi}(\bo{x})\op{\Psi}(\bo{x})\right\rangle  = \left\langle |\Psi_{\rm W}(\bo{x})|^4-2|\Psi_{\rm W}(\bo{x})|^2\delta_{\mc{P}}(\bo{x})+\frac{\delta_{\mc{P}}(\bo{x})^2}{2}\right\rangle,\qquad\label{TWAG2}\\
H_{\rm int} &=& \frac{g_0}{2} \sum_{\bo{x}} \Delta v\, G_2(\bo{x}),\label{TWAHint}\\
g^{(2)}_{\rm W}(\bo{x}) &=& \frac{\left\langle |\Psi_{\rm W}(\bo{x})|^4-2|\Psi_{\rm W}(\bo{x})|^2\delta_{\mc{P}}(\bo{x})+\frac{\delta_{\mc{P}}(\bo{x})^2}{2}\right\rangle}{n_{\rm W}(\bo{x})^2},\qquad\label{TWAg2}
}
where $\Psi_{\rm{W}}(\bo{k})$ follows the prescription \eqn{Psik}.

Both the evolution equations and the observable calculations \eqn{TWAGPE} and (\ref{TWobs}) make no distinction between contributions arising from the condensate ($j=0$ mode) and those from excited modes. Therefore, contributions of different orders of $\delta\Psi$ are not separated as the Bogoliubov description. In particular, $G_2$ and $H_{\rm int}$ involve terms of order $\mc{O}(\delta\Psi)^3$ and $\mc{O}(\delta\Psi)^4$. 

This is both an advantage and a drawback. 
The advantage is that dynamical solutions of \eqn{TWAGPE} are not limited to small condensate depletions, nor to the stationarity condition used in the Bogoliubov approximation. A practically full nonlinear dynamics can be obtained, modulo some inaccuracies related to the Wigner truncation, which can be large or small depending on the circumstances and evolution time \cite{Sinatra02}.
The drawback, at least for our purposes here, is that the interaction energy calculated via \eqn{TWAHint} will never exactly match the interaction energy in the corresponding Bogoliubov approximation. This is because the Bogoliubov energy removes $\mc{O}(\delta\Psi)^3$ and $\mc{O}(\delta\Psi)^4$ contributions by hand, whereas the observable calculations \eqn{TWAHint} performed on the Wigner fields cannot. In the initial state, one could still attempt to remove the relevant contributions manually, but this ceases to be possible once the evolution begins, especially if there is an evolution of the condensate. These higher-order contributions are responsible for the bulk of the ``extra'' terms introduced in the next section.




\section{Observables and LHY physics from bare contact interaction on a numerical lattice}
\label{OBS}

It is well known that the canonical value of the LHY correction to the mean-field energy of the system does not correspond to a simple sum of the excitation energies obtained through the 
Bogoliubov approximation applied to a bare contact interaction.
However, such a sum is all we will be able to perform once the system is represented as a numerically generated ensemble on the computational lattice. To address this issue, let us first examine how the 
observables obtained from the simple sum come out. Our initial aim is to understand what values one can expect to obtain within the Wigner representation. Therefore, where relevant, we retain higher-order terms that are absent in a self-consistent Bogoliubov treatment but implicitly present when evaluating observables from a numerical field representation, in which the distinction between the condensate field $\phi$ and the fluctuation field $\delta\op{\Psi}$ has been lost.

Analytical estimates, in the limit of a large simulation box, 
can be written in terms of the maximum (cutoff) momentum available on the lattice (in all directions)
\eq{kc}{
k_c = \frac{\pi}{\Delta x}.
}
We assume that in two and three dimensions, an isotropic momentum cutoff, $|\bo{k}|\le k_c$, is implemented, so as to exclude the spurious ``corner'' modes. Low-dimensional systems can also suffer from infrared (IR) divergences. To quantify this effect, when necessary, in the form of integral estimates 
we introduce a lower limit to the integration over $k=|\bo{k}|$, denoted $k_L$
such that 

\eqa{continuum}{
\sum_{\bo{k}\neq 0} \approx \frac{V}{(2\pi)^d}\int_{k\ge k_L}^{k\le k_c} d^d\bo{k}.
\label{continuum_k}
}
For an isotropic integrand, the integration measure becomes
\eqa{ddk}{
d^d\bo{k}  =  
\left\{
\begin{array}{c@{\qquad}l}
 2\,dk     &  {\rm(1D)},\\
 2\pi k\, dk     &  {\rm(2D)},\\
 4\pi k^2\, dk     &  {\rm(3D)}.
\end{array}
\right.
\label{ddk_int}
}
In practice, one finds that using
\eq{kL=}{
k_L = \frac{\Delta k}{2} = \frac{\pi}{L} 
}
provides a reasonable match to the results of the discretised sum. In particular, this choice yields a significantly better agreement than alternative such as $k_L=\Delta k$ or $k_L\ll\Delta k$. Hence, \eqref{kL=} will be used for all integral estimates. Throughout this Sec.~\ref{OBS}, we consider the contact-interacting gas with momentum-independent coupling $g_{\bo{k}}=g_0$ as per \eqn{deltag} and we employ dimensionless units \eqn{utrans} in which $g_0n_0=1$.

\subsection{Density}
The total density of a Bose gas in the Bogoliubov ground state $\ket{B}$ is 
\eqs{n1d}{
n = \bra{B}\dagop{\Psi}(\bo{x})\op{\Psi}(\bo{x})\ket{B} &=& n_0+\delta n\label{deltan}\\
&=& n_0 + \sum_j |v_j(\bo{x})|^2\\
&=& n_0 + \frac{1}{V}\sum_{\bo{k}\neq0}|V_{\bo{k}}|^2.\label{deltandisc}
}

The integral estimation \eqref{continuum} for the depletion density $\delta n$ encounters an IR divergence in 1D, namely
\eqa{Nkmkm}{
\delta n &=&  \frac{1
}{2\pi}\left[k_L-k_c-\sqrt{4+k_L^2}+\sqrt{4+k_c^2}+\sinh^{-1}\left(\frac{2}{k_L}\right)-\sinh^{-1}\left(\frac{2}{k_c}\right)\right].
}
In the limit $k_c\to\infty$, this reduces to the expression
\eq{Nkamin}{
\delta n \condi{\leads}{k_c\to\infty} \frac{1}
{2\pi}\left[k_L-\sqrt{4+k_L^2}+\log\left(\frac{2+\sqrt{4+k_L^2}}{k_L}\right)\right],
}
and, for a large box where $k_L\to0$, to 
\eqa{Nkaminz}{
\delta n \condi{\leads}{k_L\to0} \frac{1
}{2\pi}\left[\log\left(\frac{4}{k_L}\right)-2\right].
\label{Nkaminz_1d}}

To gauge the accuracy of these estimates, consider a 1D test case with $\Delta k=0.1$, $g_0=0.2$, $n_0=5$, $k_c=15$. The discrete estimate \eqn{deltandisc} yields $\delta N=L\delta n = 23.156$ which agrees fairly well with the integral estimate \eqn{Nkmkm} when $k_L=\Delta k/2=\pi/L$, for which $\delta N=24.314$. The dependence of $\delta N$ on $k_c=15$ is of the order of $0.002$, while the large-box limit \eqn{Nkaminz} gives $\delta N=23.820$. The dependence on small $k_L$ is logarithmic, and for the above example $k_L=0.0565=0.565\Delta k$ reproduces the exact depletion \eqn{deltan}. However, the ``optimal'' value of $k_L/\Delta k$ for the integral estimates depends slightly on the lattice details and on the observable considered.

In 2D the analogous depletion estimate is
\eqa{N2d}{
\delta n &=& \frac{
k_c}{8\pi}\left(\sqrt{k_c^2+4}-k_c\right) \condi{\leads}{k_c\gg1} \frac{1}
{4\pi}\left(1-\frac{1}{k_c^2}\right),\qquad
}
with the approximation valid for $k_c\gg1$. 
Finally, in 3D
\eqa{N3d}{
\delta n &=& \frac{1}
{12\pi^2}\left[(k_c^2-2)\sqrt{k_c^2+4}+4-k_c^3\right] \condi{\leads}{k_c\gg1} \frac{1}
{3\pi^2}\left(1-\frac{3}{2k_c}\right).
}


\subsection{Kinetic energy}
\begin{figure}
\centering\includegraphics[width=0.45\columnwidth]{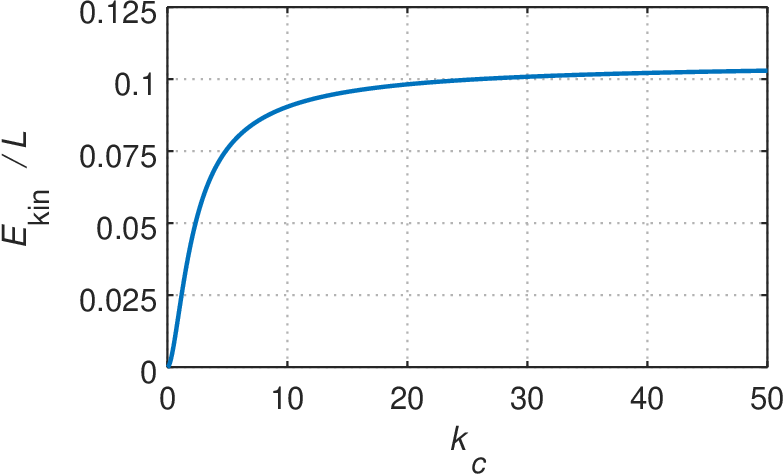}
\caption{Kinetic energy dependence on the scaled cutoff $k_c=k_{\rm max}\xi_0$ in 1D.}
\label{fig:ekinkc}
\end{figure}

The mean kinetic energy of the Bogoliubov ground state, $E_{\rm kin}=\bra{B}\op{H}_{\rm kin}\ket{B}$, is
\eqs{Hke_v}{
E_{\rm kin}&=& -\frac{1}{2}\sum_j \int d^d\bo{x}\ v_j^*(\bo{x})\nabla^2 v_j(\bo{x})\\
&=& \frac{1}{2}\sum_{\bo{k}}|V_{\bo{k}}|^2\,k^2\quad\text{(uniform case).}\label{sumkekin}
}
In 1D, in the large-box limit, it is approximated by the expression
\eqa{Hke}{
E_{\rm kin}&=& \frac{L}
{12\pi}\left[4-2\sqrt{k_c^2+4}+k_c^2\left(\sqrt{k_c^2+4}-k_c\right)\right]\nonu\\
 &\condi{\leads}{k_c\gg1}& \frac{L}
 {3\pi}\left(1-\frac{3}{2k_c}\right),
}
which 
is shown in Fig.~\ref{fig:ekinkc}. Notably, the degree of saturation depends only on the scaled cutoff, $k_c=k_{\rm max}\xi_0$, a trend that will appear repeatedly.

In 2D, there is a weak logarithmic ultraviolet (UV) divergence in the kinetic energy:
\eqa{Ekin2d}{
E_{\rm kin} &=& \frac{V}
{32\pi}\left[k_c(k_c^2-2)\sqrt{k_c^2+4}-k_c^4+8\sinh^{-1}\left(\frac{k_c}{2}\right)\right]\qquad\nonu\\
&\condi{\leads}{k_c\gg1}& \frac{V}
{4\pi}\left[\log (k_c) -\frac{3}{4} +\frac{2}{k_c^2}\right].
}
By contrast, in 3D the UV divergence in the kinetic energy is linear and hence particularly severe:
\eqa{Ekin3d}{
E_{\rm kin} &=& \frac{V}
{40\pi^2}\left[\left(16-2k_c^2+k_c^4\right)\sqrt{k_c^2+4}-32-k_c^5\right]\nonu\\
&\condi{\leads}{k_c\gg1}& \frac{V}
{4\pi^2}\left(k_c -\frac{16}{5}\right).
}

\subsection{Local two-body correlation and interaction energy}
Consider first the correlator $G_2$, which enters the interaction energy calculation,
\eqa{G2_0}{
G_2(\bo{x}) &=& \bra{B}\dagop{\Psi}(\bo{x})\dagop{\Psi}(\bo{x})\op{\Psi}(\bo{x})\op{\Psi}(\bo{x})\ket{B}.
}
 Applying \eqref{BOGpsi} and \eqref{dpsimj}, with a coherent condensate wavefunction $\phi_0$ such that $\langle\op{a}_0^{\dagger 2}\op{a}_0^2\rangle=N_0^2$, one obtains
\eq{G2_1}{
G_2^{\rm Bog}(\bo{x})= n_0^2 
+4n_0\delta n 
+2{\rm Re}\left[\wb{m}\right]n_0 
+|\wb{m}|^2+2(\delta n)^2,
}
where $n_0=N_0/V$, 
and the anomalous pair ``density'' $\wb{m}$ is defined as
\eq{mdef}{
\wb{m} = -\sum_{j} v_j(\bo{x})u_j^*(\bo{x}) = -\frac{1}{V}\sum_{\bo{k}\neq0}V_{\bo{k}}U_{\bo{k}}^*.\qquad
}
Note that $\bra{B}\op{\Psi}^2(\bo{x})\ket{B} = n_0+\wb{m}$. If a Fock state was taken for the condensate, additional finite-size corrections would appear in \eqn{G2_1}.

The last two terms in \eqn{G2_1} are of higher order than the Bogoliubov approximation, and are therefore usually discarded by hand. However, here we retain them since the purpose is to understand the values obtained from a Wigner representation of the Bogoliubov state, in which the distinction between $\phi$ and $\op{b}_j$ has already been lost.
For 1D uniform gas, the integral estimate gives
\eqa{mkmkm}{
\wb{m} &=& -\frac{1}
{2\pi}\left[\sinh^{-1}\left(\frac{2}{k_L}\right)-\sinh^{-1}\left(\frac{2}{k_c}\right)\right]\qquad\\
&\condi{\leads}{k_c\gg1,k_L\ll1}&-\frac{1}
{2\pi}\left[\log\left(\frac{4}{k_L}\right)-\frac{2}{k_c}\right].
\label{mkmkm_1d}
}
In 2D one finds
\eqa{m2d}{
\wb{m} &=& -\frac{1}
{2\pi}\sinh^{-1}\left(\frac{k_c}{2}\right)
\condi{\leads}{k_c\gg1} -\frac{1}
{2\pi}\left[\log (k_c) + \frac{1}{k_c^2}\right],\qquad
}
while in 3D
\eqa{m3d}{
\wb{m} &=& -\frac{1}
{2\pi^2}\left(\sqrt{k_c^2+4}-2\right)\\
&\condi{\leads}{k_c\gg1}&
-\frac{1}
{\pi^2}\left(\frac{k_c}{2}-1+\frac{1}{k_c}\right),
}
with negative divergences as noted by \cite{Cherny01}.

For a uniform gas system, combining the expressions for $\delta n$ in \eqn{Nkmkm} and $\wb{m}$ in \eqn{mkmkm} in 1D (and \eqn{N2d} with \eqn{m2d} in 2D, and \eqn{N3d} with \eqn{m3d} in 3D) into \eqn{G2_1} yields the local pair correlation $G_2$. This in turn leads to an estimate of the normalised second-order correlation function, which is of interest, for example, because values below 1 indicate an inherently nonclassical state:
\eq{g20def}{
g^{(2)}(0) = \frac{\bra{B}\dagop{\Psi}(\bo{x})\dagop{\Psi}(\bo{x})\op{\Psi}(\bo{x})\op{\Psi}(\bo{x})\ket{B}}{\bra{B}\dagop{\Psi}(\bo{x})\op{\Psi}(\bo{x})\ket{B}^2} = \frac{G_2}{(n_0+\delta n)^2}.
}
Importantly, the denominator must include the full density $n_0+\delta n$ rather than $n_0$ alone, since the condensate–fluctuation distinction is also lost in the TWA.


\begin{figure}
\centering
\includegraphics[width=\hfigwidth]{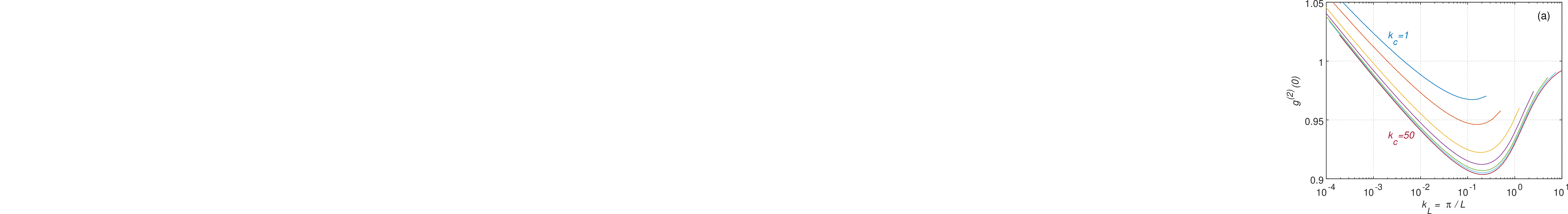}
\includegraphics[width=\hfigwidth]{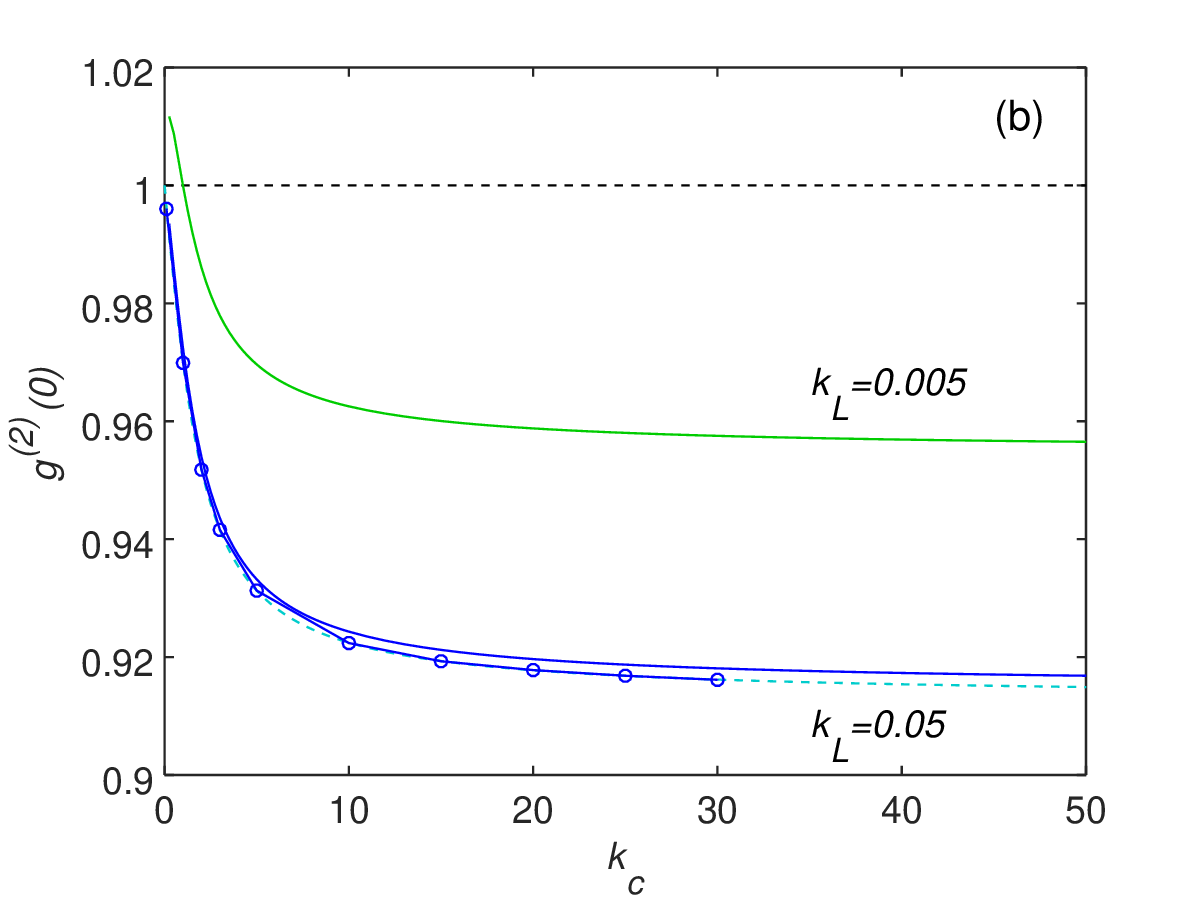}
\caption{Predicted values of $g^{(2)}(0)$ in 1D for 
$g_0=0.2, n_0=5$. Predictions use \eqn{deltandisc}, \eqn{G2_1}, \eqn{mdef}, and \eqn{g20def}.
(a) Dependence on $k_L=\Delta k/2=k_{\rm min}\xi_0/2$ (horizontal axis) and on $k_c=k_{\rm max}\xi_0$ (curves with values $k_c=1, 2, 5, 10, 20, 30, 
50$, from top to bottom).   
(b) Dependence on $k_c$ for two values of $k_L=\Delta k/2$, namely $0.005$ (green) and $0.05$ (blue). 
Integral estimates using \eqn{Nkmkm}, \eqn{G2_1}, \eqn{mkmkm} and \eqn{g20def} 
are shown with solid lines. Also shown are numerical results from a TWA representation of the Bogoliubov state with $P=10^7$ trajectories and $\Delta k=0.1$ 
($g_0=0.2, n_0=5$; blue circles). 
The cyan dashed line shows the discrete on-lattice Bogoliubov calculation using \eqn{deltandisc} and \eqn{mdef}.}
\label{fig:g2o-kmkm}
\end{figure}

What we obtain in 1D is shown in Fig.~\ref{fig:g2o-kmkm}. 
Somewhat unexpectedly, 
a minimum appears as a function of the infrared cutoff $k_L=\Delta k/2=\pi/L$ (Fig.~\ref{fig:g2o-kmkm}(a)).
For very low $\Delta k$ (i.e., very low $k_L$ in Fig.~\ref{fig:g2o-kmkm}(a)), we start seeing large fluctuations, which are due to the infrared divergence of the overall depletion and the loss of condensation, signalling a breakdown of the Bogoliubov assumptions. This might be remedied by using the quasicondensate variant of Bogoliubov \cite{Mora03}. Conversely, at very high $\Delta k=k_{\rm min}\xi_0=2\pi/L$, there are fewer low-energy excitations due to finite size effects. 
With regard to the ultraviolet cutoff (Fig.~\ref{fig:g2o-kmkm}(b)), $k_c=20$ essentially saturates the dependence. 
The minimum does not necessarily reach the Bogoliubov thermodynamic limit given in \cite{Kheruntsyan03} in the context of the Lieb-Liniger gas, namely $g^{(2)}(0)\to 1-2\sqrt{gn}/n\pi$, which takes the value 0.877 for Fig.~\ref{fig:g2o-kmkm}(b). 
The discrepancy arises from the ``extra'' contributions beyond Bogoliubov discussed below, as well as from the breakdown of the Bogoliubov approximation in 1D when the box becomes too large.
The green line in Fig.~\ref{fig:g2o-kmkm}(b) clearly shows the large difference that the choice of $\Delta k$ (i.e., the box size $L$) makes in 1D. 
Finally, the expression \eqn{g20def} provides a very good match to calculations from a Wigner numerical ensemble generated as described in Sec.~\ref{WIG1}. 

For a contact interaction, the expectation value of the interaction energy is 
\eq{Eint}{
E_{\rm int} = \langle\op{H}_{\rm int}\rangle = \frac{g_0}{2}\,g^{(2)}(0)\, nN = \frac{g_0\,G_2 V}{2}.
}
For a better understanding, it is useful to decompose this quantity into a regular and an extra contribution, based on the inspection of \eqn{G2_1}:
\eq{Eintregextra}{
E_{\rm int}=E_{\rm int}^{\rm reg}+E_{\rm int}^{\rm extra}.
}
The regular part, consistent with a self-consistent 2nd order Bogoliubov treatment, is
\eqa{Eintreg}{
E_{\rm int}^{\rm reg} &=& \frac{g_0 n_0^2 V}{2} + g_0n_0V(2\delta n + {\rm Re}[\wb{m}])\nonumber\\
 &=&\frac{g_0n_0N_0}{2}+g_0n_0\sum_{\bo{k}\neq0}\left(2|V_{\bo{k}}|^2 - {\rm Re}\left[V_{\bo{k}}U^*_{\bo{k}}\right]\right),\quad
}
where $U_{\bo{k}}$ and $V_{\bo{k}}$ are the Bogoliubov coefficients defined in \eqn{BOG_uv}.
The extra contribution consists of terms higher order than the standard Bogoliubov approximation. We retain these terms to maintain direct correspondence with the corresponding higher-order observables that appear in the Wigner representation,
\eq{Eintextra}{
E_{\rm int}^{\rm extra}   =  \frac{g_0V}{2}\left(|\wb{m}|^2+2(\delta n)^2\right).
}
Returning to Fig.~\ref{fig:g2o-kmkm}(b), at $k_c=30$ the prediction becomes $g^{(2)}(0)=0.883$ when the same higher-order terms, $|\wb{m}|^2+2(\delta n)^2$, are discarded from $G_2^{\rm Bog}(\bo{x})$ \eqn{G2_1} in the evaluation of the local pair correlation $G_2$; this value is very close to the Bogoliubov thermodynamic limit $1-2\sqrt{gn}/n\pi=0.873$. Therefore, the difference observed in the TWA is attributable to the ``extra'' terms.

\subsection{LHY correction}

\subsubsection{Regular, dressed, and extra parts}
Taking into account the decomposition of the mean-field energy based on the 2nd Born correction and associating the decomposed energy components with the Bogoliubov theory, we combine \eqn{Emf01}, \eqn{sumkekin}, and \eqn{Eintregextra}-\eqn{Eintextra}, the excess energy above the mean-field ground state \eqn{Emf} is then given by:
\eq{Elhy-def}{
E_{LHY} = E_{\rm kin}+E_{\rm int}-E_{\rm mf} =
 E_{LHY}^{\rm reg} - E_{\rm mf}^{\rm dress}+ E_{LHY}^{\rm extra}.
}
This consists of a ``regular'' part, as expected from quantum fluctuations described by the Bogoliubov approximation,  
\eqa{E0reg}{
E_{LHY}^{\rm reg} &=& E_{\rm kin} + g_0n_0V(\delta n + {\rm Re}[\wb{m}])\\
 &=&\sum_{\bo{k}\neq0}\left[ \frac{|V_{\bo{k}}|^2\,k^2}{2} + g_0n_0\left(|V_{\bo{k}}|^2 - {\rm Re}\left[V_{\bo{k}}U^*_{\bo{k}}\right]\right)\right],\nonu
}
together with the dressed mean-field contribution $E_{\rm mf}^{\rm dress}$ \eqn{Emf01}, which is related to the difference between the mean-field energy calculated using the bare coupling $g_0$ and the dressed coupling $g$, 
and finally an ``extra'' contribution, 
\eqa{Elhy_extra}{
E_{LHY}^{\rm extra}  = 
\frac{g_0V}{2}\left(|\wb{m}|^2+(\delta n)^2\right).
}
Note that $E_{LHY}^{\rm extra}$ originates purely from the algebraic expansion of the decomposed energy components within the lattice Bogoliubov description prior to any Wigner phase-space mapping.
The sum in \eqn{Elhy-def} should, in principle, equal the LHY energy correction $E_{LHY}^{\infty}$ given in \eqn{LHY}, if all contributions are working as expected.

\begin{figure}[ht]
\centering
\includegraphics[width=0.3\columnwidth]{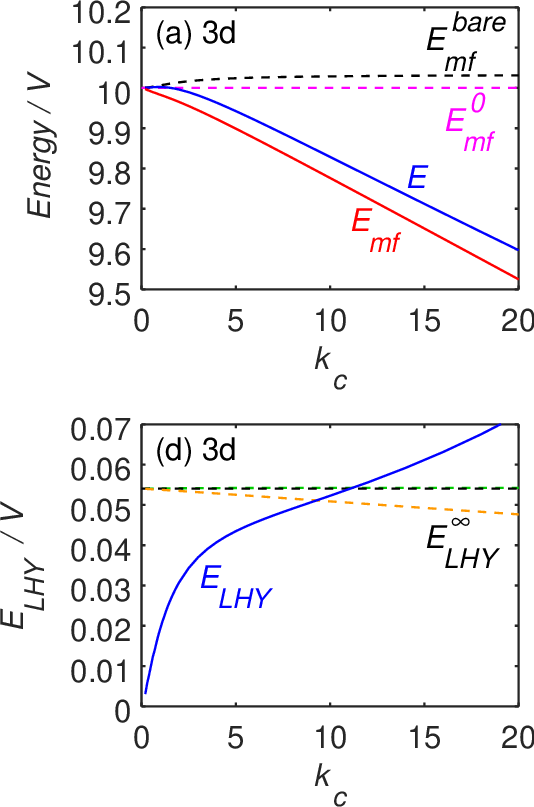}
\includegraphics[width=0.3\columnwidth]{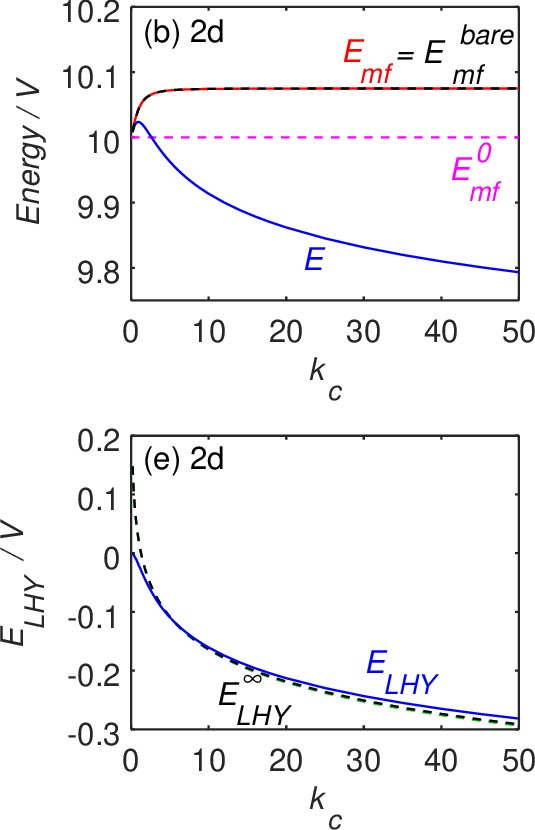}
\includegraphics[width=0.3\columnwidth]{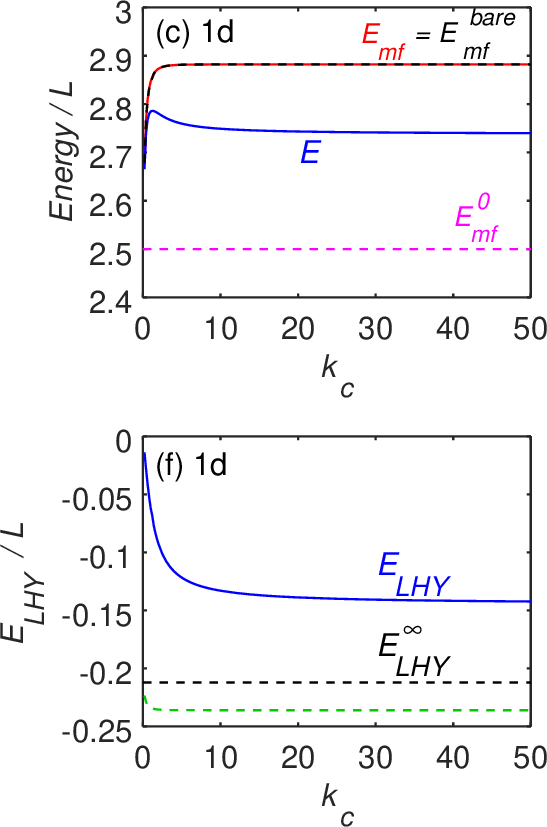}
\caption{
Dependence of the decomposed energy densities (determined analytically from Bogoliubov theory) and mean-field energy densities on $k_c$ for uniform systems in 3D (1st column), 2D (2nd column), and 1D (3rd column) when the bare interaction strength $g_0$ is kept constant.
Top row (a-c): total energy $E=E_{\rm kin}+E_{\rm int}$ calculated microscopically from all modes using \eqn{sumkekin} and \eqn{Eint} in blue, and the dressed mean-field estimate using full 
density 
$E_{\rm mf}=E_{\rm mf}^{\rm bare}+E_{\rm mf}^{\rm dress} =(g_0+g_1)n^2V/2$
in red ($g_1$ is used only in 3D). Also shown are $E_{\rm mf}^{\rm bare}=g_0n^2/2$ (black dashed) and the condensate-only  bare mean-field energy $E_{\rm mf}^0=g_0n_0^2/2$ (magenta dashed). Bottom row (d-f): the difference $E_{LHY}=E-E_{\rm mf}$, as per \eqn{Elhy-def}, with expected LHY correction \eqn{LHY} shown dashed. Three variants using $g_0n_0$ (black), $g_0n$ (green), and $gn$ (orange, 3D) are displayed.
Here, $g_0=0.05$ in 3D and 2D, $g_0=0.2$ in 1D, $n_0=1/g_0$, and the box dimensions are $L=20\pi$ in each dimension.
}
\label{fig:Ekc}
\end{figure}

Evaluating $E_{\rm kin}+E_{\rm int}$ in 3D for the case of $g=0.05$, one finds the dependence shown in blue in Fig.~\ref{fig:Ekc}(a). Despite allowing ever higher kinetic energies, the total calculated energy actually decreases strongly.
Indeed, expanding the ``regular'' part \eqn{E0reg} at large $k$, one finds
\eqa{E0reg-k}{
E_{LHY}^{\rm reg} &=& \sum_{\bo{k}\neq0}\left[ -\frac{(g_0n_0)^2}{2k^2} +\frac{(g_0n_0)^3}{k^4}+\mc{O}\left(\frac{(g_0n_0)^4}{k^6}\right)\,\right].\qquad
}
This expression is, unfortunately, UV divergent in 3D (and slightly in 2D) because of the $1/k^2$ term.
However, as pointed out by \cite{Weiss04} and others, in 3D the $g_1$ contribution \eqn{g1Weiss} in $1/k^2$ is actually part of the mean-field energy. 
Expanding the ``dressed'' mean-field contribution \eqn{g1Weiss} at large $k$, using $g_{\bo{k}}=g_0$, we find
\eqa{E0dress-k}{
E_{\rm mf}^{\rm dress} &=& -\sum_{\bo{k}\neq0}\left[ \frac{(g_0n_0)^2}{2k^2}
+ \frac{\mc{O}\left(g_0^2n_0\delta n\right)}{k^2} 
\right ].
}
The numerical evaluation of the mean-field energy from its components, incorporating the $g_1$ correction \eqn{Emf01}, is shown as the red line in Fig.~\ref{fig:Ekc}(a). The $g_1$ term actually contains the dominant contribution to the deviations from the naive condensate value $E_{\rm mf}^0$.

\begin{figure}
\centering
\includegraphics[width=\figwidth]{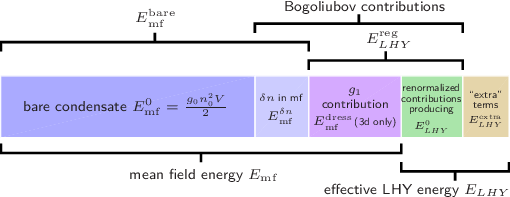}
\caption{An illustration of the contributions to ground-state energy, $E=E_{\rm kin}+E_{\rm int}$, in a numerical implementation. 
The $g_1$ contribution, $E_{\rm mf}^{\rm dress}=g_1n^2V/2$, is included 
only in 3D.}
\label{fig:gs-schema}
\end{figure}

Here we therefore observe a renormalisation ``miracle'': the leading divergent term in \eqn{E0dress-k} cancels the leading divergent term in \eqn{E0reg-k}. 
In other words, the divergent part of the Bogoliubov excitation contribution to the ground-state energy is absorbed, via a relabelling, into the mean-field coupling. 
As shown, for example, in \cite{Weiss04}, the remaining terms of $E_{LHY}^{\rm reg}$,
\eqa{alphWeiss}{
E_{LHY}^0  &=& E_{LHY}^{\rm reg} - E_{\rm mf}^{\rm dress} \nonu\\
&=& 
\sum_{\bo{k}\neq0}\frac{1}{2}\left[\ve_{\bo{k}}-\frac{k^2}{2}-g_{\bo{k}}n_0 +\frac{(g_{\bo{k}}n_0)^2}{k^2}\right],
}
integrate in 3D to the usual LHY correction \eqn{LHY3d}, $E_{LHY}^{\infty}(g_0,n_0)$.
This is encouraging; however, this knowledge is not immediately usable for our target of numerical implementation because (i) in the Wigner representation there is no provision for subtracting any divergent terms (the energy and any resulting dynamics are computed directly from just $E_{\rm kin}+E_{\rm int}$), and (ii) the ``extra'' terms, which can also be divergent, have not been addressed.

Indeed, Fig.~\ref{fig:Ekc}(d) shows the remaining difference between the calculated and mean-field energies, and its 
only partial convergence to $E_{LHY}^{\infty}$ for any choice of $g$ and $n$.
As will be seen later, e.g., in Fig.~\ref{fig:ELHY-2d3d-est}(b), the culprit responsible for the failure to converge properly to the canonical value is the ``extra'' terms, which grow as $k_c^2$.

Fig.~\ref{fig:Ekc} also shows the behaviour of the excess energies above the condensate mean-field $E_{\rm mf}^0$ for 2D and 1D. 
The match to the canonical LHY energy $E_{LHY}^{\infty}$ in 2D is remarkably good, presumably because \eqn{LHY2d} already incorporates the leading logarithmic cutoff contribution. 
In 1D, there is convergence but to 
an inaccurate value. Again primarily due to the ``extra'' terms, as will be demonstrated below and seen, e.g., in Fig.~\ref{fig:ELHY-kmkm}. 

An overall schematic of the contributions to the ground-state energy in a numerical implementation is shown in Fig.~\ref{fig:gs-schema}.
Let us now examine 
how the various contributions to the energy shift play out in practice, before proposing a solution to obtain the correct LHY energy in the numerical representation in Sec.~\ref{MATCH}.

\subsubsection{Behaviour in 1D}
The limiting behaviour of the LHY energy contributions in the $V\to\infty$ 
limit in 1D is as follows. 
The high-$k_c$ limit (with $k_L\to 0$) of the regular contribution \eqn{E0reg}, recalling that $g_0n_0=1$ in our units for the observables \eqn{Nkamin}, \eqn{Hke}, and \eqn{mkmkm_1d}, is: 
\eq{1dlimskc}{
\frac{E_{LHY}^{\rm reg}}{L} \condi{\leads}{k_c\gg1, k_L\ll1} 
-\frac{2}{3\pi} +\frac{1}{2\pi k_c} + \frac{k_L}{2\pi}. 
}
This expression converges to \eqn{LHY1d} in the limits $k_c\to\infty$ and $k_L\to0$, without the need for renormalisation. 
In fact, in 1D the quantity $g_1$ is not well behaved as the box size becomes large, since integrating over the isotropic momentum using \eqn{continuum_k} and \eqn{ddk_int} for the 2nd term in the Born expansion \eqn{g1Weiss} yields
\eqa{g1_1d}{
g_1^{(1d)}= \frac{g_0^2}{\pi}\left(\frac{1}{k_c}-\frac{1}{k_L}\right).
}
For example, the dressed mean-field contribution evaluates to $E_{\rm mf}^{\rm dress}/L=-0.255$ for the system in Fig.~\ref{fig:Ekc}(c), a rather distant and unhelpful value. This divergence of $g_1$ as $k_L\to0$ is related to the breakdown of the condensate in a large system. Consequently, a renormalisation using this $g_1$ is not useful in 1D, and therefore we do not 
separate out the $E^{\rm dress}_{\rm mf}$ contribution from the mean-field or LHY energies as shown in Figs.~\ref{fig:Ekc}(c) and~\ref{fig:Ekc}(f).

On the other hand, despite the well-behaved nature of $E_{LHY}^{\rm reg}$, the ``extra'' terms \eqn{Elhy_extra} can become very significant as the box size grows. For the 1D observables \eqn{Nkaminz_1d} and \eqn{mkmkm_1d}, the low $k_L$ (and high $k_c$) limit of the ``extra'' terms is  
\eqa{1dlimskL}{
\frac{E_{LHY}^{\rm extra}}{L} 
&\condi{\leads}{k_c\gg1, k_L\ll1}&
\frac{g_0}{4\pi^2}\left[
\mc{L}^2-2\mc{L}+2
-\frac{2}{k_c}\mc{L}\right],\qquad
\text{with}\qquad
\mc{L}=\log\left(\frac{4}{k_L}\right).\label{Ldef}
}
To leading order, with $E_{LHY}^{\rm reg}/L\approx-2/3\pi$ and $E_{LHY}^{\rm extra}/L\approx g_0\mc{L}^2/4\pi^2$, we then have:
\eq{exreg1d}{
E_{LHY}^{\rm extra}\approx -
\sqrt{\gamma_L}
\frac{3}{8\pi}\left[\log\left(\frac{4}{k_L}\right)\right]^2
E_{LHY}^{\rm reg},
}
in terms of the Lieb 
parameter 
$\gamma_L=g/n\approx g_0/n_0$, which reduces to $\gamma_L=g_0^2$ in our units $g_0n_0=1$. 
Thus, the extra terms become less important as $\gamma_L$ becomes sufficiently small, but increasingly important as the box size $L\approx\pi/k_L$ grows.

\begin{figure}[htb]
\centering
\includegraphics[width=\hfigwidth]{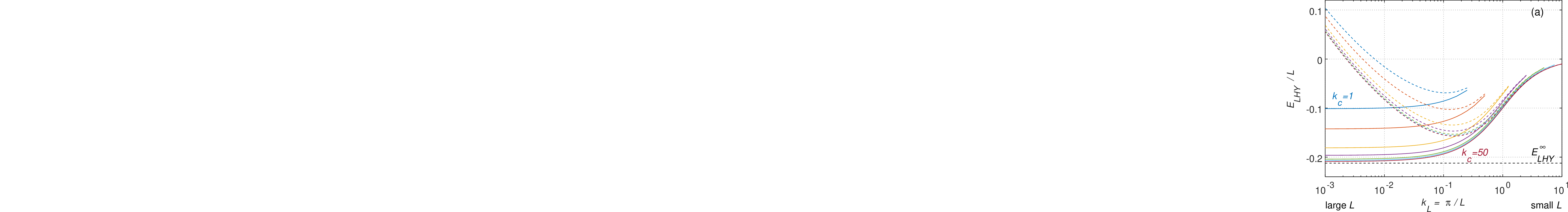}
\includegraphics[width=\hfigwidth]{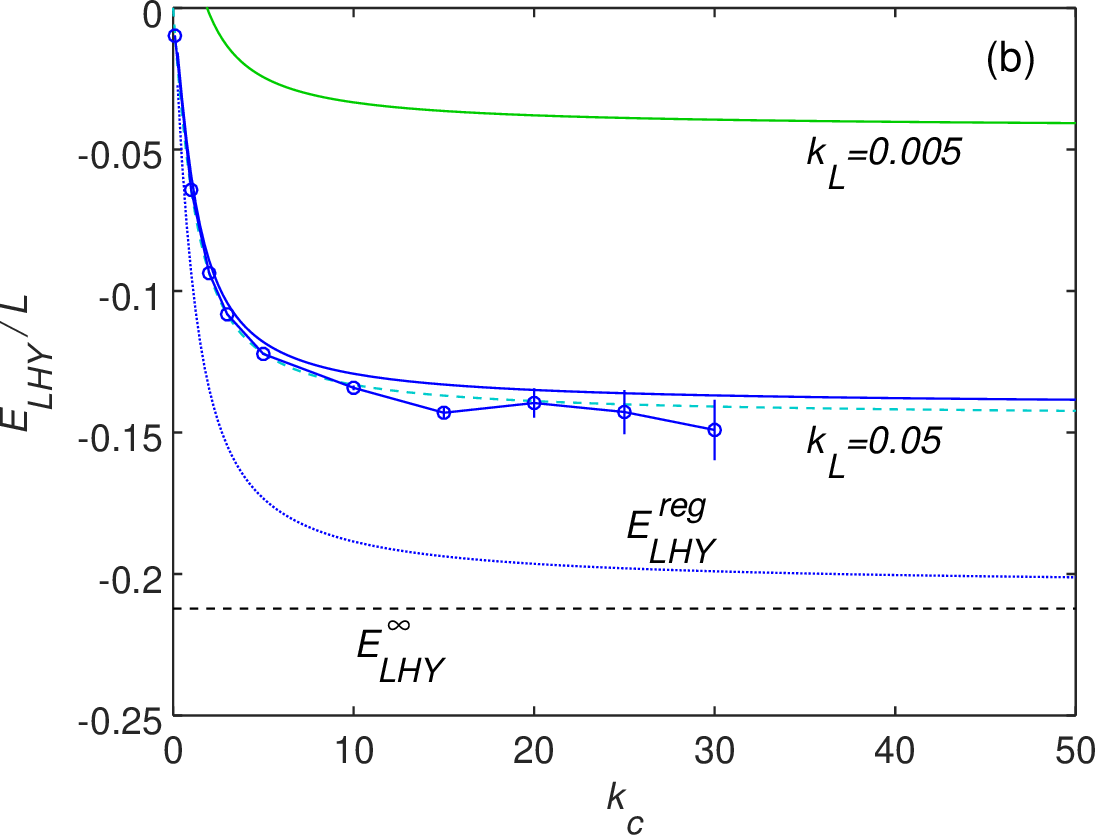}
\caption{Predicted values of the energy shift relative to the mean-field value $E_{\rm mf}$ for a uniform 1D system.
(a): $E_{LHY}^{\rm reg}$ (solid) and the full shift $E_{LHY} =E_{LHY}^{\rm reg} + E_{LHY}^{\rm extra}$ (dashed, colour-matched) as functions of 
$k_L=\pi/L=k_{\rm min}\xi_0/2$ (horizontal axis) and $k_c=k_{\rm max}\xi_0$ (curves with values $k_c=1, 2, 5, 10, 20, 30, 
50$, from top to bottom), for $g_0=0.2, n_0=5$. Also shown as a black dashed line is the standard LHY correction $E_{LHY}^{\infty}(g_0,n_0)$ from \eqn{LHY1d}.
Data here were calculated on a discrete lattice with $M=4,\dots,2^{12}$ $k$-space modes, from top to bottom. 
(b): Dependence on $k_c$ for two values of $k_L=\Delta k/2 = 0.005$ (green) and $0.05$ (blue), for $g_0=0.2, n_0=5, L=20\pi$. 
The solid lines use the large box estimates \eqn{Nkmkm}, \eqn{Hke}, \eqn{G2_1}, \eqn{mkmkm}, and \eqn{Eint}, while the cyan dashed line shows the full lattice-based prediction using \eqn{deltandisc}, \eqn{sumkekin}, and~\eqn{mdef}.
Also shown are values from a numerical realisation of the TWA ensemble with $P=10^7$ trajectories and $k_L=0.05$ (blue circles), the standard LHY correction \eqn{LHY1d} $E_{LHY}^{\infty}(g_0,n_0)/L = -2/3\pi$ from [\citenum{Rakshit19b}] (black dashed), and the ``regular'' part of the LHY energy $E_{LHY}^{\rm reg}$ from \eqn{E0reg} (blue dotted). Fig.~\ref{fig:g2o-kmkm} shows the corresponding $g^{(2)}(0)$ values for the same parameters.} 
\label{fig:ELHY-kmkm}
\end{figure}

Here we consider two examples. First, a case with relatively strong interactions ($\gamma_L=0.04$). Fig.~\ref{fig:ELHY-kmkm}(a) shows the dependence of the regular contribution (solid) and the total calculated LHY energy (dashed) on the lattice parameters $k_c$ and $k_L=\pi/L$. 
Inspecting the plot, the regular terms converge nicely to the standard 1D LHY value, $E_{LHY}^{\infty}$ in \eqn{LHY1d}, when $k_L$ is small and $k_c$ is large. 
However, the extra terms introduce a very large systematic deviation and cannot be removed if one wants to use a dynamical field theory such as the TWA. There is an optimum at large $k_c$ and $k_L\approx 0.2$, but the resulting magnitude of $E_{LHY}$ is only about 2/3 of the required value of $E_{LHY}^{\infty}$. As such, this representation is unusable for generating correct LHY physics at this value of $\gamma_L$. A solution to this problem will be provided in Sec.~\ref{MATCH}.

A second view of this behaviour for the same $\gamma_L=0.04$ is shown in Fig.~\ref{fig:ELHY-kmkm}(b), where the convergence with the ultraviolet cutoff $k_c$ is shown for two box sizes $L=2\pi/\Delta k$, both for the regular contribution and for the total energy correction $E_{LHY}$. 
The latter is also compared with a numerical realisation of the truncated Wigner ensemble with $P=10^7$ samples, implemented as described in Sec.~\ref{WIG1}. From this figure we can confirm that the full estimate $E_{LHY}=E_{LHY}^{\rm reg}+E_{LHY}^{\rm extra}$ matches the TWA, but also that the ``extra'' terms are crucial for reproducing any reasonable prediction of the TWA behaviour.

\begin{figure}[htb]
\centering
\includegraphics[width=\hfigwidth]{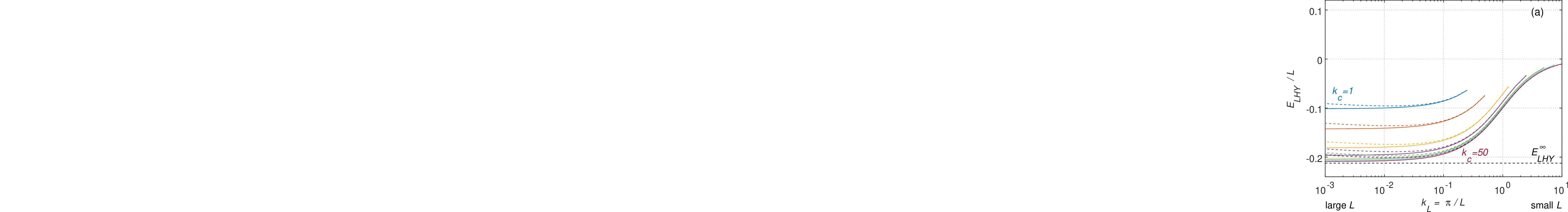}
\includegraphics[width=\hfigwidth]{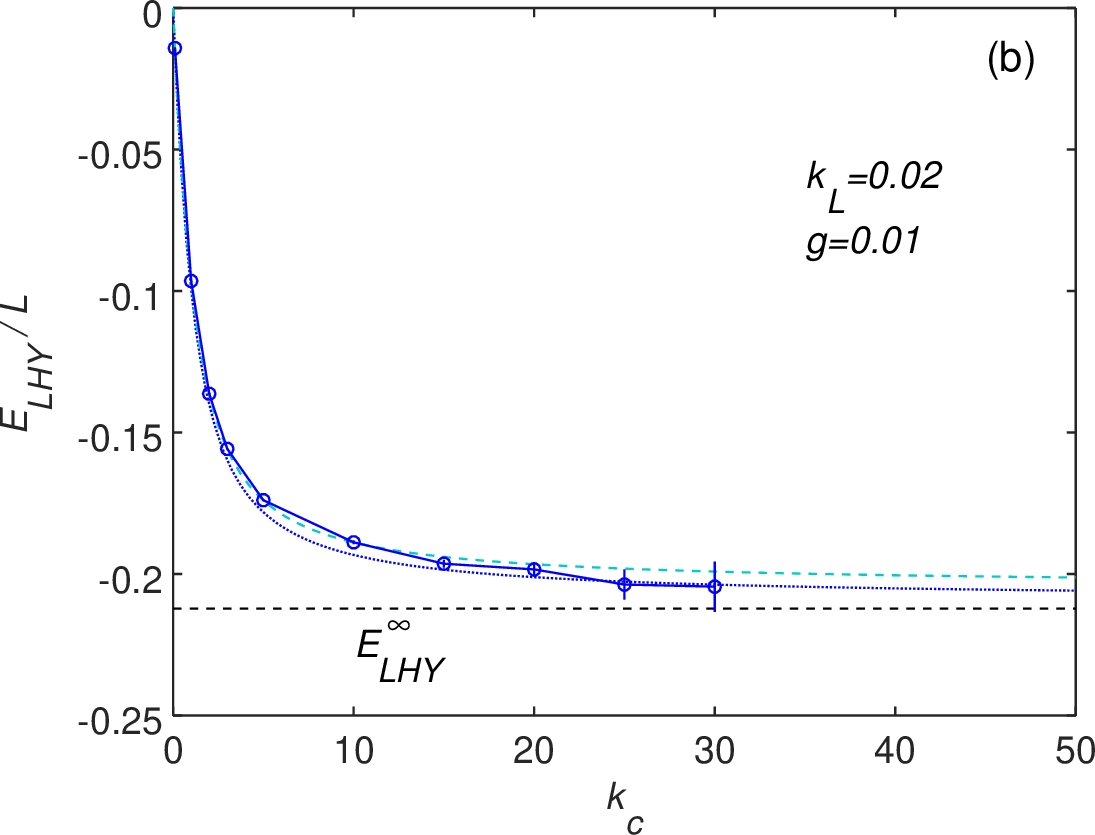}
\caption{(a) Data as in Fig.~\ref{fig:ELHY-kmkm}(a), but for $g_0=0.01$, $n_0=100$, corresponding to $\gamma_L=10^{-4}$. (b) Data for $k_L=\Delta k/2 = 0.02, L=2\pi/\Delta k$, presented in the same format as Fig.~\ref{fig:ELHY-kmkm}(b).}
\label{fig:ELHY-kmkm-001}
\end{figure}

For smaller $\gamma_L$, the representation of the LHY energy is much more forgiving. Fig.~\ref{fig:ELHY-kmkm-001} shows the behaviour for $\gamma_L=10^{-4}$, a typical value for dilute ultracold Bose gases. In this case, a cutoff $k_c\gtrsim 10$ is sufficient to obtain a reasonable match to the required LHY energy over a wide range of $k_L$. 
Fig.~\ref{fig:ELHY-kmkm-001}(b) also shows the explicit $k_c$ dependence of the terms, similarly to Fig.~\ref{fig:ELHY-kmkm}(b), but now with an improved agreement. Nevertheless, the full value of $E_{LHY}$ remains unattainable in this direct approach, which evaluates the Bogoliubov excitations around the bare interaction $g_0$ alone.

\subsubsection{Behaviour in 2D}

In 2D, the behaviour of the LHY energy contributions involves logarithmic dependence on the lattice parameters. Applying the 2D observables \eqn{N2d}, \eqn{Ekin2d}, and \eqn{m2d} into the contributions \eqn{E0reg} and \eqn{Elhy_extra}, the asymptotic behaviours are
\eqa{Ereglim2d}{
\frac{E_{LHY}^{\rm reg}}{V} &\condi{\leads}{k_c\gg1, k_L\ll1}& -\frac{1
}{4\pi}\left[\log (k_c) -\frac{1}{4} +\frac{1}{k_c^2}\right],
}
\eqa{Ereglim2d_2}{
\frac{E_{LHY}^{\rm extra}}{V} &\condi{\leads}{k_c\gg1, k_L\ll1}& \frac{g_0
}{16\pi^2}\left[2[\log (k_c)]^2+\frac{1}{2}+\frac{4\log (k_c) -1}{k_c^2}\right].
}
The leading two terms of the regular part agree with the canonical expression in \eqn{LHY2d}, although that result itself is weakly divergent as $k_c\to\infty$. We can see that the agreement shown in Fig.~\ref{fig:Ekc}(e) is markedly better than the agreement in 1D shown in Fig.~\ref{fig:Ekc}(f).
Following the integration over momentum as in \eqn{g1_1d}, the 2nd Born term here is
\eqa{g1_2d}{
g_1^{(2d)} = -\frac{g_0^2}{2\pi}\log\left(\frac{k_c}{k_L}\right),
}
so that the resulting mean-field contribution, $E_{\rm mf}^{\rm dress}=g_1n^2V/2$, if subtracted from $E_{LHY}^{\rm reg}$ \eqn{Elhy-def} to incorporate it into $E_{\rm mf}$, would remove the weak UV divergence but instead leave a supposedly weak IR divergence proportional to $\log k_L$. 
Incorporating $E_{\rm mf}^{\rm dress}$ turns out to be completely unhelpful. For example, in the system shown in Fig.~\ref{fig:Ekc}(b), one finds $E_{\rm mf}^{\rm dress}/V=-0.550$, which would completely break the reasonably good agreement obtained without it (note that in Figs.~\ref{fig:Ekc}(a–c), $g_1$ is not used in 1D or 2D). As in 1D, the IR divergence turns out to be the more onerous one.

Looking at the ``extra'' term, one finds 
the curious relationship that, to leading order,
\eq{exreg2d}{
E_{LHY}^{\rm extra}\approx -\frac{g_0}{2\pi}\log (k_c)\ E_{LHY}^{\rm reg} = -
\frac{\sqrt{\gamma_L}\log (k_c)}{2\pi}\,E_{LHY}^{\rm reg}.\qquad
}
Thus, while the leading term becomes less important as $g_0$ decreases, it nevertheless retains a logarithmic UV divergence. This indicates that, as the numerical resolution (i.e., $k_c$) is increased, the ``extra'' terms slowly become more dominant.

\begin{figure}
\centering
\includegraphics[width=\hfigwidth]{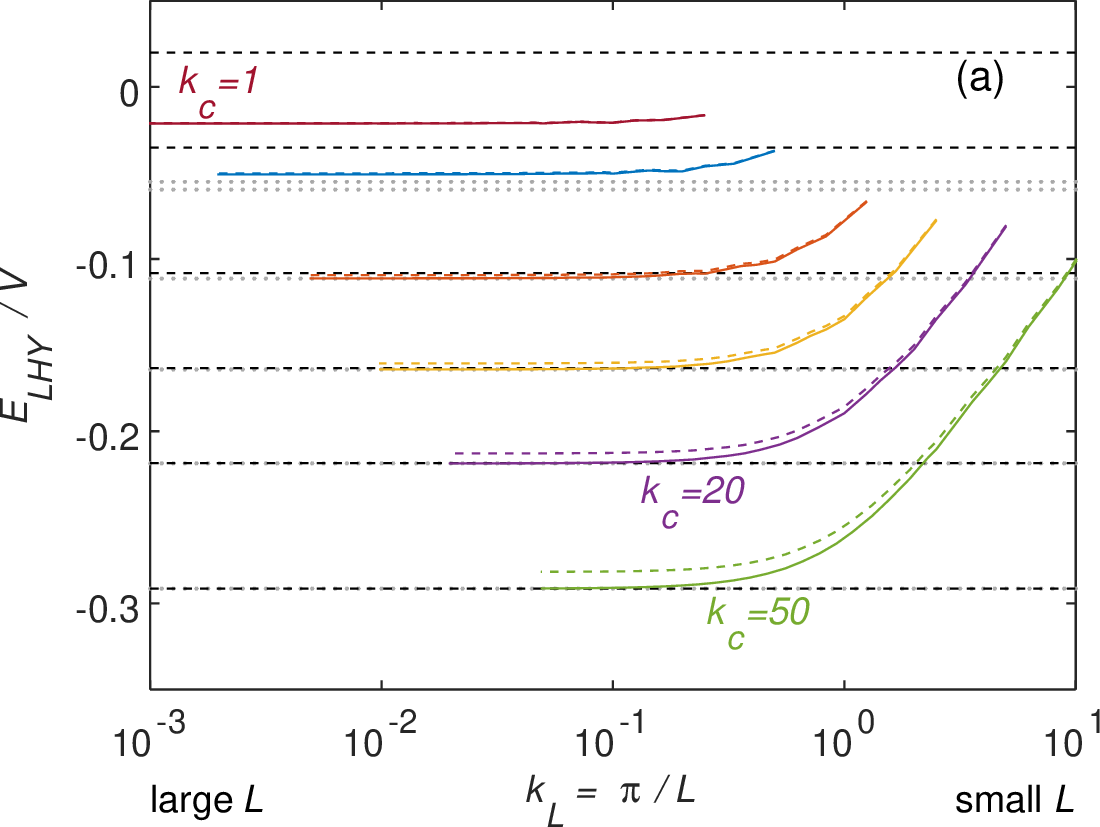}
\includegraphics[width=\hfigwidth]{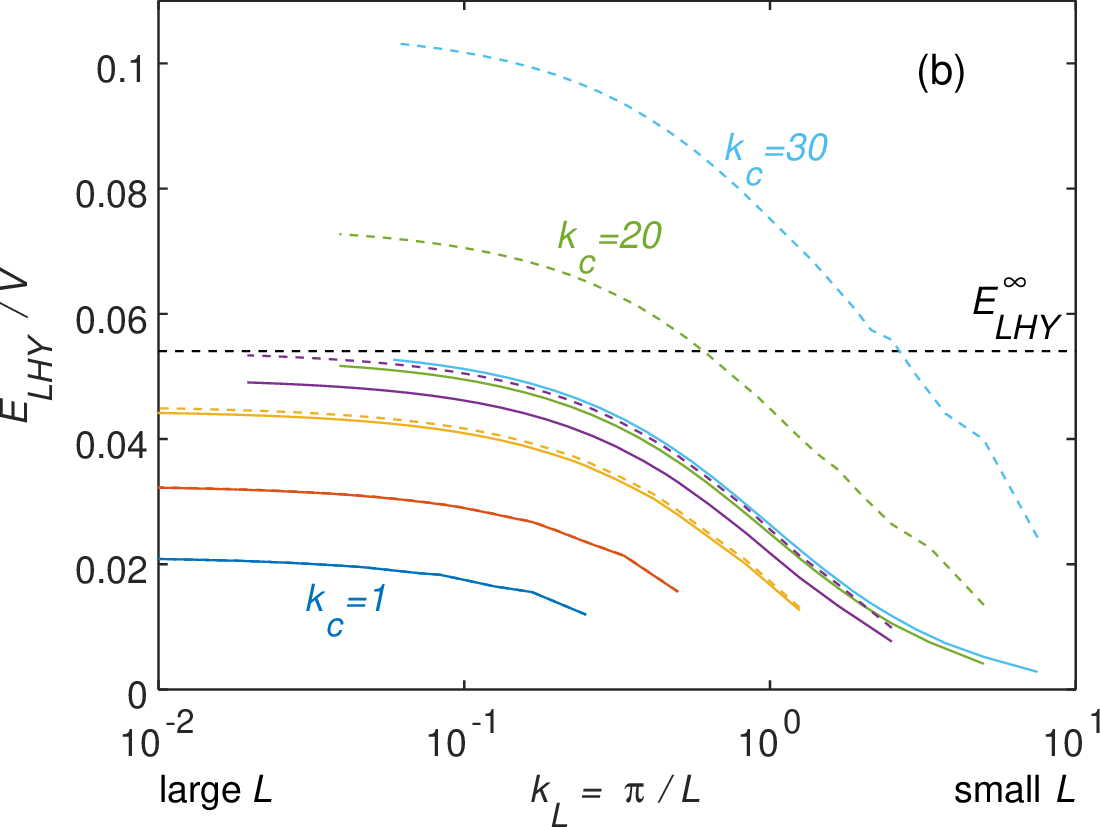}
\caption{
Predicted values of $E_{LHY}^{\rm reg}$ (solid) and $E_{LHY} =E_{LHY}^{\rm reg} + E_{LHY}^{\rm extra}$ (dashed, colour-matched) in a 2D (a) and a 3D (b) system, presented in a format similar to Fig.~\ref{fig:ELHY-kmkm}(a). Here, $g_0=0.05, n_0=20$. Standard LHY values $E_{LHY}^{\infty}(g_0,n_0)$ are shown as black dashed lines. These are $k_c$-dependent in 2D 
\eqn{LHY2d} and single-valued \eqn{LHY3d} in 3D. (a) In 2D, $k_c=1, 2, 5, 10, 20, 50$ (top to bottom), with even modes $M=4,\dots,2^{10}$, and ``regular'' large-box estimates $\lim_{k_L\to0} E_{LHY}^{\rm reg}$ \eqn{Ereglim2d} are shown as grey dots. (b) In 3D,
$k_c=1, 2, 5, 10, 20, 30$ (bottom to top), with even modes $M=4,\dots,2^{9}$.}
\label{fig:ELHY-2d3d-est}
\end{figure}

Fig.~\ref{fig:ELHY-2d3d-est}(a) shows the full Bogoliubov calculations on 2D lattices with a circular cutoff $|\bo{k}|\le k_c$, in a style similar to Fig.~\ref{fig:ELHY-kmkm}(a). The logarithmic cutoff dependence, $\sim\log k_c$, is seen and, in fact, is quite strong. We also see that the ``extra'' terms become more visible at \emph{high} cutoff, although they are not very large in an absolute sense for the parameters shown. The strong modification of $E_{LHY}$ (dashed lines) at small $k_L$ observed in 1D does not recur here. Note that 
at low $k_c$, the LHY estimate in \eqn{LHY2d} does not provide a good approximation, because the expression $\log(\sqrt{e}/k_c^2)$ becomes inaccurate in this regime.

\subsubsection{Behaviour in 3D}
\label{3D}

Finally, consider the classic case of 3D. 
Here the UV divergence in the kinetic energy \eqn{Ekin3d} is linear in the regular part \eqn{E0reg} of the LHY energy contributions: 
\eqa{Ereglim3d}{
\frac{E_{LHY}^{\rm reg}}{V} &\condi{\leads}{k_c\gg1, k_L\ll1}& -\frac{1}
{4\pi^2}\left(k_c-\frac{32}{15}\right).
}

The divergent term can be identified as equal to the contribution of the 2nd Born term arising from $E^{\rm dress}_{\rm mf}=g_1n^2V/2$, where 
\eq{g13d}{
g_1^{(3d)} = -\frac{g_0^2}{2\pi^2}\left(k_c-k_L\right),
}
and can therefore be incorporated into the mean-field energy \eqn{Emf01}, $E_{\rm mf}=(g_0+g_1)n^2V/2$.

Fig.~\ref{fig:ELHY-2d3d-est}(b) shows the full Bogoliubov calculations on 3D lattices (with a spherical cutoff $|\bo{k}|\le k_c$), where $E_{LHY}$ (dashed colour lines) is obtained as the sum of the LHY energy contributions, $E_{LHY}^{\rm reg}$ and $E_{LHY}^{\rm extra}$. The regular part $E_{LHY}^{\rm reg}$ (solid lines) converges quite nicely to the canonical value $E_{LHY}^{\infty}(g_0,n_0)$; however, the contribution of the ``extra'' terms, which cannot be separated out from the regular Bogoliubov contributions in a numerical TWA implementation, introduces a significant discrepancy (dashed lines). 
This is because, in 3D, the extra terms grow more rapidly with $k_c$ than the regular ones -- quadratically:
\eq{Eextralim3d}{
\frac{E_{LHY}^{\rm extra}}{V} \condi{\leads}{k_c\gg1, k_L\ll1} \frac{g_0}
{8\pi^4}\left[k_c^2-4k_c+\frac{76}{9}\right]. 
}
As a result, in the limit of very fine numerical resolution ($k_c\to\infty$), the extra terms come to dominate the total LHY energy even if $g_1$ (i.e. $E^{\rm dress}_{\rm mf}$) is used to regularise the regular term $E_{LHY}^{\rm reg}$, as in \eqn{Elhy-def}.
The relative scale between the regular Bogoliubov and extra terms is, to leading order,
\eq{exreg3d}{
E_{LHY}^{\rm extra}\approx -\sqrt{\gamma_L}\,\frac{k_c}{2\pi^2}\,E_{LHY}^{\rm reg}.
}
The conclusion is that, in 3D, the bare energy calculated from a Bogoliubov representation around the bare interaction $g_0$ and condensate density $n_0$ is, in practice, unusable for capturing LHY corrections within a numerical TWA implementation.

\section{Bare interaction corresponding to dressed coupling}
\label{MATCH}

The results in the previous section gave detailed insight into the behaviour of the energies and observables when implemented numerically with a bare contact interaction $g_0$. The match with the accepted LHY values \eqn{LHY} is imperfect, and the values obtained depend strongly on the lattice parameters. Except perhaps for some special cases, it does not appear that any choice of lattice parameters $k_c$ and $\Delta k$ provides a satisfactory match -- contrary, perhaps, to common expectations. However, in the spirit of renormalisation, and of contact-interaction models in general, what is really desired is for physical predictions to match the \emph{dressed} quantities, while the microscopic details underneath may vary. What we are aiming for, then, is to correctly include the LHY correction to the energy (and with luck -- corrections to other observables as well) within the TWA. That is, 
upon rearranging \eqn{Elhy-def}, we require 
\eq{Elhy4}{
E_{\rm kin}+E_{\rm int} = E_{LHY}^{\infty}+E_{\rm mf}.
}
Here, the left-hand side ($\mathrm{LHS}$) is calculated numerically based on a bare interaction $g_0$, condensate fraction $n_0$, and lattice parameters $V, \Delta k$, etc., while the right-hand side ($\mathrm{RHS}$) is postulated using set values of $g=g_{\rm set}$ and $n=n_{\rm set}$ that are desired. 

Proceeding in this vein, we aim to determine the bare interaction $g_0$ that makes the numerical model match the postulated dressed interaction $g$ for a postulated density $n$. We take the lattice parameters to be set by the practicalities of the model and by the physical question under consideration. Also, since there are two independent postulated quantities -- $g$ and $n$, we require two bare quantities (inputs to the microscopic Bogoliubov and subsequent TWA formulation) to be matched. One of the two is $g_0$; for the other, it is convenient to choose the condensate fraction $n_0$. 

Throughout this section and Sec.~\ref{DYN}, units are scaled according to the scheme of \eqn{utrans}, but instead of $\xi_0$ we use a length unit based on the ``set'' values, 
\eq{xiset}{
\xi_{\rm set}=\hbar/\sqrt{mg_{\rm set}n_{\rm set}}.
}



\subsection{Determination of matched bare interaction}
\label{g0steps}

\subsubsection{Procedure}

\begin{figure}[h]
\centering
\includegraphics[width=\hfigwidth]{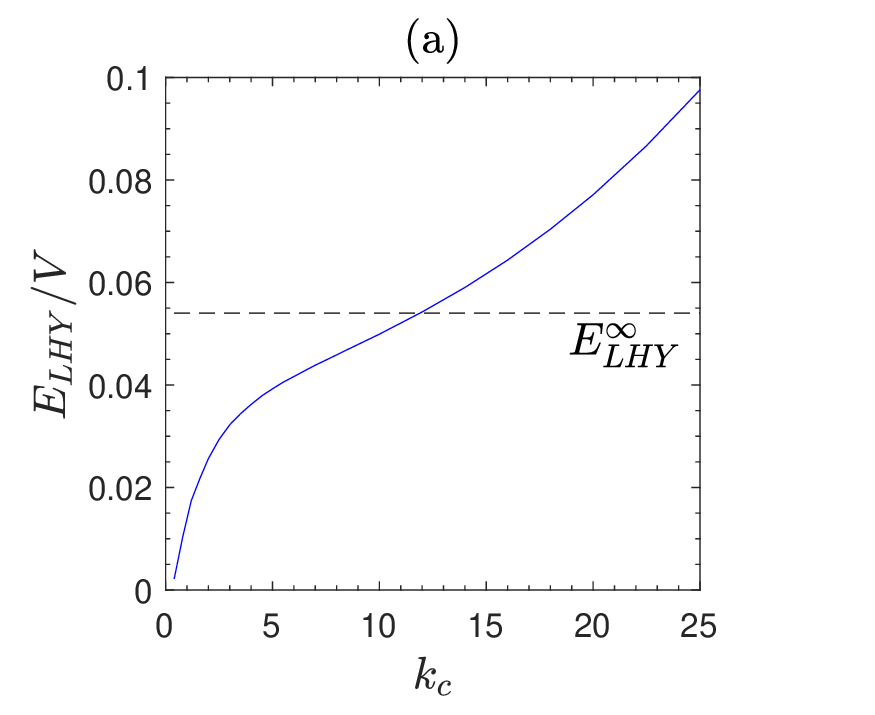}
\includegraphics[width=\hfigwidth]{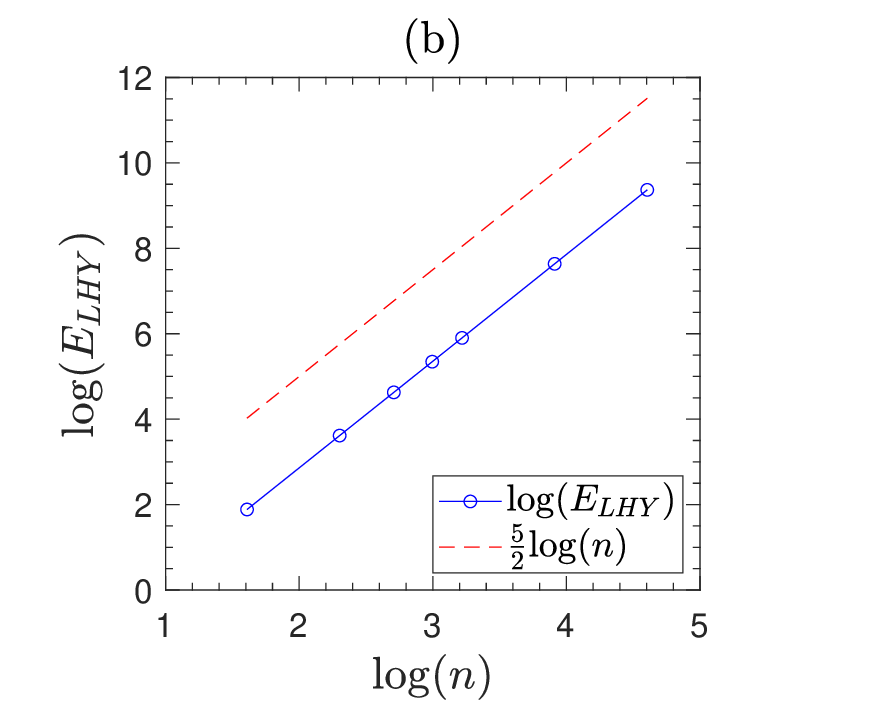}
\caption{(a): Dependence of $E_{LHY}$ on $k_c$ in a 3D homogeneous system when the interaction and density are optimised ``directly'' as in \eqn{O1}, with $g_{\rm set}=0.05, n_{\rm set}g_{\rm set}=1$, and $L=5\pi$. 
The dependence is similar, but not identical, to Fig.~\ref{fig:Ekc}(d).
(b): Dependence of $E_{LHY}$ on the density $n$ in a 3D homogeneous system with $g_{\rm set}=0.05$, $n_{\mathrm{set}}=5,10,15,20,25,50,100$, $L=5\pi$, and $M^3=25^3$ numerical modes. The blue solid line with circles shows the numerically determined data,
where 
$E_{LHY}=E^\mathrm{BOG}_\mathrm{tot}(g_0,n_0,\bo{k})-E_\mathrm{mf}(g_\mathrm{set},n_\mathrm{set},V)$ 
as defined in \eqn{O2sig}, with $n=n_{\rm set}$. The red dashed line indicates 
a $\frac{5}{2}\mathrm{log}(n)$ scaling for reference, 
confirming that $E_{LHY} \propto n^{5/2}$ within our energy-matching framework.
}
\label{fig:3D_LHY}
\end{figure}

Explicitly, we have a target interaction strength $g_{\rm set}$ and density $n_{\rm set}$, together with a particular lattice configuration (values of $\bo{k}$, box size $L$). Choosing appropriate values of $g_0$ and $n_0$ can be formulated as equating $\mathrm{LHS}$ and $\mathrm{RHS}$ of an equation:
\eq{LHSRHS}{
\mathrm{LHS}(g_{\rm set},n_{\rm set}) = \mathrm{RHS}(g_0,n_0,\bo{k}).
}
Here, the $\mathrm{LHS}$ contains the target values $g_{\rm set}=g$ and $n_{\rm set}=n$ based on 
overall theory, while the $\mathrm{RHS}$ is a calculation based on the Bogoliubov description on a concrete lattice $\bo{k}$. The matching is generally performed numerically. In principle, various quantities could be matched. One possibility would be to match the dressed interaction and density directly: 
\eqa{O1}{
g_{\rm set}&=&g_0+g_1(g_0,\bo{k})\nonu,\\
n_{\rm set}&=&n_0+\delta n(g_0,n_0,\bo{k}),
}
which would give perfect agreement on the mean-field energy. However, as shown later in Fig.~\ref{fig:3D_LHY}(a), this does not provide the best match to the LHY energy, which is our main target here. Therefore, instead, we directly match the total calculated energy based on the Bogoliubov description, $E_{\rm tot}^\mathrm{BOG}$, to the expected beyond-mean-field energy, $E_{\rm tot}^\mathrm{BMF}$, involving \eqn{Emf} and \eqn{LHY} as per:
\eqa{O2sig}{
\mathrm{LHS} &=& E_{\rm tot}^{\rm BMF}= E_{\rm mf}(g_{\rm set},n_{\rm set},V)+E_{LHY}^{\infty}(g_{\rm set},n_{\rm set},k_c),\nonu\\
\mathrm{RHS} &=& E_{\rm tot}^{\rm BOG} =E_{\rm kin}(g_0,n_0,\bo{k})+E_{\rm int}(g_0,n_0,\bo{k}).\qquad
}
The cutoff $k_c$ appears in $E_{LHY}^{\infty}$ if the system is 2D. 
Some choices must be made to ensure self-consistency -- in particular, the expressions for the energies on the $\mathrm{RHS}$ could, in principle, be chosen to involve either $n_0$ or $n_{\rm set}$ and similarly $g_0$ or $g_{\rm set}$ in various places. 
Here, the Bogoliubov spectrum based on $g_0$ and $n_0$ is used.
As shown in Appendix~\ref{BAREDRESS}, the difference between these choices amounts only to a higher-order correction, although it leads to small numerical differences in the ultimately matched values.

In order to carry out the matching, the $\mathrm{RHS}$ quantities are calculated directly on the chosen $\bo{k}$ lattice by summing the kinetic and interaction-energy contributions as per \eqn{deltandisc}, \eqn{sumkekin}, \eqn{G2_1}, \eqn{mdef}, \eqn{Eint}. 

To finalise the matching of $g_0$ and $n_0$ to $g_{\rm set}$ and $n_{\rm set}$, we have found that, rather than performing a two-parameter optimisation, a more convenient and stable self-consistent approach is to iterate as follows:
\begin{enumerate}
    \item Choose an initial guess $n_0^{(0)}=n_{\rm set}$.
    \item Evaluate the Bogoliubov energies for several values of the bare interaction $g_0$. This involves several substeps:
    \begin{enumerate}
        \item Using the current estimate $n_0^{(0)}$ and chosen $g_0$, compute the Bogoliubov coefficients $U_{\bo{k}}$, $V_{\bo{k}}$, as well as $E_{\rm kin}$, the depletion $\delta n$, and the anomalous pair density $\wb{m}$.  These are all functions of $n_0^{(0)}$ and $g_0$. 
        \item  Using the $\delta n(n_0^{(0)},g_0)$ from the previous step, obtain an improved estimate of the condensate density $n_0(n_0^{(0)},g_0)$ $= n_{\rm set}-\delta n(n_0^{(0)},g_0)$, and substitute this value into the explicit $n_0$ factors in \eqn{G2_1} to thus calculate $G_2$ and $E_{\rm int}$. Omitting this step leads to convergence problems.
    \end{enumerate}
    \item Interpolate between the resulting values of $E_{\rm tot}^\mathrm{BOG}(n_0^{(0)},g_0) = E_{\rm int}+E_{\rm kin}$ obtained above for different $g_0$ values to estimate the value $g_0=g_0^{\rm fit}$ that satisfies the energy-matching condition  \eqn{LHSRHS}. 
    \item Compute also the corresponding best estimate of the condensate fraction, 
    \eq{n0est}{
    n_0^{\rm fit}=n_{\rm set}-\delta n(n_0^{(0)},g_0^{\rm fit}).
    }
    \item Return to step 2 for the next iteration by using the latest condensate-density estimate $n_0^{(0)}=n_0^{\rm fit}$, repeat the procedure until one is satisfied with the convergence.
\end{enumerate}

\begin{figure}[h]
\centering
\includegraphics[width=\hfigwidth]{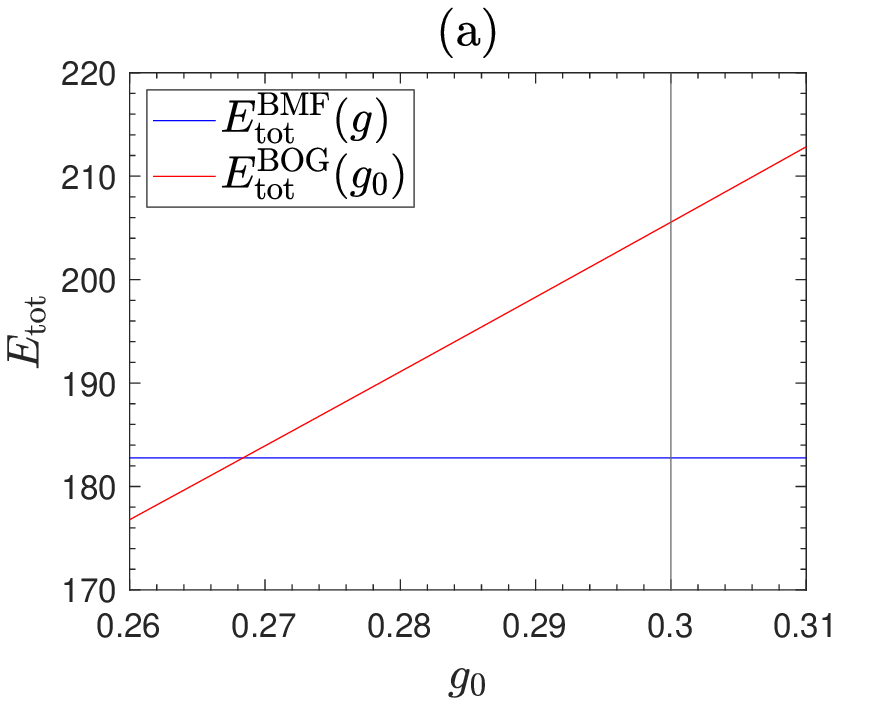}
\includegraphics[width=\hfigwidth]{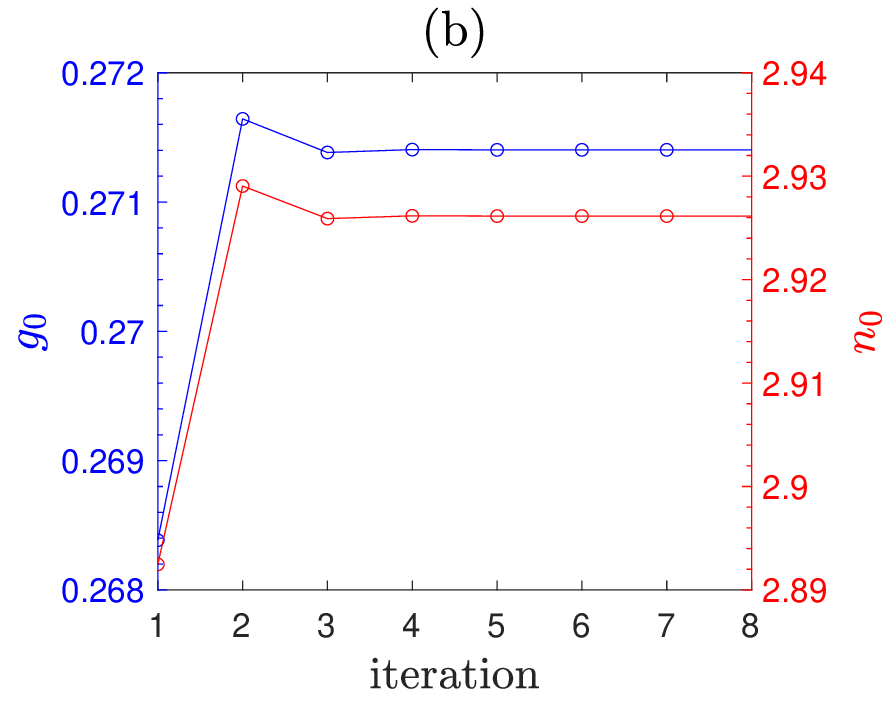}
\caption{
Numerical determination of the bare interaction $g_{0}$ that matches the energy of a uniform 1D Bogoliubov ground state with $g_{\rm set}=0.3$ and $n_{\rm set}=1/g_{\rm set}\approx3.3333$. 
Here $L=40\pi$, 
$M=301$, and therefore $k_c=7.5, k_L\approx0.025$. The corresponding $E^\mathrm{BMF}_\mathrm{tot}\approx182.773$.
(a)
First iteration. To satisfy $E^\mathrm{BMF}_\mathrm{tot}=E^\mathrm{BOG}_\mathrm{tot}$, the matched value after the first iteration is $g_{0}^{\rm fit}\approx 0.2684$. 
%
(b)
Self-consistency: convergence of $g_{0}$ and $n_0$ under further iterations following the initial matching presented in (a). 
After 8 iterations, the condensate density converges to $n_{0}\approx2.9261$ and the depletion density to $\delta n\approx0.4072$ at a value of $g_{0}\approx0.2714$. 
}
\label{fig:get_bare_g0}
\end{figure}

The $g_0$ determined from the above numerical algorithm differs from the standard continuum coupling (i.e., the dressed coupling $g$ or $g_{\text{set}}$) because the Bogoliubov matching framework must explicitly account for lattice-cutoff effects and the contributions of the ``extra'' cross-correlations. 
However, because it is derived entirely from analytical Bogoliubov theory, 
it functions as a systematically derived microscopic parameter for the lattice theory. In summary, the bare coupling ($g_0$) is the unphysical, ultraviolet-divergent interaction parameter used as an input in the microscopic lattice Hamiltonian. The value of $g_0$ depends directly on the lattice grid spacing (momentum cutoff $k_c$) and is determined through the above matching algorithm. The dressed coupling ($g$ or $g_{\text{set}}$) corresponds to the physical, regularised interaction strength of the continuous quantum field, which is independent of the numerical grid and represents the observable low-energy scattering properties of the Bose gas.
Fig.~\ref{fig:get_bare_g0}(a) shows an example of the total energy $E_{\rm tot}^{\rm BOG}$ of a uniform 1D system as a function of the bare interaction $g_0$ 
after the first iteration of the algorithm above (i.e., after carrying out step 2).  
The matched value of $g_0$ is 
$0.2684$. 
Subsequent iterations of the algorithm above generate the converging sequence shown in Fig.~\ref{fig:get_bare_g0}(b). Three iterations are sufficient for complete convergence, with the first iteration already providing a very good match.

Finally, in principle, the above algorithm can be applied without significant modification to non-uniform systems with Bogoliubov modes $j$ and coefficients $U_j$, $V_j$, and using the total condensate occupation $N_0$ instead of $n_0=N_0/V$ (for the uniform case) as the second matching parameter on the $\mathrm{RHS}$ of \eqn{LHSRHS}, together with a spatially averaged mean-field and LHY energy density to construct the $\mathrm{LHS}$.

\subsubsection{Estimated bare interaction}

The functional estimates from Sec.~\ref{OBS} can also be used for the evaluation of the $\mathrm{RHS}$ in step 2 of the algorithm, but there is one subtlety to keep in mind. The expressions for 2D and 3D given there assume a spherical cutoff at $|\bo{k}|=k_c$, and therefore those estimates only provide a good choice of $g_0$ and $n_0$ for TWA simulations whose initial conditions and subsequent dynamics are carried out in a similarly projected mode space. For the case of an unprojected ``square'' $\bo{k}$-space, Appendix~\ref{SQUARE} provides an appropriately tailored large-box integral. 

These formulae are useful both for obtaining explicit estimates of the best-matched quantities and for understanding the dominant contributions. 
Obtaining them turns out to be more involved than one might expect, but an outline is provided in Appendix~\ref{A:DG}. An estimate for the shift of the bare interaction relative to the dressed one can be obtained (when small) to leading orders $\mc{O}(n)$ and $\mc{O}(g)$ in the regime $n\gg1$, $k_c\gg1$, via evaluating the first iteration of the algorithm. In 1D, it comes out as
\eqa{dgest1d}{
g_0&=&g_{\rm set}(1+\delta_g),\\
\delta_g^{(1d)} &\approx& -\frac{1+\frac{gk_c}{2\pi}\left(\mc{L}^2-2-\frac{4\left(\mc{L}-1\right)}{k_c}\right)}{n\pi k_c+2-2k_c+\frac{gk_c}{2\pi}\left(2\mc{L}^2+\mc{L}-4-\frac{2\left(5\mc{L}-4\right)}{k_c}\right)}.\nonu
}
Here $\mc{L}=\log(4/k_L)$, $n=n_{\rm set}$, $g=1/n$, and assuming $\delta_g\ll 1$. For the case shown in Fig.~\ref{fig:get_bare_g0},
this estimate gives $\delta_g=-0.107$, which is 
close to the full calculation value of $-0.0953$. Notably, in this case the lower-order contributions explicitly proportional to $g$ are dominant (without them, the estimate becomes $\delta_g=-0.015$). All of these contributions are absent at the usual Bogoliubov order -- specifically, cross terms such as $\delta^2+\wb{m}^2$ in $E_{LHY}^{\rm extra}$, as well as corrections arising from $n_0$ being smaller than $n$.

For smaller values of $g_{\rm set}$ and/or box size $L$, the $g$-dependent terms in the estimate \eqn{dgest1d} become much less important, and the tendency is toward $\delta_g^{(1d)}\to-1/(n\pi k_c)$. The corresponding expressions in 2D and 3D are given by \eqn{dgest2d} and \eqn{dgest3d}.

\begin{figure}[h]
\centering
\includegraphics[width=\hfigwidth]{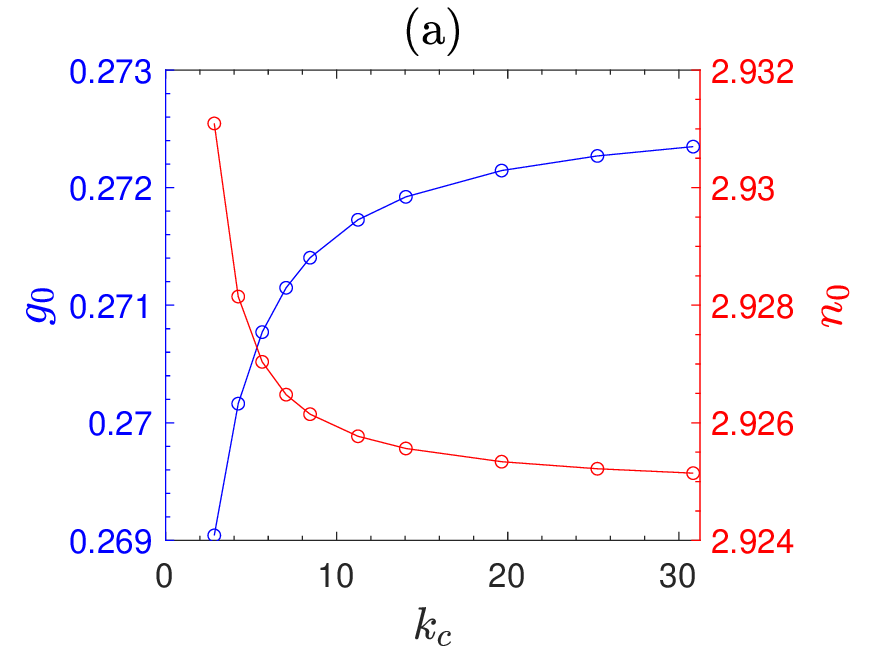}
\includegraphics[width=\hfigwidth]{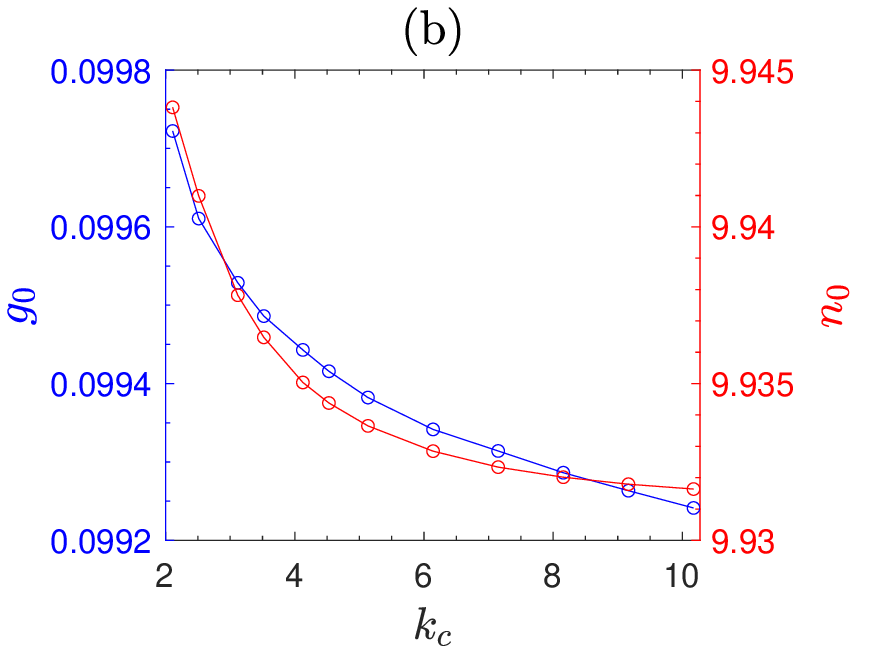}\\
\includegraphics[width=\hfigwidth]{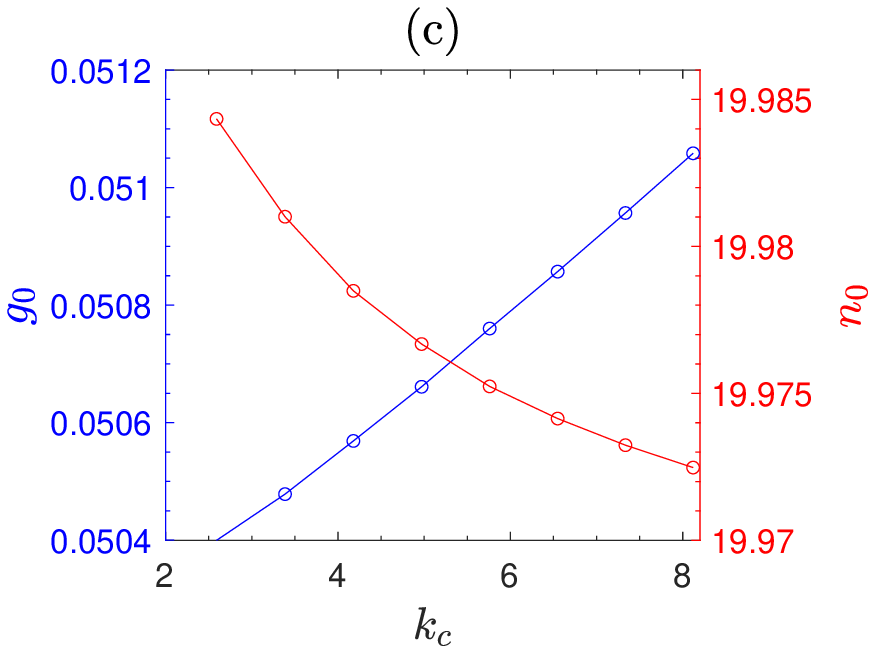}
\includegraphics[width=\hfigwidth]{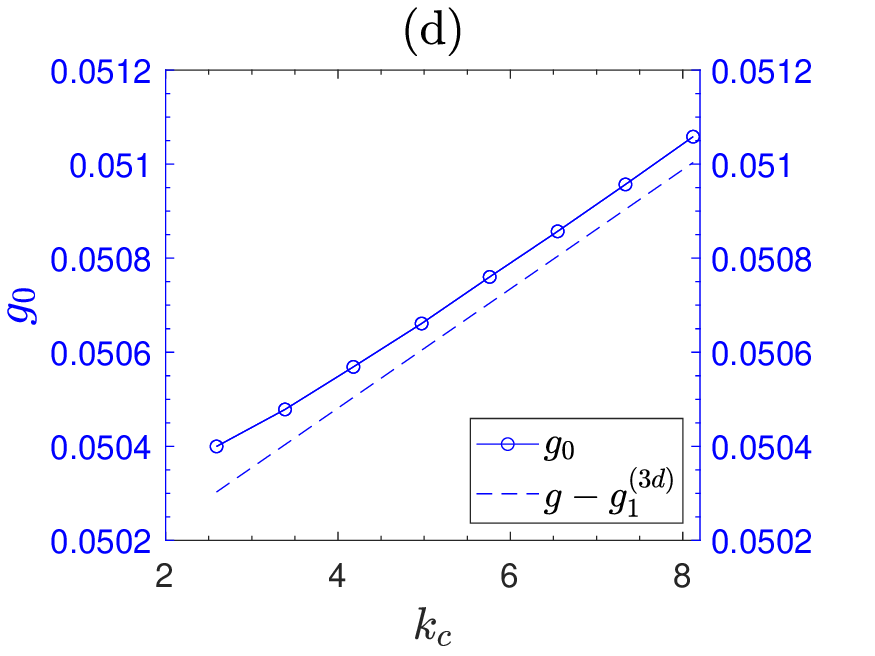}
\caption{(a-c): Dependence of properly matched $g_0$ and $n_0$ on the numerical cutoff momentum $k_c=k_{\rm max}\xi_{\rm set}$ for a uniform gas in 1D (a), 2D (b), and 3D (c). In 2D and 3D,
a spherical cutoff $|\bo{k}|\le k_c$ was imposed. 
Numerical parameters are as follows:
1D~(a): $L=40\pi$ and $g=0.3$, 
$M=101$, $151$, $201$, $251$, $301$, $401$, $501$, $701$, $901$, $1101$ numerical modes;
2D~(b): 
$L=10\pi$, $V=L^2$, and $g=0.1$. $M^2=21^2$, $25^2$, $31^2$, $35^2$, $41^2$, $45^2$, $51^2$, $61^2$, $71^2$, $81^2$, $91^2$, $101^2$;
3D~(c): 
$L=5\pi$, $V=L^3$,  $g=0.05$. $M^3=13^3$, $17^3$, $21^3$, $25^3$, $29^3$, $33^3$, $37^3$, $41^3$ numerical modes.   
(d): Comparison between the matched $g_0$ in 3D (solid line, same data as in Fig.~\ref{fig:g0n0123D}(c)) and the coupling constant $g-g^{(3d)}_1$ (dashed line) from the regularisation \eqn{g=g1g2} and \eqn{g13d} in 3D.
}
\label{fig:g0n0123D}
\end{figure}

\subsubsection{Cutoff dependence}
\label{KCD}
As we have seen, the numerically produced Bogoliubov ground state and the resulting observable estimates are, in general, cutoff- and lattice-dependent, while the mean-field and LHY energy densities (apart from the 2D case) are not. Therefore, the best-matched values of $g_0$ and $n_0$ will depend on the cutoff. 

Fig. \ref{fig:g0n0123D}(a) shows the $g_0$ and the corresponding $n_0$, determined following our scheme, at various momentum cutoffs $k_c=k_{\rm max}\xi_{\rm set}$ in a one-dimensional homogeneous system. Since the Bogoliubov coefficients converge as ${k\rightarrow\infty}$ and the fluctuations are dominated by low-energy modes, both $g_0$ and $n_0$ exhibit cutoff convergence at large $k_c$. 
For a two-dimensional system, the LHY correction \eqn{LHY2d} has a logarithmic dependence on the maximum allowed momenta $k_c$, and therefore one does not expect convergence of $g_0$ in the $k_c\to\infty$ limit. 
Fig. \ref{fig:g0n0123D}(b) shows the resulting dependence of the matched $g_{0}$ and $n_{0}$ in 2D.
%
%
An example of the situation in 3D is shown in Fig.~\ref{fig:g0n0123D}(c). 
Interestingly, in this case the variation in $g_0$ and $n_0$ is very minor. 



\subsubsection{Matching density and interaction instead of energy}

Wrapping up the investigation of technical options, Fig.~\ref{fig:3D_LHY}(a) shows the behaviour of the LHY correction in 3D when \eqn{O1} is used to determine $g_0$ and $n_0$ directly, instead of our usual energy-matching procedure employed elsewhere.

One observes that the energy shift is of approximately the right magnitude, but it is not properly matched except for one particular, rather high, cutoff choice of $k_c\approx 12$ in this case. As we will see in the next section, such high cutoffs lead to a very poor signal-to-noise ratio in the TWA. It therefore appears that the energy-matching of \eqn{LHSRHS} is much more practical. Fig.~\ref{fig:3D_LHY}(b) verifies the energy-matching procedure reproducing the LHY energy density scaling law in 3D homogeneous systems, where in 3D the renormalised LHY energy \eqn{LHY3d} is proportional to the density, $n^{5/2}$, the energy-matching framework valid across a wide range of density.





\subsection{Observables in the TWA}
\label{cutoff}

Now that we are equipped with values of the bare interactions $g_0$ and condensate fractions $n_0$ matched to the correct dressed mean-field and LHY energies, in this section we investigate how well other observables are represented, and how closely TWA stochastic realisations follow the underlying Bogoliubov theory in practice (i.e., how does the signal-to-noise ratio look like?). 


\subsubsection{One-dimensional homogeneous system}


\begin{figure}[h]
\centering
\includegraphics[width=\hfigwidth]{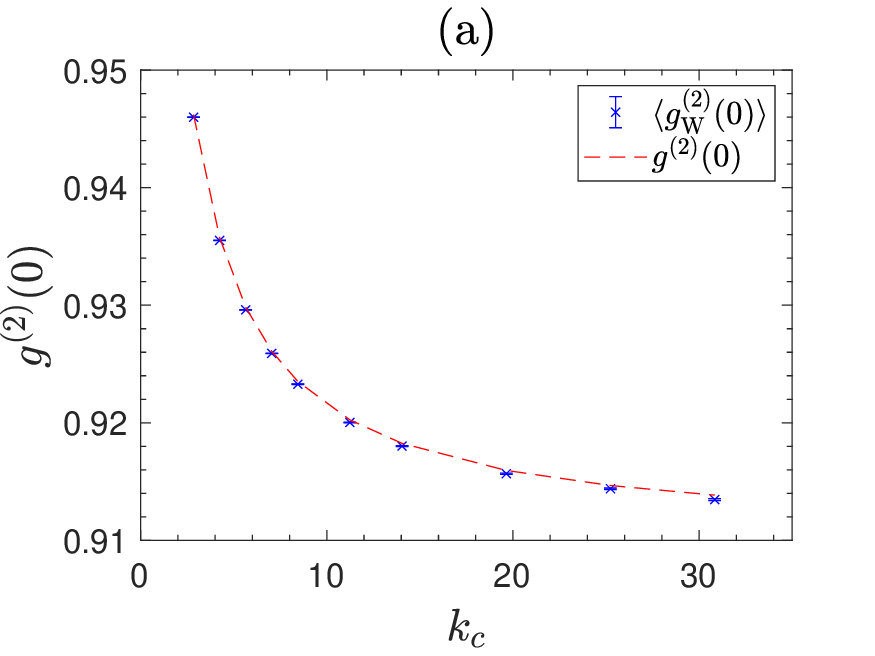}
\includegraphics[width=\hfigwidth]{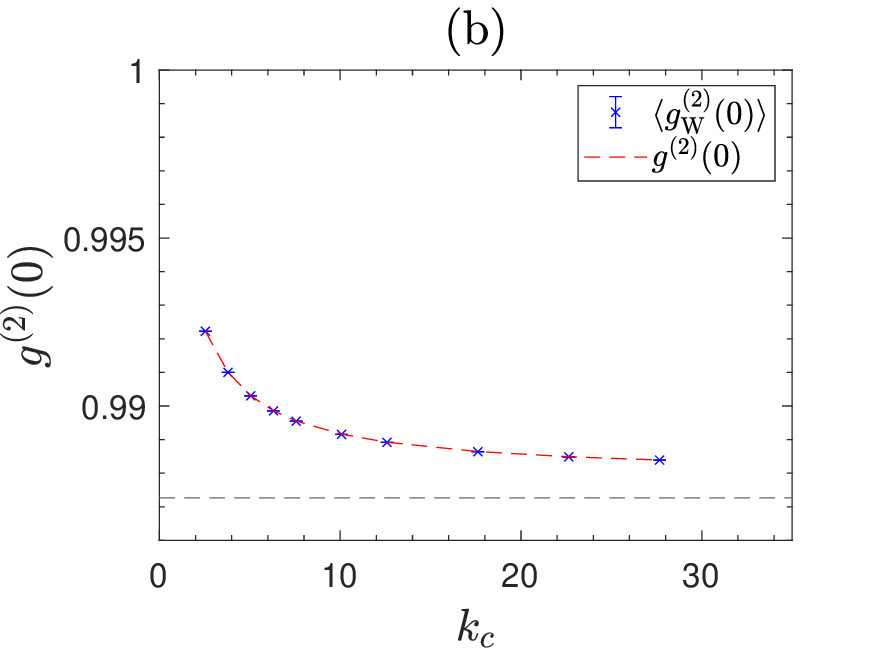}
\caption{
TWA predictions of the local density-density correlation $g^{(2)}_{\rm W}(0)$ in a uniform 1D gas and its dependence on the maximal lattice momentum $k_c$ (blue data points). $10^7$ TWA trajectories were used, with $1$ standard deviation error bars (negligible). 
Red dashed lines indicate Bogoliubov-based estimates that include the ``extra'' cross terms using  \eqn{G2_1}. Box size $L=40\pi$, and 
(a): $g=0.3$ ($\gamma_L=0.09$) as per the system in Fig.~\ref{fig:g0n0123D}(a); 
(b): $g=0.02$ at much lower $\gamma_L=0.0004$, hence with a much lower influence of $E_{\rm int}^{\rm extra}$. The grey dashed line here shows the thermodynamic value $1-2g/\sqrt{gn}\pi=0.98727$. The value in the full Lieb-Liniger model is $0.98723$ [\citenum{Kheruntsyan03}].}
\label{fig:g20_1d_natural}
\end{figure}


\begin{figure}[htb]
\centering
\includegraphics[width=\hfigwidth]{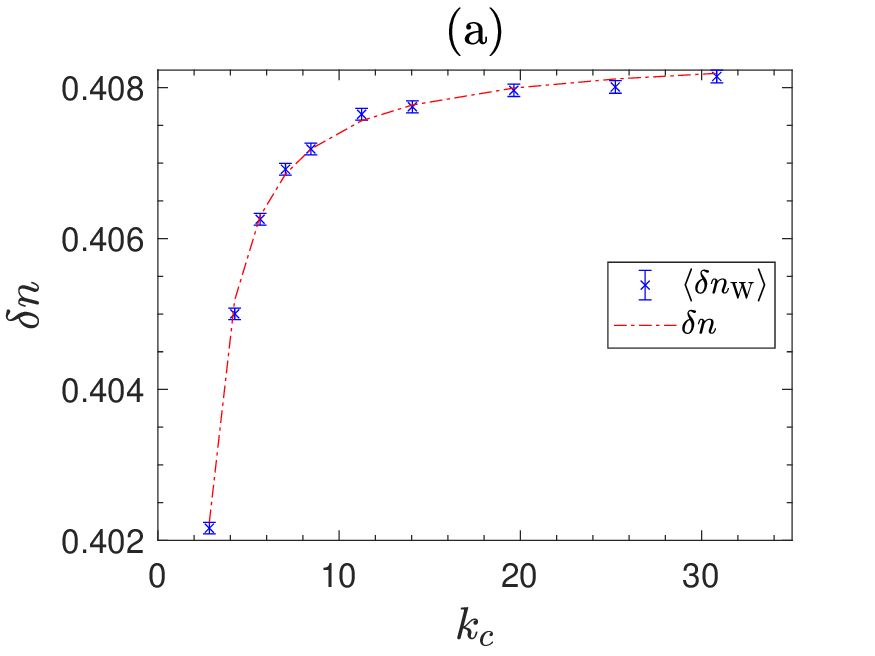}
\includegraphics[width=\hfigwidth]{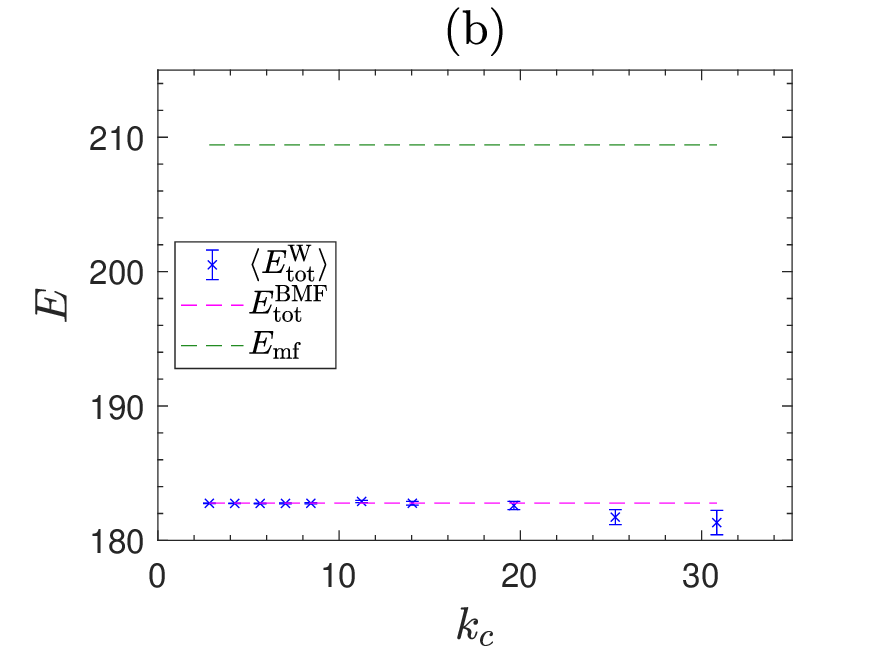}\\
\includegraphics[width=\hfigwidth]{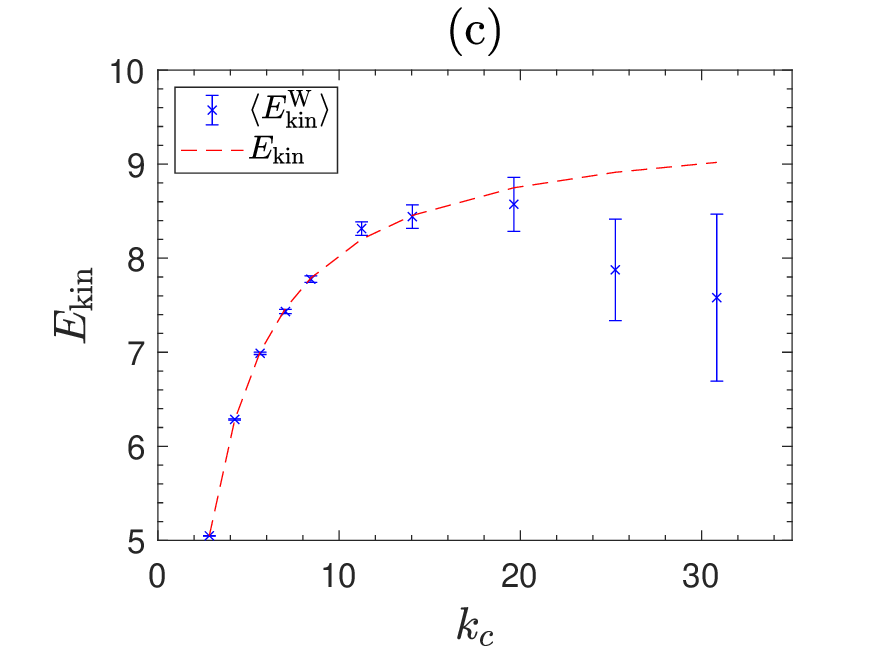}
\includegraphics[width=\hfigwidth]{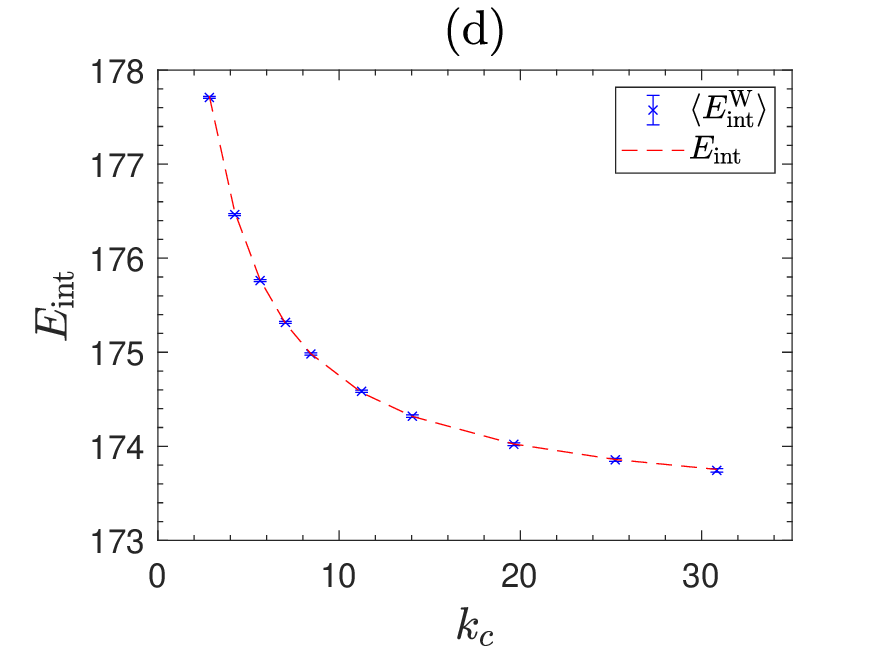}
\caption{TWA predictions (blue) of several observables in a uniform 1D gas and their dependence on the maximal lattice momentum $k_c=k_{\rm max}\xi_{\rm set}$, for the parameters of Figs.~\ref{fig:g0n0123D}(a). $10^7$ TWA trajectories were used, error bars show statistical uncertainty of $1$ standard deviation. 
Bogoliubov-based estimates that include the ``extra'' cross terms are shown as red dashed lines.  
(a): Condensate depletion;
(b): Total energy, $E^\mathrm{BMF}_{\mathrm{tot}} =182.8$ from \eqn{O2sig}, while $E_\mathrm{mf}=209.4$ is the mean-field energy without LHY corrections from \eqn{Emf}.
(c-d): Kinetic and interaction energies. 
}
\label{fig:1d-4obs}
\end{figure}

\begin{figure}[htb]
\centering
\includegraphics[width=0.6\columnwidth]{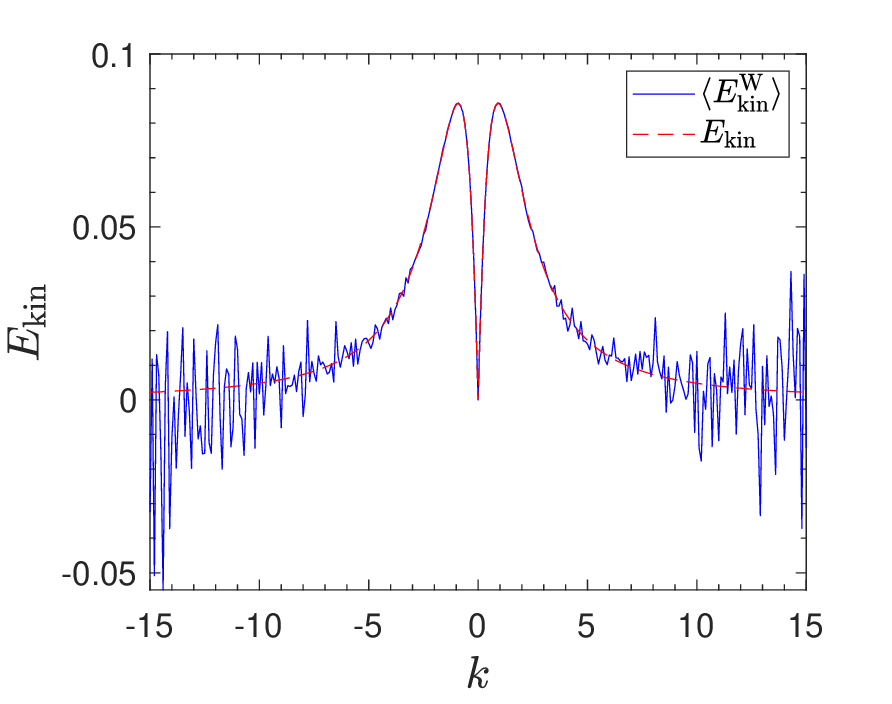}
\caption{Kinetic energy integrand for $\Delta k=0.1$, $g_0=0.2$, $n_0=5$, and $L=2\pi/\Delta k$. Blue: sum over $10^7$ Wigner trajectories; red: exact contributions $k^2E_{\bo{k}}/2$ from \eqn{sumkekin}.} 
\label{fig:ekin-k}
\end{figure}

We have calculated the normalised local pair-correlation $g^{(2)}(0)$ and $g^{(2)}_{\rm W}(0)$ with the TWA using LHY-matched $g_0$ and $n_0$ values for two homogeneous systems: one as in Fig.~\ref{fig:g0n0123D}(a), with strong beyond-mean-field effects due to the large $g=0.3$ value, shown in Fig.~\ref{fig:g20_1d_natural}(a), and one at much lower $\gamma_L$, shown in Fig.~\ref{fig:g20_1d_natural}(b).
The TWA observable expression uses \eqn{TWAg2}.
%
%
Unlike a 
mean-field description, 
a very strong antibunching is observed, and happily the $k_c$ dependence is very small once values $k_c\gg1$ are reached. 
As per the previous discussion and in \eqn{exreg1d}, we know that the normalised pair-correlation values will be distorted compared to a pure Bogoliubov description because of the ``extra'' cross terms when $\gamma_L$ is large. Fig.~\ref{fig:g20_1d_natural}(b) shows that this effect does indeed become small 
as $g$ decreases.



The depletion dependence on the
cutoff is shown in Fig.~\ref{fig:1d-4obs}(a), again for the $g=0.3$ case with strong effects. Here, the depletion $\delta n\approx0.4$ is about $13\%$ of the condensate density $n_0$ (Fig.~\ref{fig:g0n0123D}(a)). In the TWA, we can calculate the depletion by subtracting the condensate mode occupation, 
\eq{depletion_W}{
\delta n_\mathrm{W}=n_\mathrm{W}-n_0,
}
which is straightforward in a uniform system since the condensate is merely the $k=0$ mode in $k$-space.
The corresponding Bogoliubov-based prediction is given in \eqn{n1d}.
The signal-to-noise ratio in the TWA is satisfactory for both $g^{(2)}_{\rm W}(0)$ and $\delta n_{\rm W}$.


Fig.~\ref{fig:1d-4obs}(b) displays the total energy as a function of $k_c$. As prescribed and hoped for, the TWA results match the targeted energy $E_{\rm tot}^{\rm BMF}=E_{\rm mf}+E_{LHY}^{\infty}$, with no visible bias. They also resolve the difference between the mean-field and LHY-corrected value well. However
the energy uncertainty increases as the cutoff grows.

It is instructive to examine how energy is distributed between kinetic and interaction components. Figs.~\ref{fig:1d-4obs}(c) and \ref{fig:1d-4obs}(d) show TWA values for the now ``standard'' case discussed above. 
The growing statistical uncertainty lies in the kinetic energy part. Inspection of the contributions of individual modes into \eqn{sumkekin}, shown in Fig.~\ref{fig:ekin-k}, reveals that the noisiness grows very substantially as $k$ increases (due to the $k^2$ factor). This is a primary reason why very high cutoffs $k_c$ are detrimental in practice. It also suggests that attempting to obtain LHY corrections through \emph{including a properly resolved potential} without dressing is likely to fail in practice -- since this would lead to the requirement of a very high $k_c$, and result in enormous kinetic noise in the TWA.



\begin{figure}[htb]
\centering
\includegraphics[width=\hfigwidth]{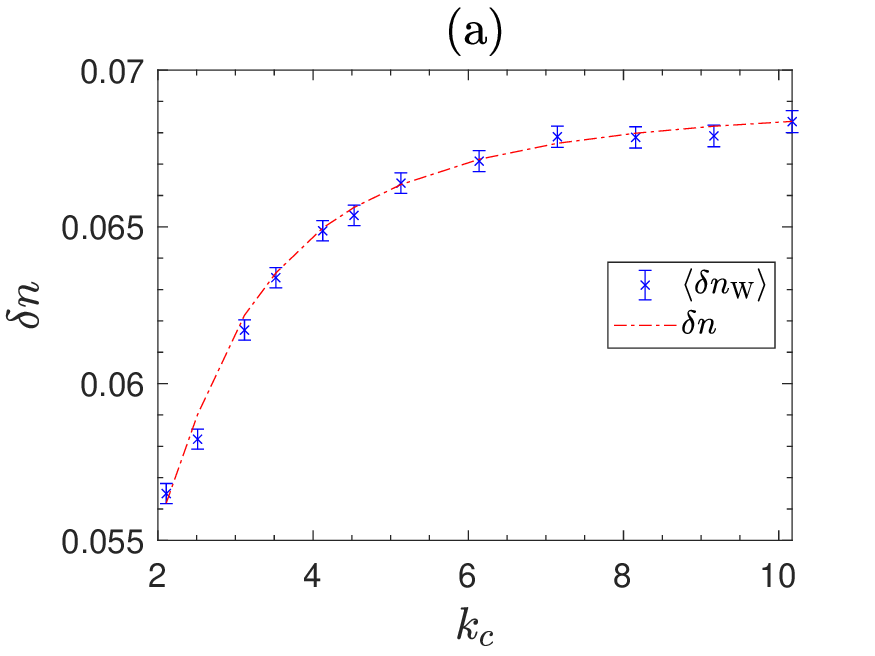}
\includegraphics[width=\hfigwidth]{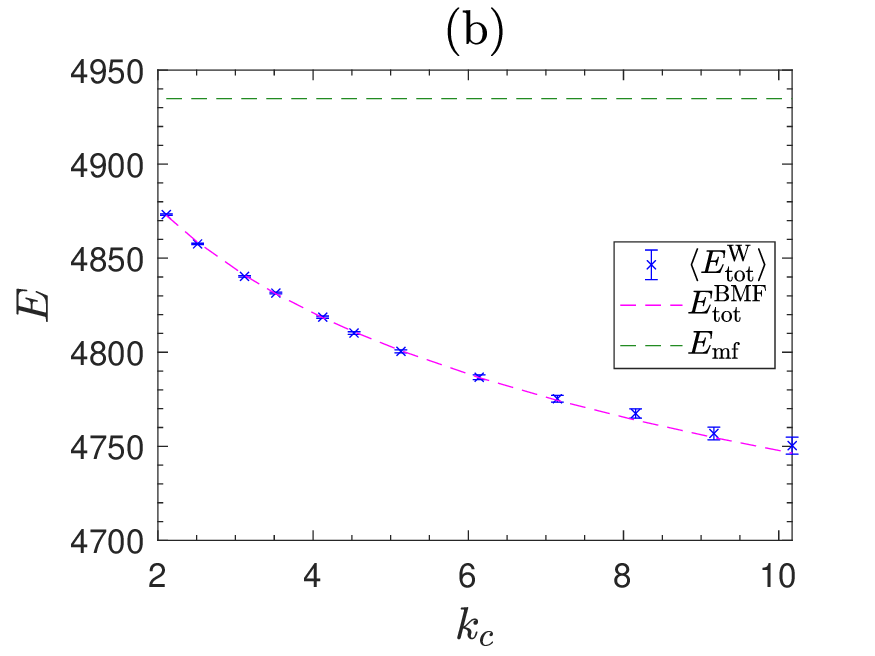}\\
\includegraphics[width=\hfigwidth]{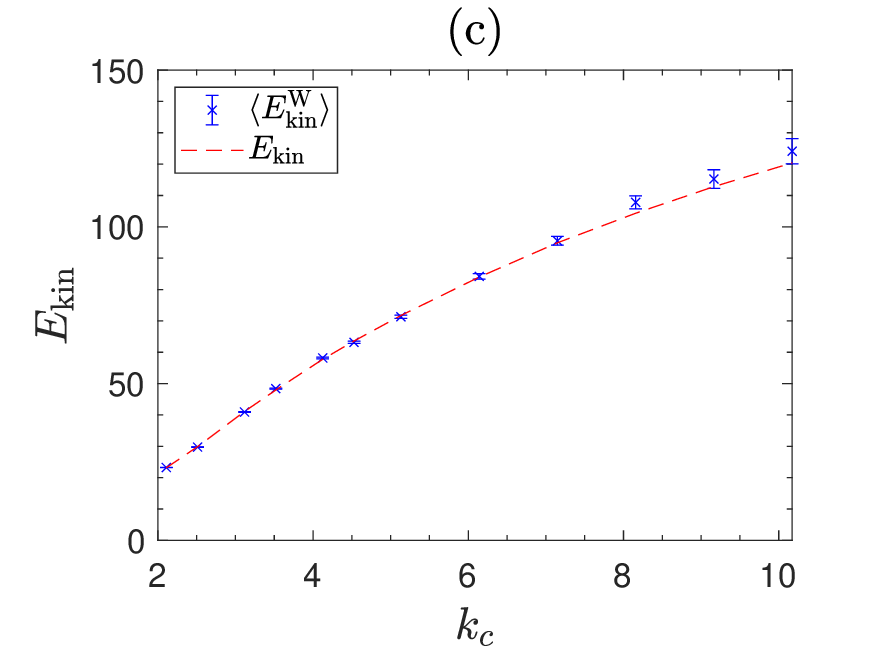}
\includegraphics[width=\hfigwidth]{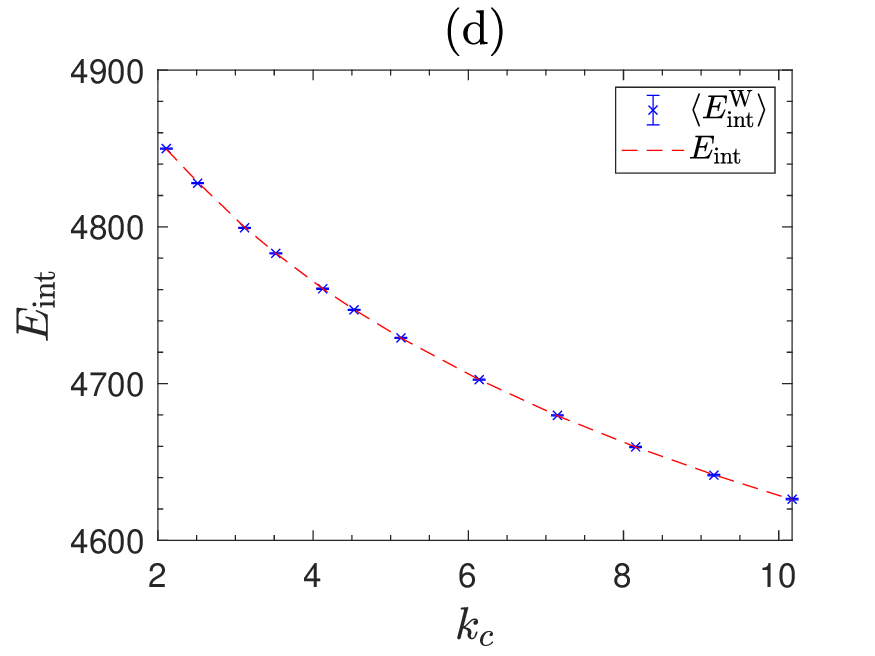}
\caption{TWA predictions (blue) of observables in a uniform 2D gas, for the parameters of Fig.~\ref{fig:g0n0123D}(b). A spherical cutoff in $\bo{k}$-space at $|\bo{k}|\le k_c=k_{\rm max}\xi_{\rm set}$ was used. Figure format is the same as in Fig.~\ref{fig:1d-4obs}. TWA simulations employed $10^5$ trajectories. In (b), $E_\mathrm{mf}=4935$ is the mean-field energy without LHY corrections.
}
\label{fig:2d-4obs}
\end{figure}

\subsubsection{Two-dimensional homogeneous system}


Turning to the 2D case, Figs.~\ref{fig:2d-4obs}(a) and~\ref{fig:2d-4obs}(b) show the behaviour and relative noisiness of the depletion and total energy, respectively. 
The Bogoliubov-based prediction, shown in red, is evaluated from the direct sum. The total energy remains $k_c$-dependent in this dimensionality. The agreement is excellent, while the distinction between mean-field and LHY-inclusive energies is even more clearly resolved. Notably, the increase in noisiness with cutoff is much less pronounced than in 1D.
Figs.~\ref{fig:2d-4obs}(c,d) show the kinetic and interaction energy components. In contrast to the 1D case, the growth of statistical uncertainty in the kinetic energy is much reduced in 2D.



\begin{figure}[htb]
\centering
\includegraphics[width=\hfigwidth]{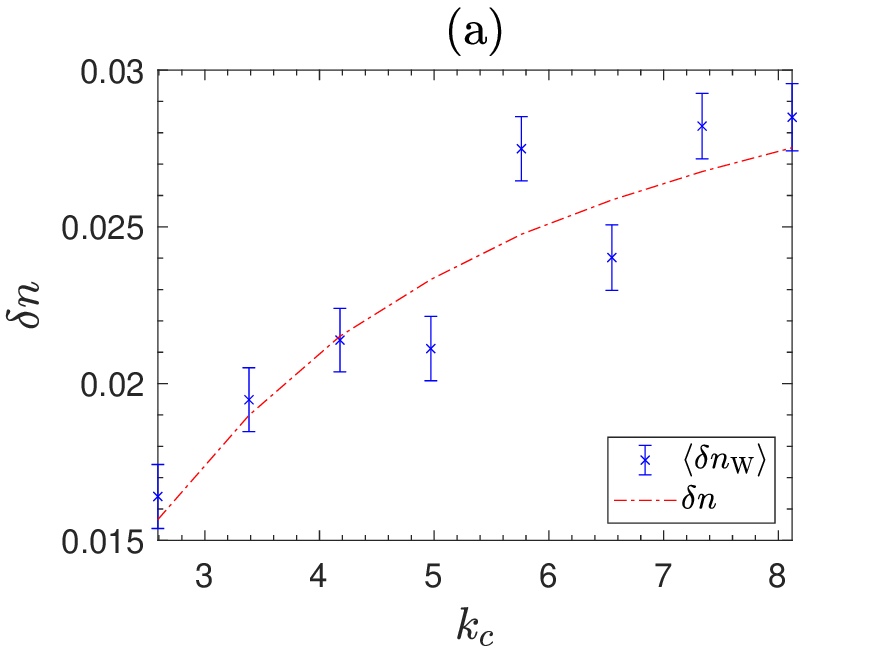}
\includegraphics[width=\hfigwidth]{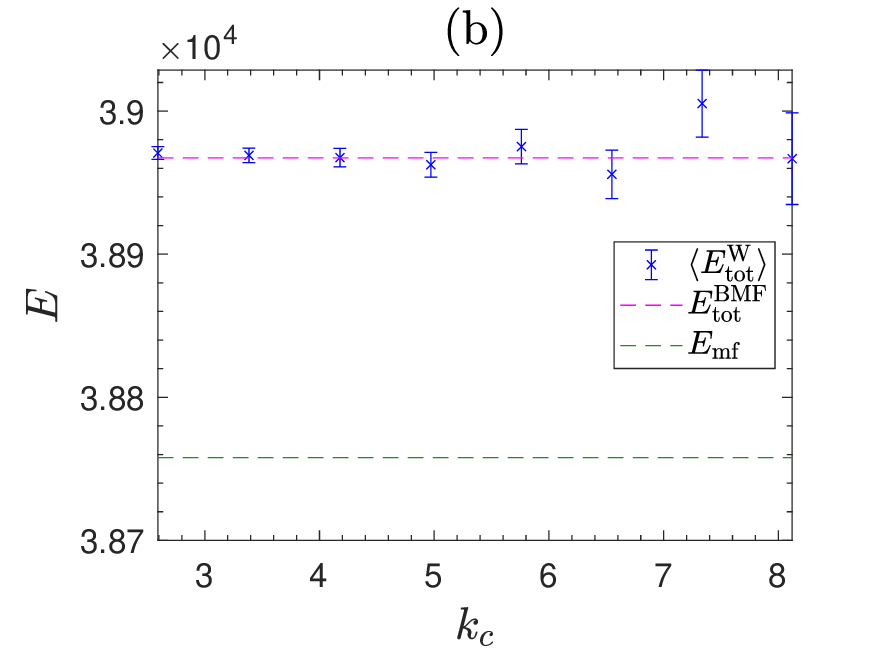}\\
\includegraphics[width=\hfigwidth]{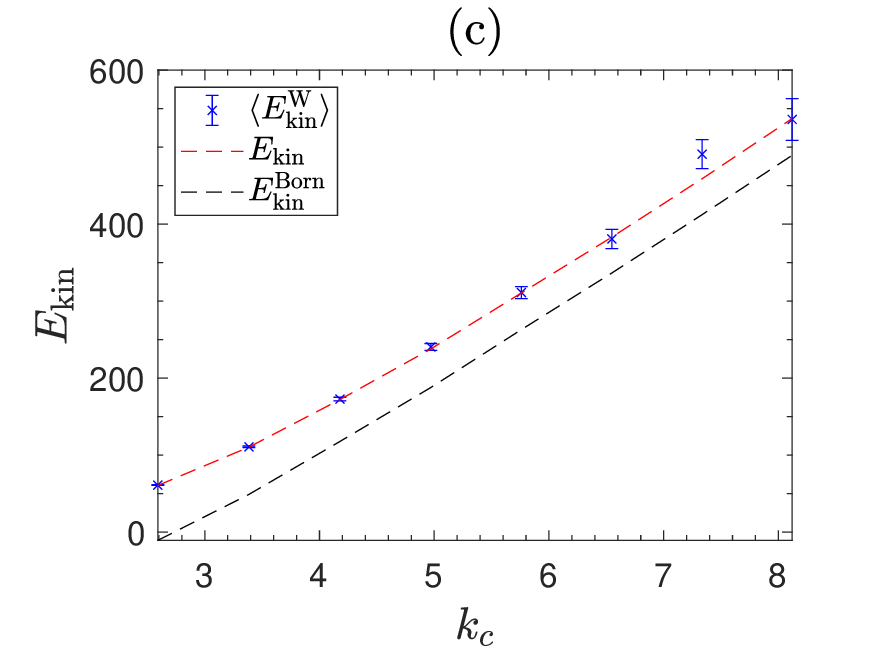}
\includegraphics[width=\hfigwidth]{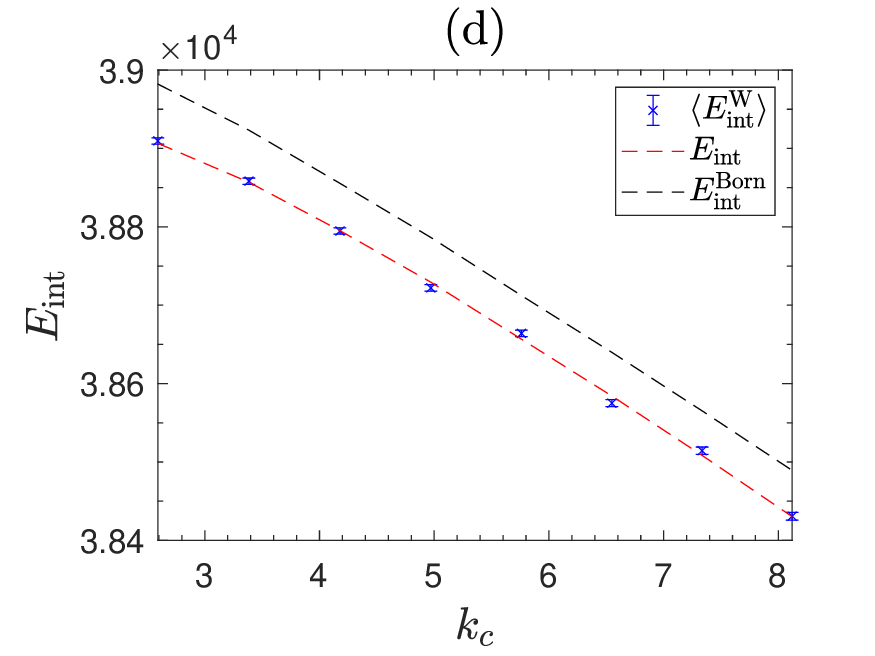}
\caption{TWA predictions (blue) of observables in a uniform 3D gas, for the parameters of Fig.~\ref{fig:g0n0123D}(c) in the same figure format as Fig.~\ref{fig:1d-4obs}. A spherical cutoff in $\bo{k}$-space at $|\bo{k}|\le k_c=k_{\rm max}\,\xi_{\rm set}$ was used. The TWA simulations employed $5000$ trajectories. In (b), $E_\mathrm{mf}=3.876\times10^4$ is the mean-field energy without LHY corrections, while $E^\mathrm{BMF}_\mathrm{tot}(g)=3.897\times10^4$ includes LHY corrections. In (c,d), $E^{\rm Born}_{\rm kin}$ and $E^{\rm Born}_{\rm int}$ are the kinetic and potential energies from the 2nd Born approximation (black), as described by \eqn{Echerny}. 
}
\label{fig:3d-4obs}
\end{figure}

\subsubsection{Three-dimensional homogeneous system}

Finally, for the 3D system, $\delta n$ and the total energy are shown in Figs.~\ref{fig:3d-4obs}(a) and~\ref{fig:3d-4obs}(b). 
As in 2D, the 3D system exhibits a well-resolved difference between mean-field and LHY-inclusive energies, with good signal-to-noise characteristics. 
The quantum depletion $\delta n$ follows \eqn{N3d}, which converges slowly as $1/k_c$ to the limiting value of $1/3/\pi^2=0.03377$, as reflected also in the relatively slow convergence of $n_0$ values in Fig.~\ref{fig:g0n0123D}(c). It therefore appears that it is numerically much easier to achieve accurate emulation of the LHY energy corrections than to reproduce the details of the quantum depletion.



Figs.~\ref{fig:3d-4obs}(c,d) show the kinetic and interaction energy components and additionally compare them to the 2nd Born approximation estimates \cite{Cherny01} for these partial energies:
\eq{Echerny}{
E^{\rm Born}_{\rm kin} = \frac{nN}{2}\left(-g_1-\frac{8}{5\pi^2}\sqrt{ng_{0}^5}\right) \qquad,\qquad
E^{\rm Born}_{\rm int} = \frac{nN}{2}\left(g+g_1+\frac{8}{3\pi^2}\sqrt{ng_{0}^5}\right),
}
where $g_1=g-g_0$ and $g=g_{\rm set}, n=n_{\rm set}$. 
Here we can see the reappearance of the underlying divergence through $g_1$ 
in a mutually cancelling form. 
\section{Full LHY dynamics in the Wigner representation versus coherent models}
\label{DYN}

Having developed the ability to implement LHY physics in the Wigner representation, we present here a preliminary study of the dynamical consequences of the LHY contributions, and how the full TWA model differs from the standard EGPE theory, in which LHY effects appear coherently and only as an additional energy in the equation of state.
We consider three test cases with uniform initial conditions and periodic boundary conditions, for which the initial state prescriptions in Sec.~\ref{BOG} can be followed directly. 
Non-uniform starting conditions, such as bright solitons or quantum droplets, will be treated in a following work.
Note that once the system represented in the TWA begins its evolution, explicit memory of Bogoliubov modes is not retained, and nothing precludes the emergence of non-uniform and evolving density. 

However, the presence of ultraviolet vacuum fluctuations imposes a well-defined regime of validity on such dynamical simulations within the TWA. Since the Wigner representation populates every lattice mode with a half-quantum of virtual stochastic noise, the total virtual particle number scales as $N_{\text{noise}} \approx \frac{1}{2} M$, where $M$ is the total number of grid points. To prevent this unphysical vacuum noise from dominating the dynamics, we must operate in the high phase-space density regime where the physical particle number significantly exceeds the virtual noise ($N_{\text{phys}} \gg N_{\text{noise}}$) \cite{Sinatra02,Blakie08}. Furthermore, in the presence of interactions, these high-momentum modes can undergo unphysical scattering into low-energy modes, eventually leading to unphysical thermalisation \cite{Sinatra02,Deuar09b}. Consequently, the regime of validity of our dynamic simulations is bounded by: (i) a cutoff where the lattice resolution is fine enough to capture the healing length ($k_c \xi_{\text{set}} \gtrsim 1$) but coarse enough to restrict the total vacuum noise capacity, and (ii) a finite-time range prior to the start of unphysical thermalisation induced by high-momentum.

More specifically, our three test cases are one-dimensional and start at $t=0$ in a homogeneous state, after which a nonzero external potential is turned on. The system has
length $L=40\pi$ and $M=301$ momentum modes. Therefore, the simulation lattice spacing is $\Delta x\approx0.4175$ in healing length units, and $k_c=7.5$. 
Most of our examples use the dressed coupling $g=g_{\rm set}=0.3$, hence density $n=n_{\rm set}=3.3333$ in our $g_{\rm set}n_{\rm set}=1$ healing length units ($\xi_{\rm set}=1$).
We compare the TWA full quantum evolution to the mean-field GPE and the beyond-mean-field extended GPE (EGPE) \cite{Petrov15,Petrov16}, which has become the standard workhorse in the community when dealing with dynamics of systems incorporating LHY physics. 

The TWA evolution follows \eqn{TWAPGPE}, 
with matched bare interaction $g_0\approx0.2714$ and initial condensate fraction $n_0\approx2.9261$ (except for Figs.~\ref{fig:n_TWA_1}(c,d),~\ref{fig:twa_nx_4000}, and~\ref{fig:gew}(c,d)), calculated as per the algorithm in Sec.~\ref{g0steps}. 

The mean-field GPE for the condensate wavefunction $\psi(x)=\phi(x)\sqrt{N_0}$ takes the usual form:
\eq{GPE}{
i\frac{\partial \psi}{\partial t}=\left[-\frac{\nabla^{2}}{2}+V(x)+g|\psi|^{2}\right]\psi.
}
The EGPE, which includes a LHY energy term from the equation of state is, in 1D \cite{Petrov16}: 
\begin{equation}
i\frac{\partial \psi}{\partial t}=\left[-\frac{\nabla^{2}}{2}+V(x)+g|\psi|^{2}-\frac{g^{3/2}}{\pi}|\psi|\right]\psi.
\label{EGPE}
\end{equation}
Both equations use the dressed coupling $g=g_{\rm set}$.
Their order parameter has the initial condition $|\psi|^{2}=n=n_{\rm set}$, and we will use $\psi(x)=\sqrt{n_{\rm set}}$. 
The simulation results presented in this section were generated using a modified version of the xSPDE3 software package \cite{Kiesewetter23}.

%


\subsection{Sharp Gaussian potential}
\label{VSHARP}
In the first test case, 
a narrow Gaussian external potential $V(x)=4\mathrm{exp}\left[-\frac{1}{2}(\frac{x}{\sigma})^2\right]$ 
is applied instantaneously at $t=0$ to a 
homogeneous Bose gas, and the resulting disturbance is tracked. In the simulation, $\sigma=0.2\ll\xi=1$, 
so the potential is effectively a delta-function-like disturbance that is not well resolved. The maximum strength is four times the chemical potential, $\mu=g_{\rm set}n_{\rm set}=1$, so it produces a density dip that nearly reaches zero. 

\begin{figure}[h!]
\centering
\includegraphics[width=\hfigwidth]{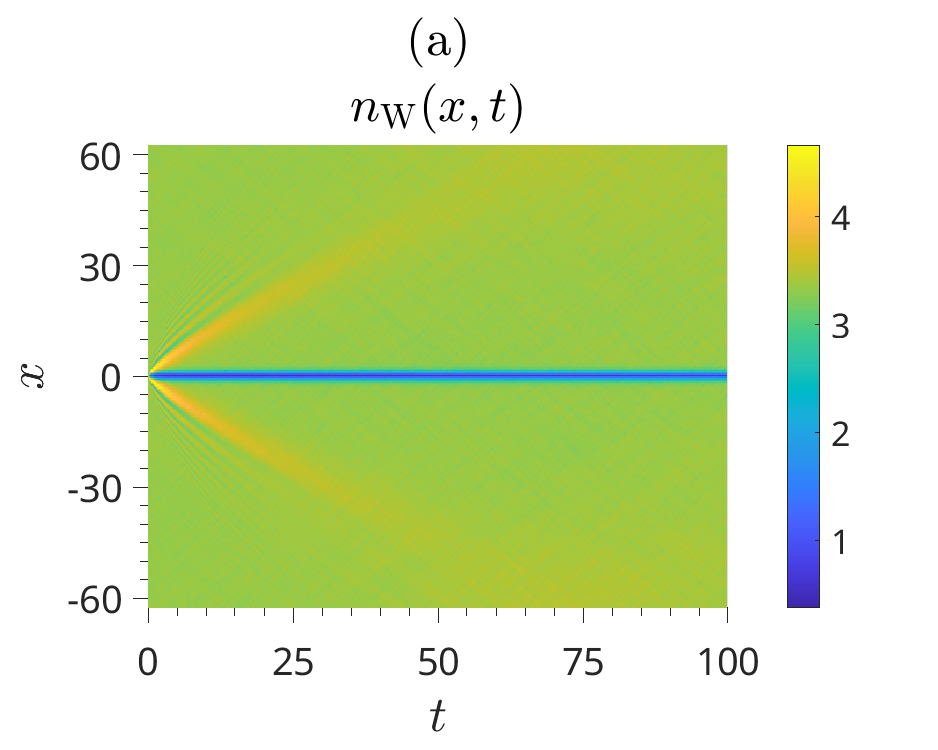}
\includegraphics[width=\hfigwidth]{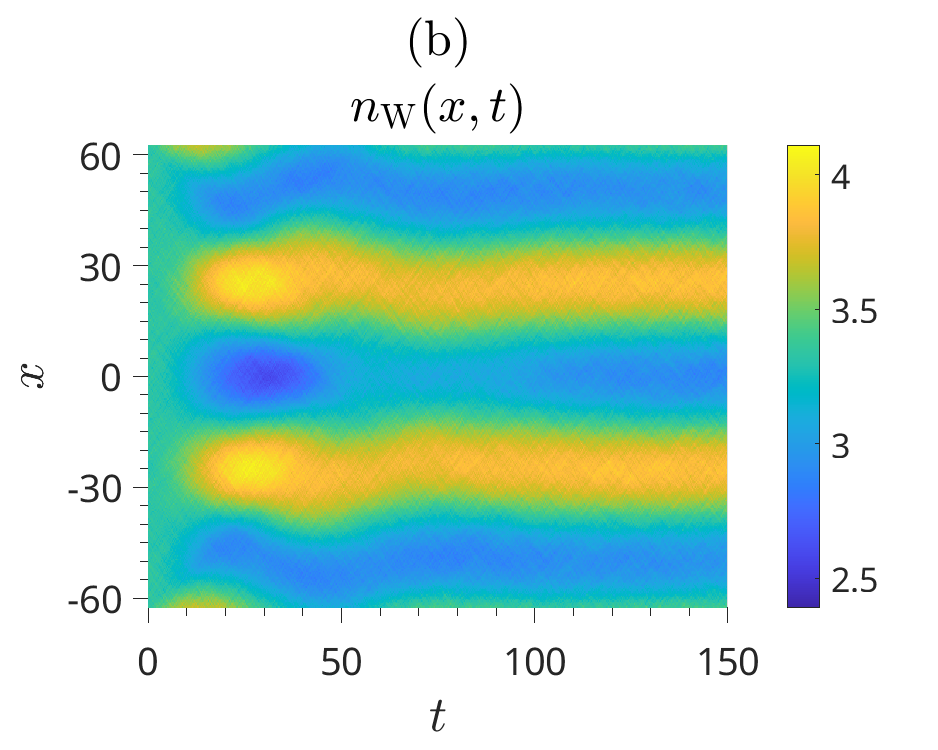}\\
\includegraphics[width=\hfigwidth]{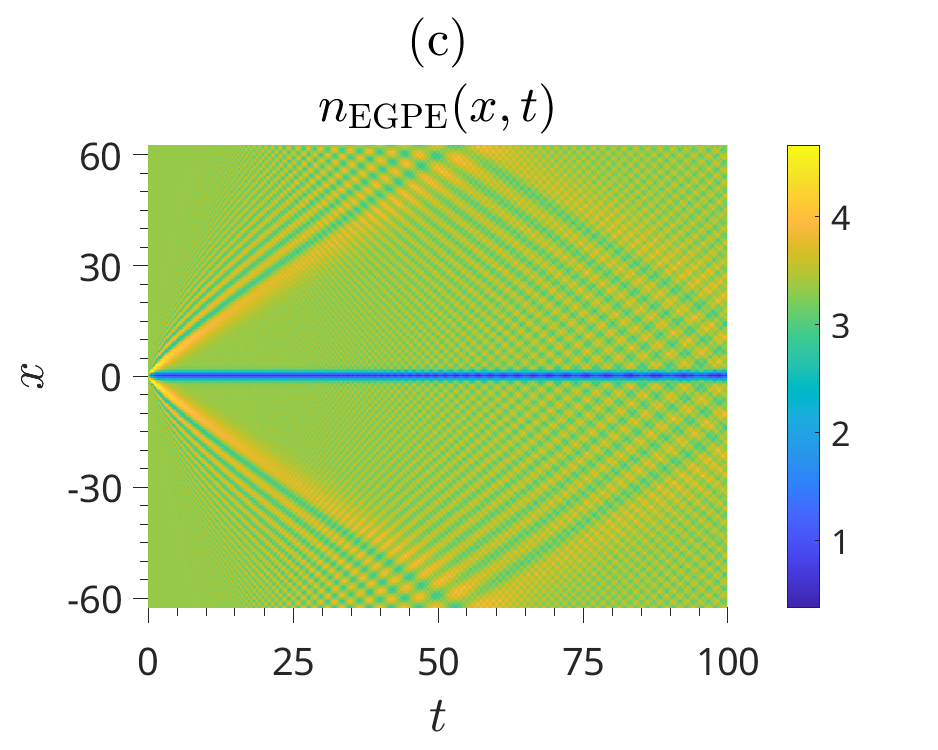}
\includegraphics[width=\hfigwidth]{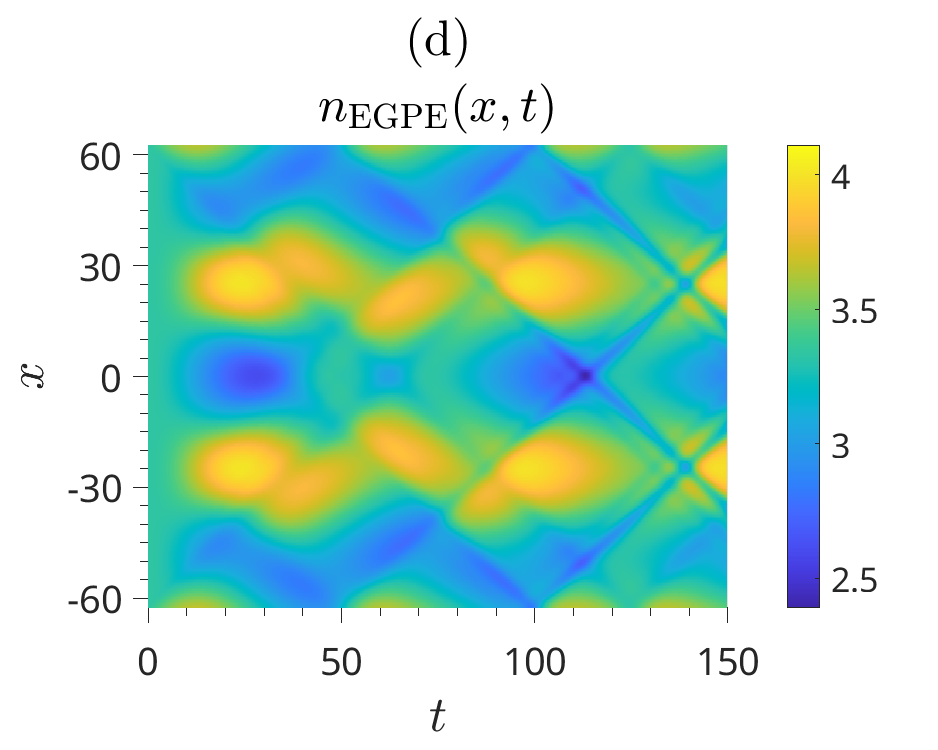}\\
\includegraphics[width=\hfigwidth]{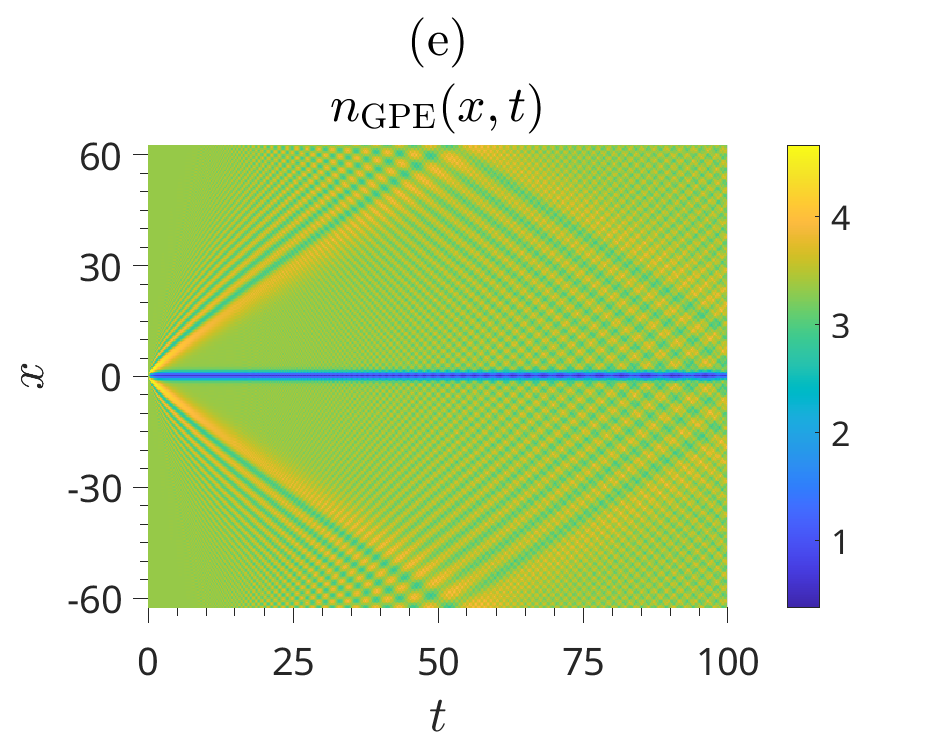}
\includegraphics[width=\hfigwidth]{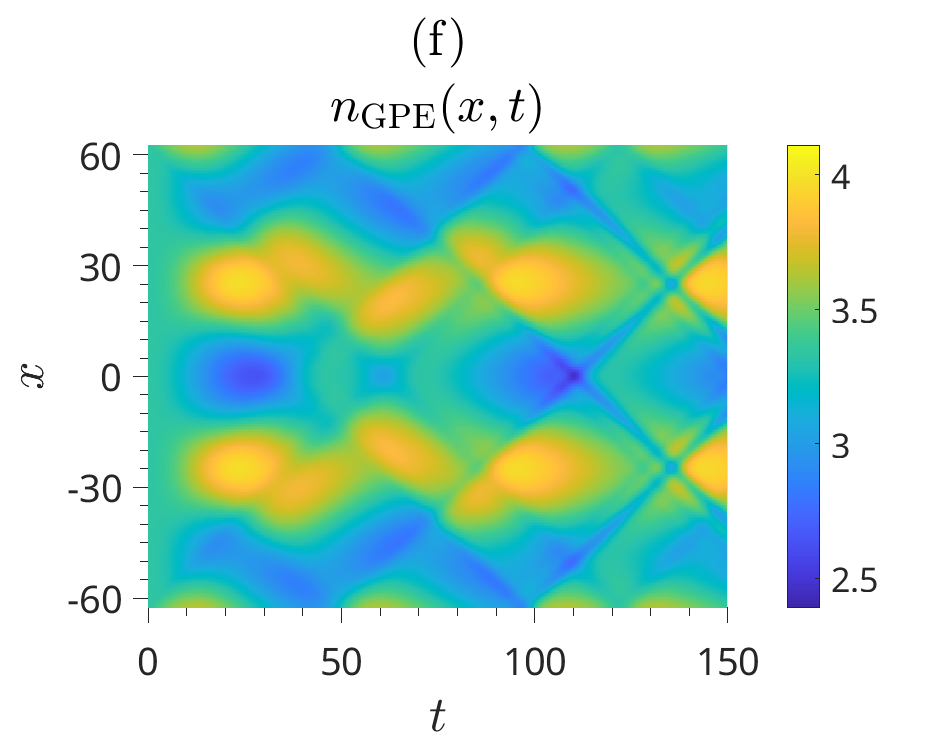}
\caption{Evolution of the density in the narrow Gaussian potential $V(x)$ of Sec.~\ref{VSHARP} with width $\sigma=0.2\xi_{\rm set}$ (left column), 
and in the weak periodic potential of Sec.~\ref{VPERIODIC} (right column), according to three models: 
Top row (a,b) -- Mean TWA field density $n_{\rm W}$ calculated via \eqn{TWAn} with 
 $g_0\approx0.2714$, averaged over 10000 trajectories. The field propagation follows \eqn{TWAPGPE}.
Middle row (c,d) -- EGPE 
density $n_{\rm EGPE}$ following \eqn{EGPE}.
Bottom row (e,f) -- Plain mean-field GPE
density $n_{\rm GPE}$ following \eqn{GPE}.
In all cases,  $L=40\pi$, $g=0.3=1/n$, and $M=301$ modes.
}
\label{fig:narrow-periodic-3x2}
\end{figure}

Figs.~\ref{fig:narrow-periodic-3x2}(a,c,e) 
show the evolution of the density in the TWA, EGPE and plain mean-field GPE, respectively. Apart from the deep trough at the location of the potential (as expected, of width $\mc{O}(\xi)$, which is much wider than the potential itself), waves travelling at $c\approx1$ are produced. The apparent reflection at $t\sim 50$ seen in the coherent models (GPE and EGPE) is an artefact of the periodic boundary conditions combined with perfect coherence.


What is observed is that 
the initial TWA evolution is similar to that of the EGPE and GPE.
However, the density peaks in the truncated Wigner simulations dissipate at $t\gtrsim25$ due to the fluctuations of the quantum depletion in the Bogoliubov ground state, and wrap-around on the periodic boundary conditions is therefore not really observed. Additionally, the interference fringes in the regions outside the sound cone (corresponding to waves travelling faster than $c$) are much less pronounced in the TWA.


\subsection{Weak periodic potential}
\label{VPERIODIC}
The next example considers a weak periodic external potential $V(x)=\frac{1}{10}\mathrm{cos}(\frac{x}{8})$, where two potential minima 
are created at the positions $x=\pm25.13$ across the simulation box. In this example, the strength of the potential is chosen weak enough that is insufficient to reduce the density close to vacuum levels, so only mild modification of the density is 
expected. The EGPE (Fig.~\ref{fig:narrow-periodic-3x2}(d)) and the truncated Wigner simulations (Fig.~\ref{fig:narrow-periodic-3x2}(b)) give similar density profiles 
until $t\approx25$. However, the simulation results from the two methods diverge noticeably at longer times.
In the EGPE, sharp interference features in the density profile with $\mc{O}(\xi)$ develop at later times and propagate, as well as becoming stronger for $t\gtrsim 100$ once waves can travel around the periodic boundary conditions and re-interfere. Similar interference patterns appear in the GPE but are slightly displaced in position upon close inspection. 
The density excursions away from the mean are also more pronounced in the coherent models (EGPE and GPE) than in the TWA. 

Most strikingly -- the system in the Wigner representation comes to a largely stationary state after $t\sim100$, whereas both coherent models become instead more and more agitated in density, with increasingly high-amplitude waves and interference patterns developing. 
This strongly suggests that the EGPE is producing spurious and very strong interference effects, likely due to its 100\% coherent assumption for the wavefunction. The fuller TWA model, which includes two-body correlations and quantum fluctuations, however, does not preserve these spurious interference.


The gas reaches an on-average-stationary solution
with density modulations that match the inverse of the potential, $-V(x)$, at least in the Wigner simulations. 

\subsection{Harmonic trap potential}
\label{VTRAP}
Our final example is a strong harmonic potential $V(x)=4(\frac{x}{L/2})^2$ turned on at $t=0$. It causes the initially uniform gas to focus into a high-density cloud (the yellow region in Figs.~\ref{fig:harmonic-3x2}(a,c,e)) around $t\approx30$, the formation of a high-density ``plateau'' around $t=40$, and subsequent dispersal. 

\subsubsection{Density}
\begin{figure}[h!]
\centering
\includegraphics[width=\hfigwidth]{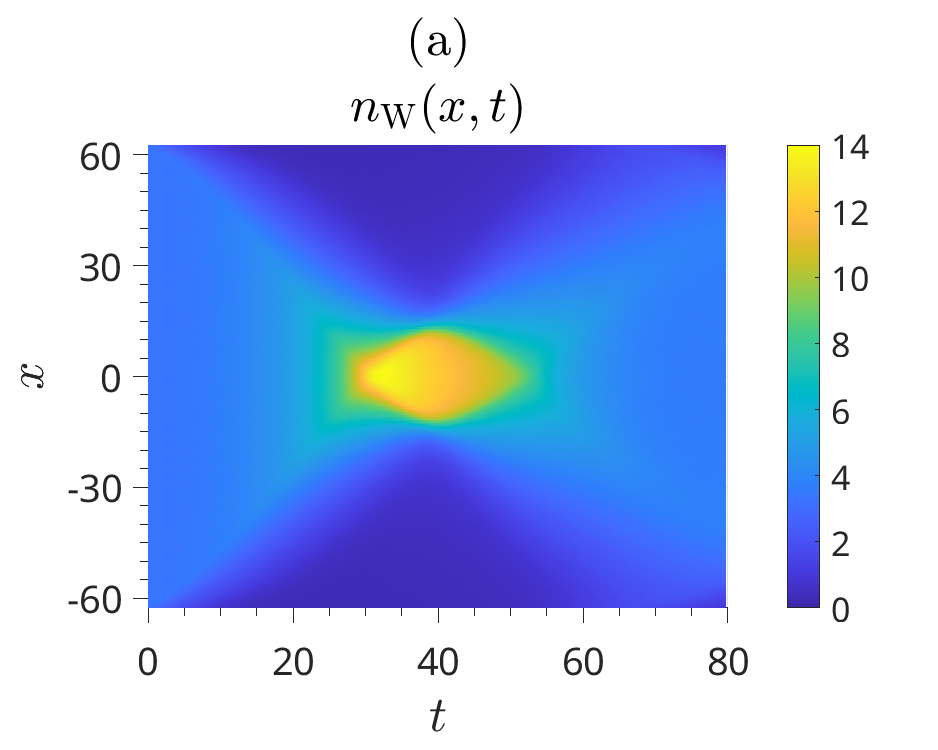}
\includegraphics[width=\hfigwidth]{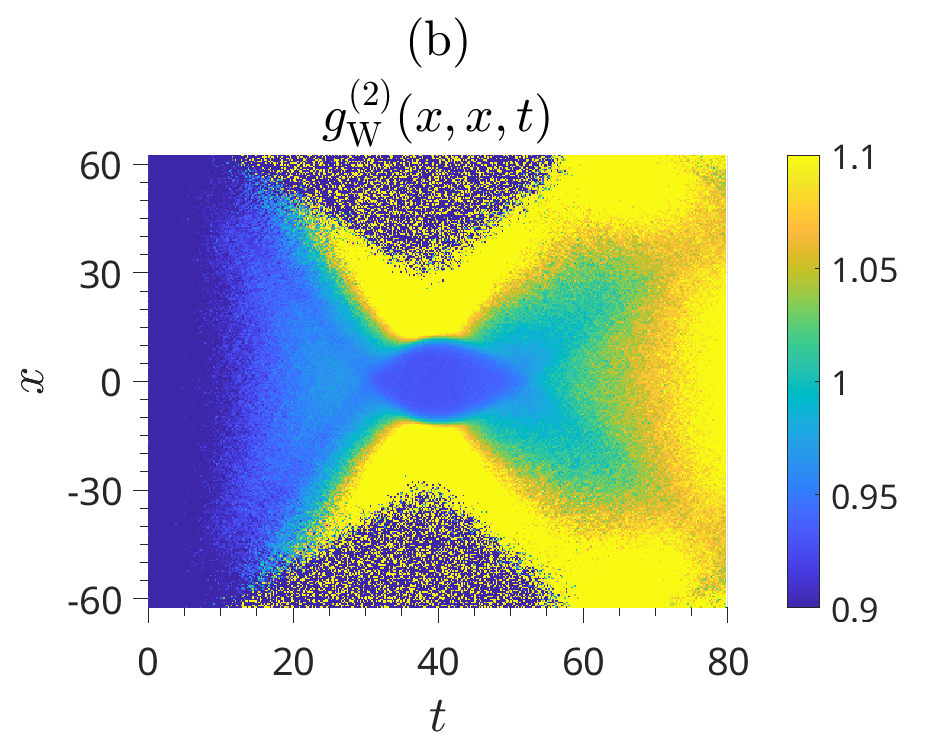}\\
\includegraphics[width=\hfigwidth]{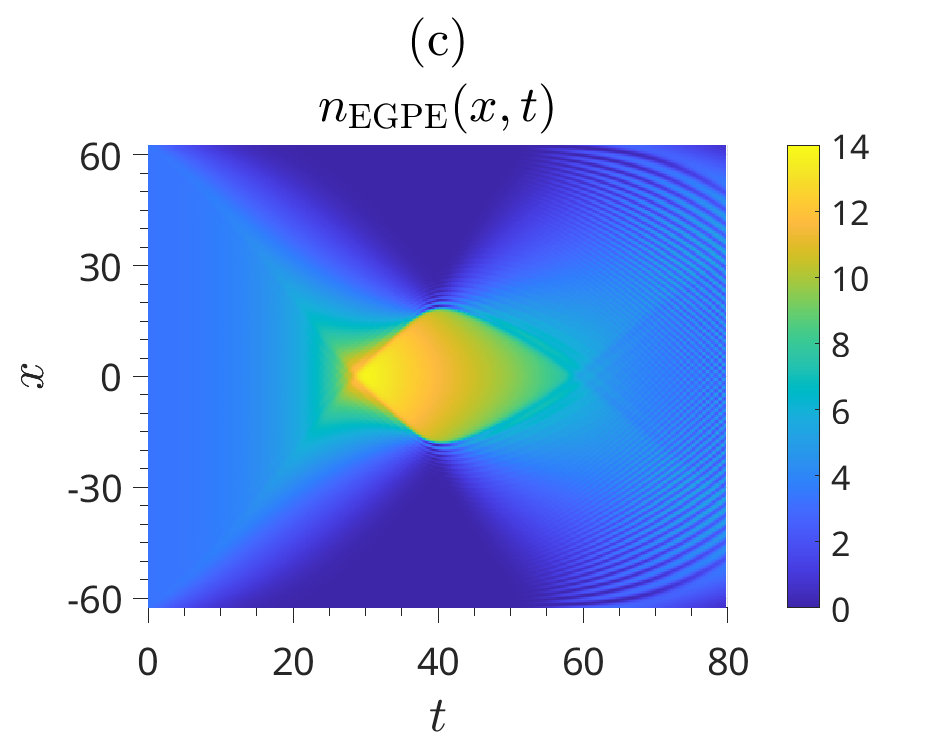}
\includegraphics[width=\hfigwidth]{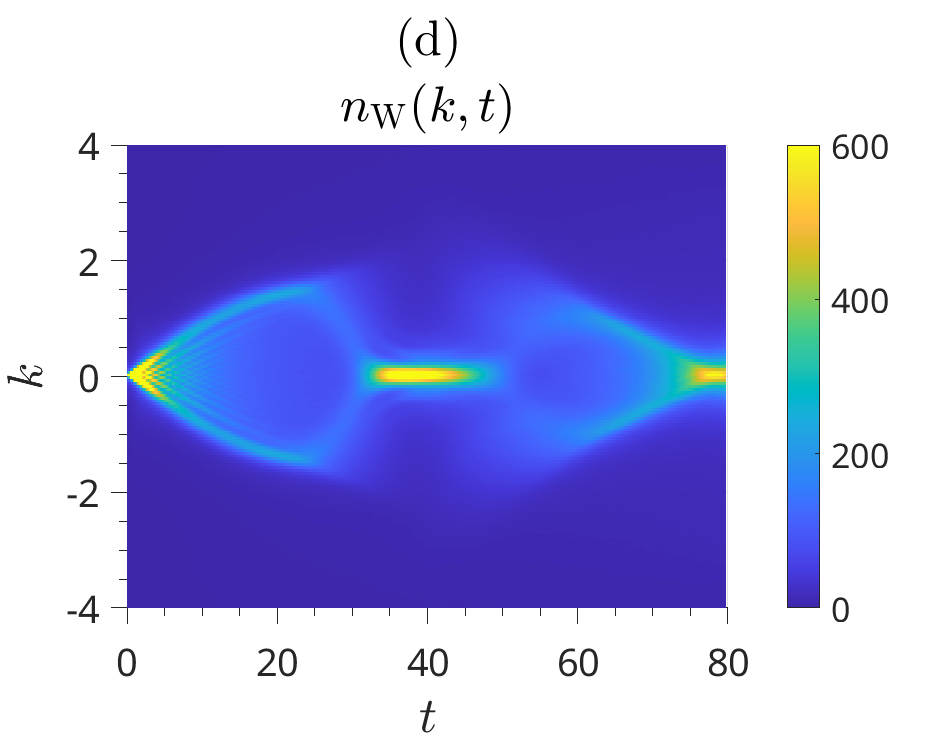}\\
\includegraphics[width=\hfigwidth]{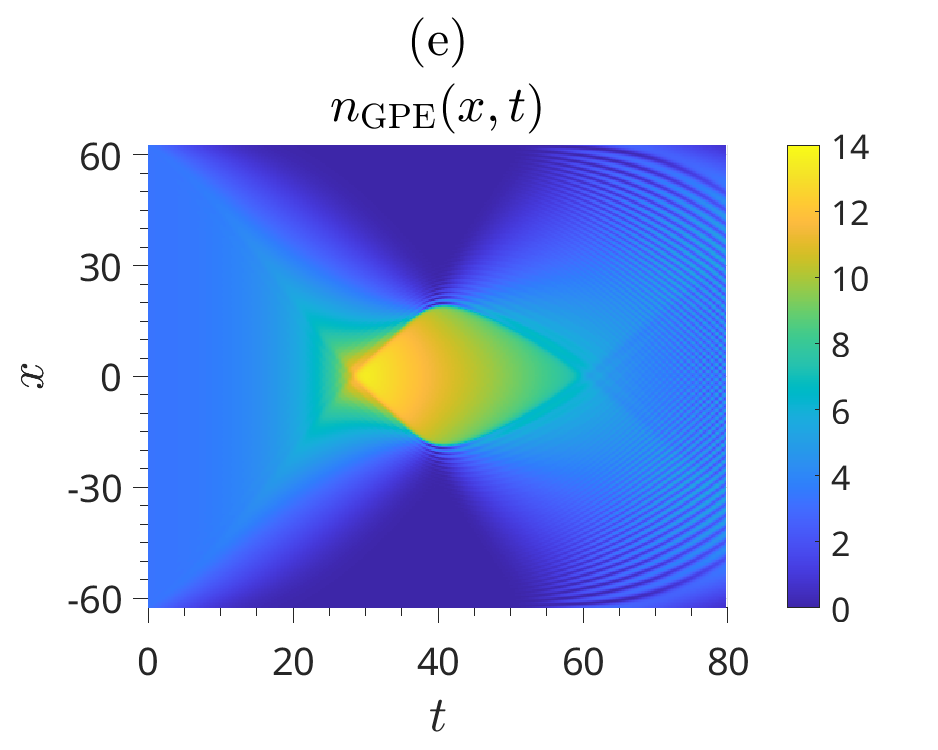}
\includegraphics[width=\hfigwidth]{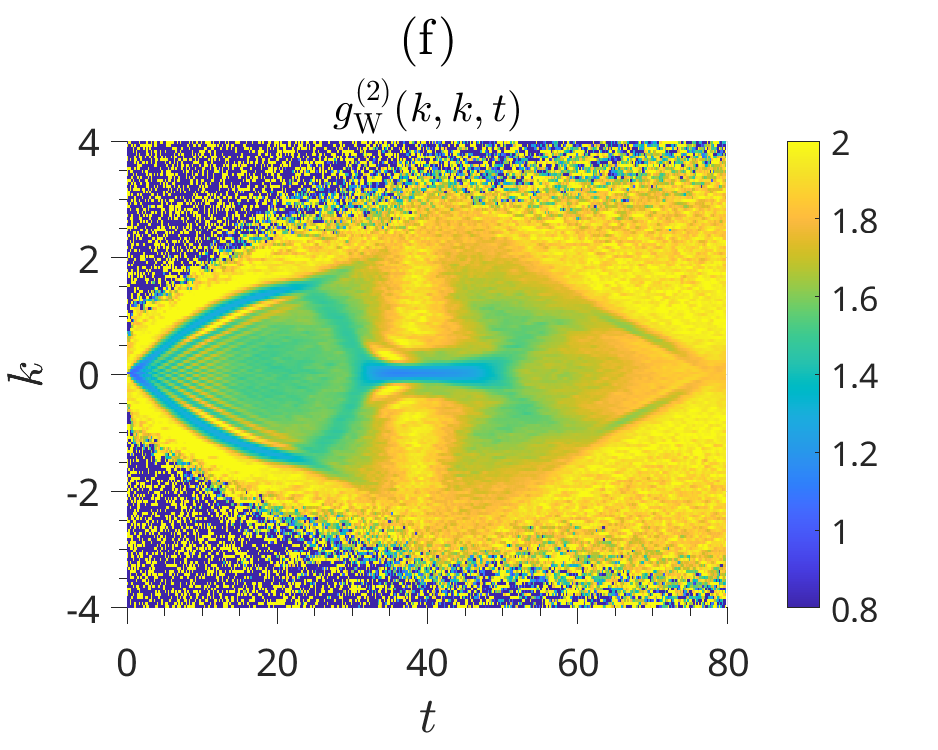}
\caption{Evolution of observables in a harmonic external potential $V(x)=4gn(\frac{x}{L/2})^2$, showing in the left column: (a) the density from the Wigner evolution, $n_{\rm W}$, averaged over 10000 trajectories; (c) the EGPE evolution; and (e) the plain mean-field GPE evolution, similarly arranged as in Fig.~\ref{fig:narrow-periodic-3x2}. 
Panels in the right column show additional observables from the TWA simulation. 
(b): Local pair correlation $g^{(2)}_{\rm W}(x,x)$;
(d): Wigner-field density in momentum space;
(f): Pair correlation $g^{(2)}_{\rm W}(k,k)$ in momentum space.
In all cases, $L=40\pi$, $g=0.3=1/n$, and $M=301$ modes as in Fig.~\ref{fig:narrow-periodic-3x2}. 
}
\label{fig:harmonic-3x2}
\end{figure}

The density evolution 
in the truncated Wigner simulations is as shown in Fig.~\ref{fig:harmonic-3x2}(a). 
Figs.~\ref{fig:harmonic-3x2}(c,e) 
display the EGPE and GPE evolutions, respectively. The comparison is very revealing.

As in the previous example, the coherent models (EGPE and GPE) exhibit evolution filled with small-wavelength coherent waves, especially after the dispersal of the plateau at $t\gtrsim 40$, while the TWA does not. It appears that proper inclusion of the quantum fluctuations (LHY physics) suppresses these spurious waves. In contrast, the addition of only the LHY energy in the EGPE, without its attendant decoherence, preserves these superfluous interference features. However, intrinsic phase-space mixing between distinct classical trajectories within the TWA can also cause or contribute to the suppression of such fine-scale wave interference \cite{Mathew_2019}. We further explore the nature of this wave suppression in Sec.~\ref{single_realisation}.

To further understand the dynamics in this example, it is also instructive to examine the momentum distribution and local density correlations. The former is shown in Fig.~\ref{fig:harmonic-3x2}(d), where 
\eq{nWk}{n_\mathrm{W}(k)=\left\langle|\psi_\mathrm{W}(k)|^2-\frac{1}{2\Delta k}\right\rangle
}
for a momentum spacing of $\Delta k=2\pi/L$, with the transformation to $k$-space wavefunction given by \eqn{Psik}. In the plot, we find that the initially almost pure condensate at $t=0$ (with nearly all occupation at $k=0$) transforms into a focusing cloud dominated by travelling waves with opposite momenta until the density maximum near $t=28$. Surprisingly, a very narrow momentum distribution around $k=0$, close to that of the pure condensate, reappears during the ``plateau'' phase until $t\sim 45-50$, after which momentum is dispersed over a wide range of $k$ again as the plateau dissipates.



\subsubsection{Correlations}

The TWA allows access to the density-density correlations in the gas, a feature which is hardwired to be completely absent in both mean-field GPE and beyond-mean-field EGPE coherent approaches.

Fig.~\ref{fig:harmonic-3x2}(b) shows the local pair correlations of the harmonic trap system calculated from the truncated Wigner simulations via \eqn{TWAg2} as $g^{(2)}_{\rm W}(x,x)=G_2(x)/\big( n_{\rm W}(x)\big)^2$. The Bose gas is initially in a coherent condensate state, exhibiting antibunching effects identified by the correlations $g^{(2)}_\mathrm{W}(x,x)\approx0.90<1$. The gas is strongly confined during the plateau phase at the times $t\sim28-50$, creating a large region of vacuum state (a region of pure speckle noise in Fig.~\ref{fig:harmonic-3x2}(b)) at the edges of the harmonic trap. The confined Bose gas itself reforms into a coherent condensate cloud (blue region between $28<t<50$ in Fig.~\ref{fig:harmonic-3x2}(b)), which matches the contour profile of the high-density cloud shown in Fig.~\ref{fig:harmonic-3x2}(a). Interestingly, this cloud also exhibits antibunching, though to a lesser degree than the initial state at $t=0$ with $g^{(2)}_\mathrm{W}(x,x)\approx0.93$. The antibunching strengthens again after being weakened during the focusing stage (with $g^{(2)} _\mathrm{W}(x,x)$ growing to about $\sim0.96$ before the plateau appears). At later times, this antibunched region disappears and the cloud disperses with growing $g^{(2)}_{\rm W}(x,x)$, particularly in the low-density wings of the cloud where it begins to rise to thermal values.



We can also examine the fluctuation properties and coherence of the 
modes directly. To this end, we can
determine the pair correlations $g^{(2)}_\mathrm{W}(k,k)$ in momentum space, calculated from the truncated Wigner simulations using an expression very similar to \eqn{TWAG2}:
\eq{g2W_kk}{
g^{(2)}_\mathrm{W}(k,k)=\frac{1}{n_\mathrm{W}(k)^{2}}\left\langle|\psi_\mathrm{W}(k)|^{4}-2|\psi_\mathrm{W}(k)|^{2}\delta_{\mc{P}}(k)+\frac{\delta_{\mc{P}}(k)^2}{2}\right\rangle.
}
Here $\delta_{\mc{P}}(k)=1/\Delta k$ takes the place of $\delta_{\mc{P}}(x)$ that appears in real space.
Fig.~\ref{fig:harmonic-3x2}(f) shows the evolution of $g^{(2)}_{\rm W}(k,k)$. The features seen here match the $k-$space density features in Fig.~\ref{fig:harmonic-3x2}(d). The correlation plot confirms that the main travelling waves during the focusing stage and the modes around $k=0$ in the plateau state are not only narrow in $k$-space but also have strongly suppressed density fluctuations, with $g^{(2)}_{\rm W}(k,k)\approx 1$. Thus, the state in the plateau phase is confirmed to be a condensate or quasicondensate. The modes in the dispersal phase evolve towards thermal states with $g^{(2)}_{\rm W}(k,k)\to2$ at times $t\gtrsim40$. 
Interestingly, some higher-momentum modes that are still occupied in the plateau phase exhibit thermal behaviour during this period, later losing it (seen as the vertical yellow feature in the range of $-2\lesssim k\lesssim2$ at times $t\sim 30-45$). In addition, the low-momentum modes during the early focusing phase show large fluctuations, at times again approaching the thermal value of $g^{(2)}_{\rm W}(k,k)\approx2$.



The above analysis demonstrates that the TWA formulation of the LHY state is amenable to the natural and efficient simulation of highly complex correlation and quantum depletion dynamics.

\begin{figure}[h]
\centering
\includegraphics[width=\hfigwidth]{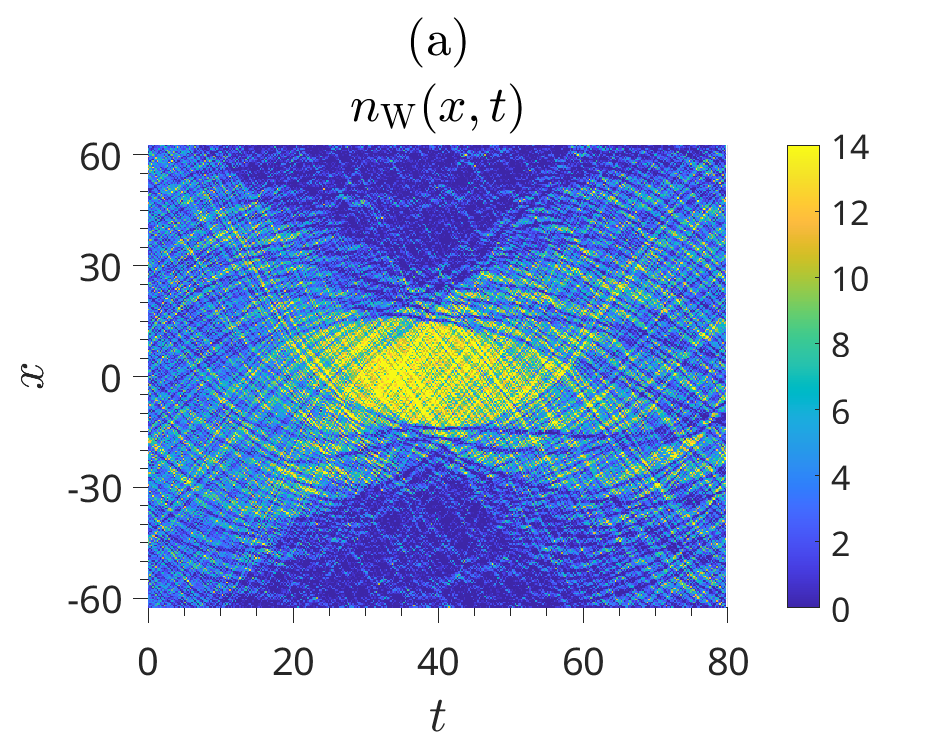}
\includegraphics[width=\hfigwidth]{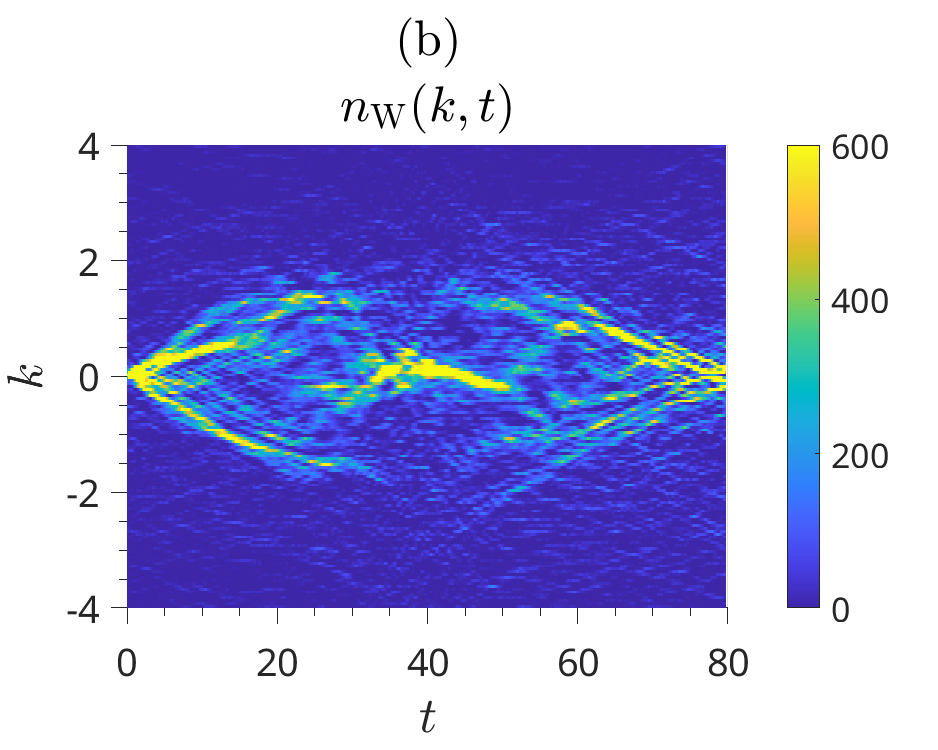}\\
\includegraphics[width=\hfigwidth]{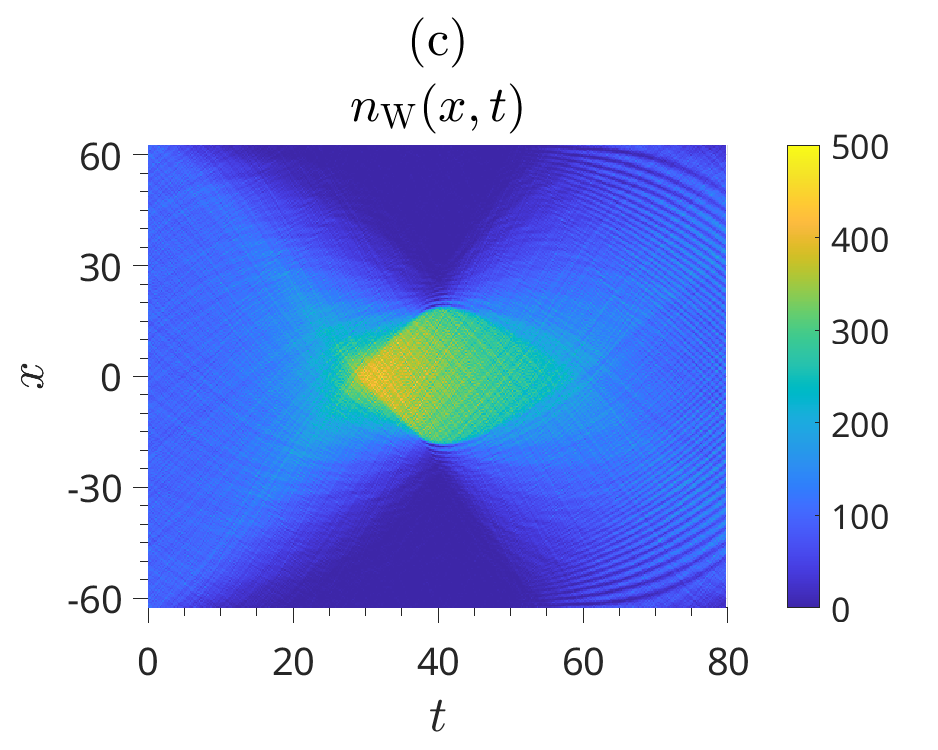}
\includegraphics[width=\hfigwidth]{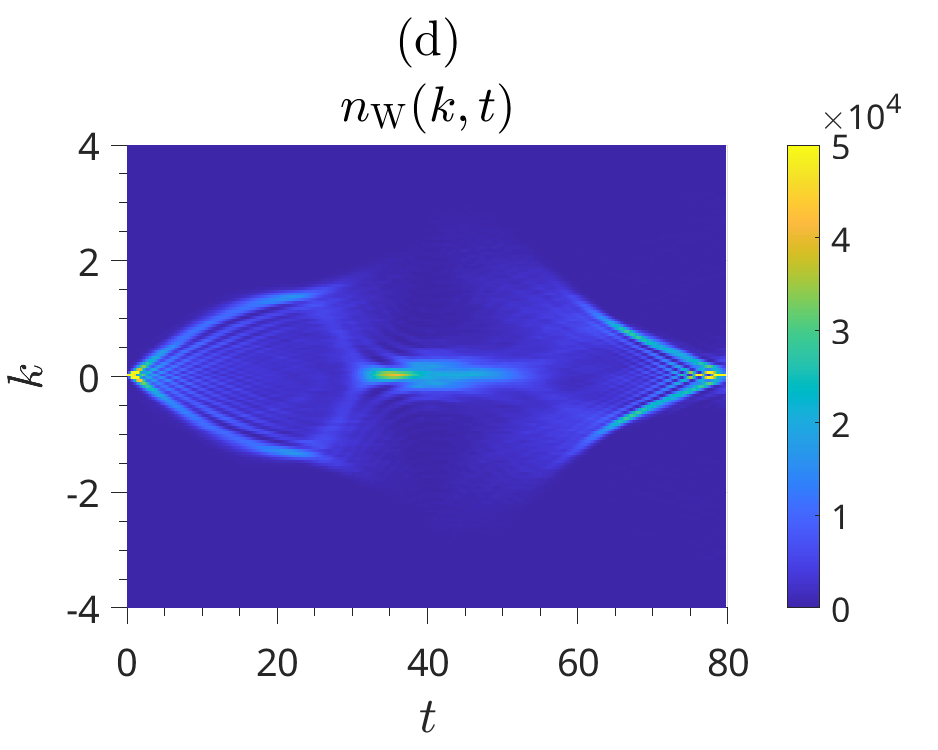}
\caption{Density evolution in a single realisation of the TWA in the trap potential of Sec.~\ref{VTRAP}. (a,c): $x$-space; (b,d): $k$-space.
(a,b) -- parameters are the same as in Fig.~\ref{fig:harmonic-3x2} with $g_0\approx0.2714$; (c,d) --  similar to (a,b) but with a lower interaction strength, the matched bare values in (c,d) are $g_0\approx0.009994$, $n_0\approx99.5270$.} 
\label{fig:n_TWA_1}
\end{figure}

\subsubsection{Single realisation}\label{single_realisation}
A single Wigner trajectory from the fuller TWA model, shown in Figs.~\ref{fig:n_TWA_1}(a,b), provides deeper insight into how the physics differs from that of the fully coherent GPE and EGPE models. We can see that the individual fluctuations are strong enough to dominate the coherent interference fringes present in the GPE, suppressing them entirely, and in fact in this case are strong enough to produce dark soliton excitations (at least for this very strong interaction $g=0.3$). The most macroscopic feature -- the formation of a plateau around $t\sim40$ -- is, however, retained even in a single trajectory. 
The $k$-space plot gives an indication of what would be observed after expansion in single experimental realisations: the main features, consisting of the plateau around $t\sim40$ and two dominant travelling waves, are retained, but not the fine interference fringes suggested by the EGPE as shown in Fig.~\ref{fig:harmonic-3x2}(c) at $t\gtrsim40$. Therefore, in this case, it is not that the full model produces EGPE-like interference with phases differing between individual realisations that are then washed out upon averaging; rather, such fringes never form in any realisation and are instead replaced by stronger nonlinear and incoherent fluctuations.

\begin{figure}[h]
\includegraphics[width=\hfigwidth]{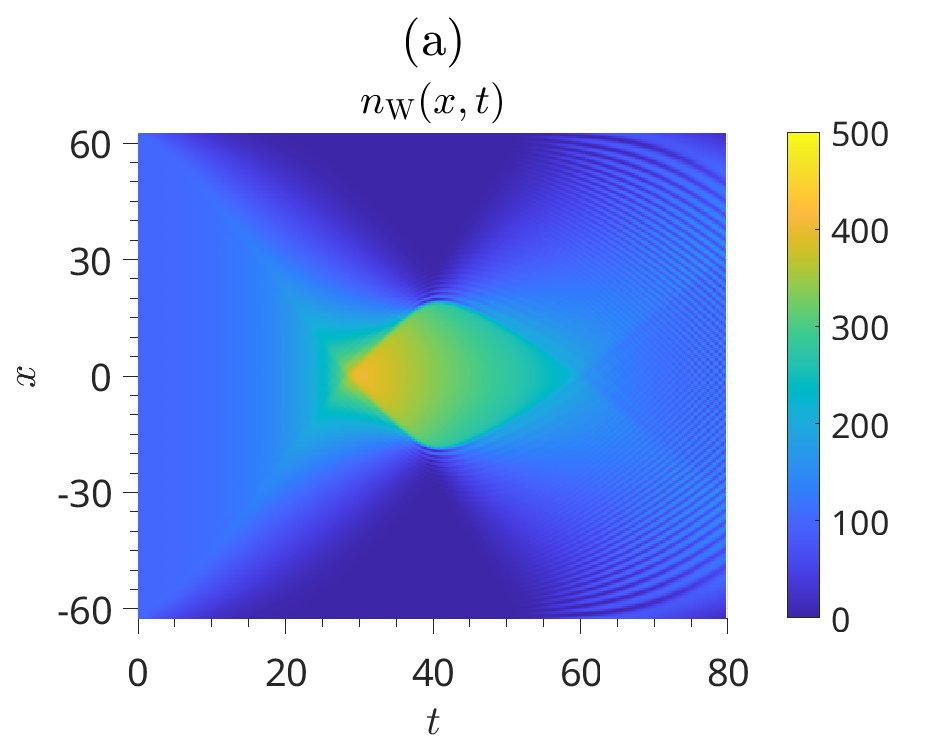}
\includegraphics[width=\hfigwidth]{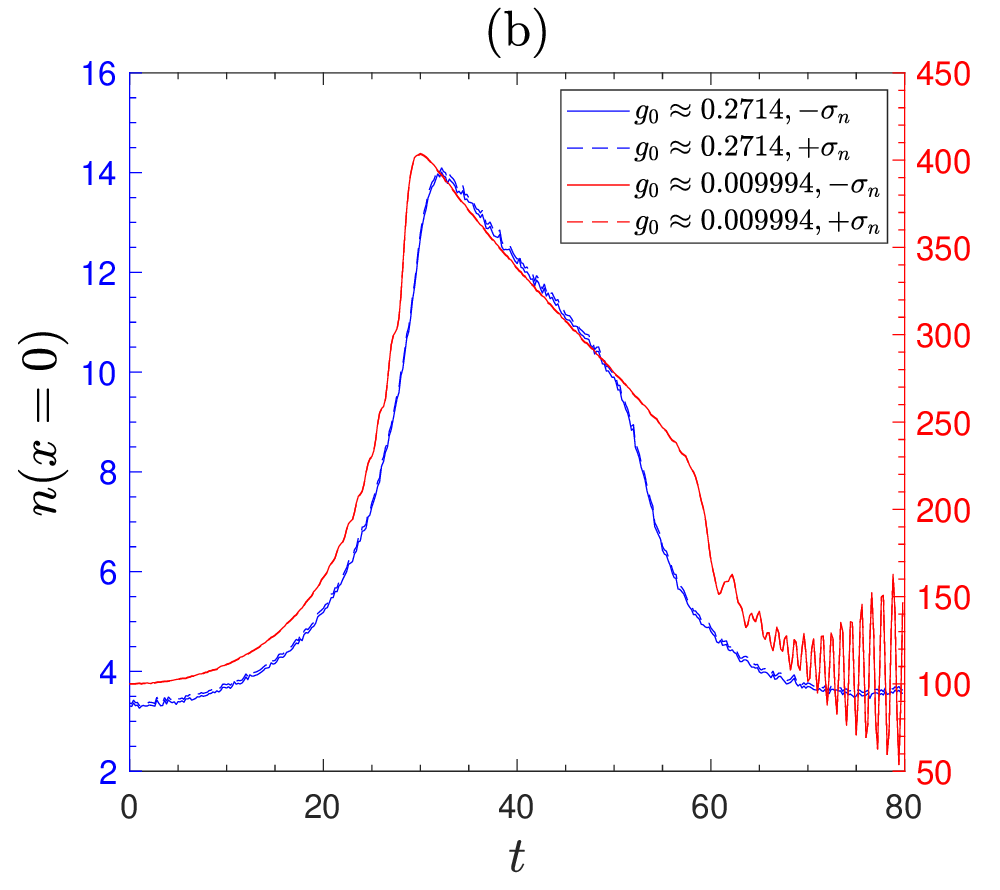}
\caption{(a) Mean density evolution (averaged over $10000$ trajectories) of the TWA in the harmonic external potential with $g_0\approx0.009994$, using the same parameters as in Figs.~\ref{fig:n_TWA_1}(c,d). Note the return of interference fringes for this low value of $g_0$. (b) Statistical convergence of the TWA central density $n_\mathrm{W}(x=0,t)$ using $10000$ trajectories for both weak interaction (red lines, $g_0\approx0.009994$) and strong interaction (blue lines, $g_0\approx0.2714$). The upper (dashed) and lower (solid) curves represent $\pm1$ standard error of the mean ($\bar{n}(t) \pm \sigma_{n}(t)$). In the weak-interaction case, the error bounds tightly overlap with the mean, while the strong-interaction case exhibits a narrow, highly converged error envelope. This confirms that the observed results reflect statistically stable physical dynamics. 
}
\label{fig:twa_nx_4000}
\end{figure}

One must consider whether the intrinsic phase-mixing in the TWA over a collapse time could artificially suppress the interference fringes observed in Figs.~\ref{fig:harmonic-3x2}(c,e). To verify that the observed wave suppression in our TWA simulations (Fig.~\ref{fig:harmonic-3x2}(a)) is a genuine physical effect rather than a numerical artefact of path dephasing, we examined the harmonic trap system under a significantly weaker interaction strength. A very interesting contrast is revealed when a smaller interaction strength, $g=0.01$, is used. Figs.~\ref{fig:n_TWA_1}(c,d) show single trajectories in this case, while Fig.~\ref{fig:twa_nx_4000}(a) shows the corresponding TWA averaged density. Here, the interference fringes reappear both in the averaged density and within individual trajectories (although a small residual asymmetry due to quantum fluctuations is still present in single trajectories). 
Since the coherent interference patterns are recovered cleanly in weak interactions ($g=0.01$), it indicates that our TWA simulations are conducted within the physical validity regime of the semiclassical approximation with regard to phase-space mixing. Therefore, the suppression of interference fringes observed in the strong interaction regime ($g=0.3$) appears to be physical, arising from strong quantum fluctuations.
From the results in Fig.~\ref{fig:n_TWA_1}, we can see that the TWA formulation allows one to simulate both strongly and weakly interacting regimes, and to uncover physically relevant differences between them which would be completely obscured in the EGPE.

\begin{figure}[h!]
\centering
\includegraphics[width=\hfigwidth]{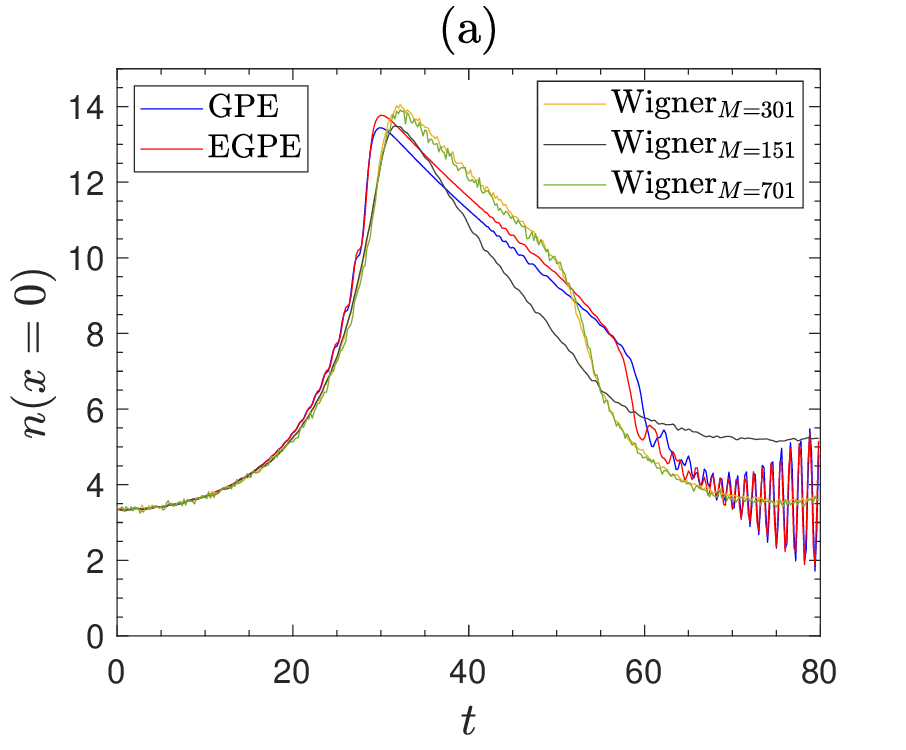} 
\includegraphics[width=\hfigwidth]{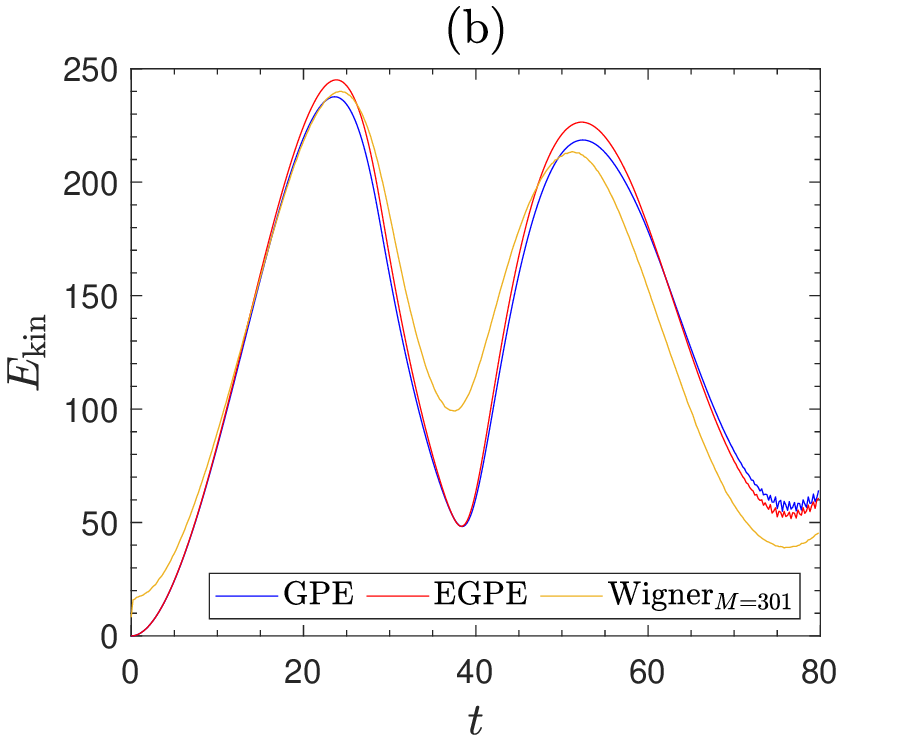}\\
\includegraphics[width=\hfigwidth]{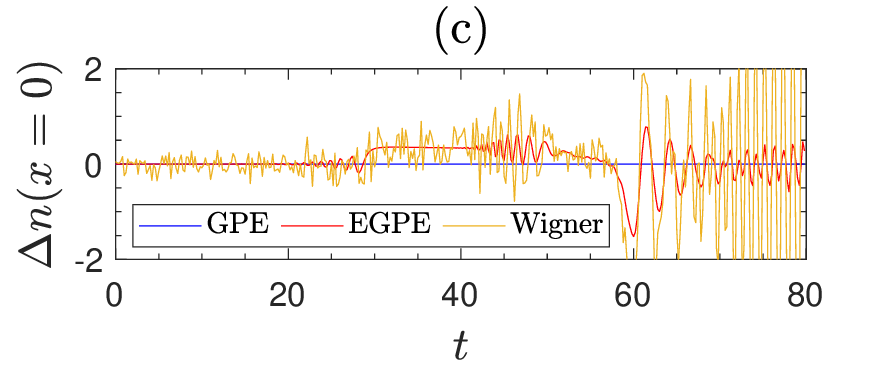} 
\includegraphics[width=\hfigwidth]{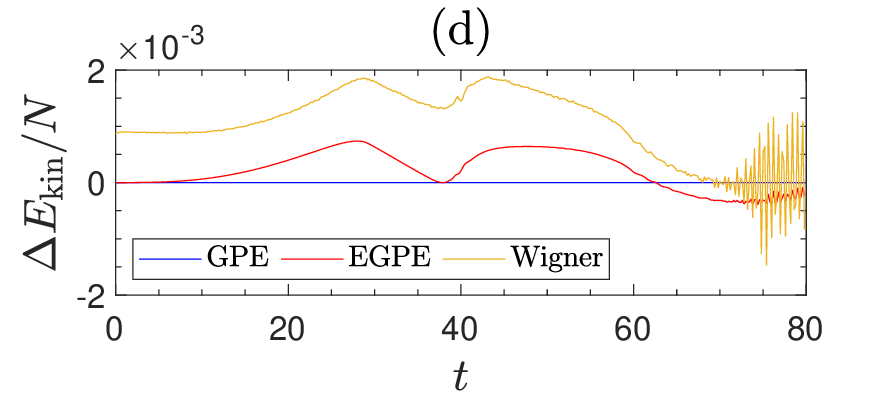} 
\caption{
Quantitative comparison of the TWA at several cutoffs with
the two coherent models, GPE (blue) and EGPE (red), for the harmonic trap case with $L=40\pi$ and $g=g_{\rm set}=0.3$ (units $g_{\rm set}n_{\rm set}=1$). (a-b): central density and kinetic energy; yellow data show the TWA for $M=301$, $k_c=7.5$, $g_0=0.2714$, $n_0=2.9261$ as in Fig.~\ref{fig:get_bare_g0}, averaged over $10\,000$ trajectories. Panel (a) further shows a lower cutoff $k_c=3.75$, $M=151$, $g_0=0.2702$, $n_0=2.9282$ (grey data), and a higher cutoff $k_c=17.5$, $M=701$, $g_0=0.2722$, $n_0=2.9253$ (green data). 
(c-d): differences relative to the GPE values for the smaller $g_{\rm set}=0.01$ of Figs.~\ref{fig:n_TWA_1}(c-d). The TWA data (yellow) use $k_c=7,5$, $M=301$, $g_0\approx0.009994$, $n_0\approx99.5270$, averaged over $50\,000$ trajectories. 
}
\label{fig:gew}
\end{figure}

\subsubsection{Comparison of models}


Fig.~\ref{fig:gew} shows quantitative details of the predictions for several quantities, compared across the three models. 
Somewhat unexpectedly, for the strong interactions with $g=0.3$ (Figs.~\ref{fig:gew}(a,b)), the TWA predictions do not fall clearly closer to the EGPE than to the GPE. If anything, the GPE and EGPE give results that are closer to each other than to those of the fuller model. The statistical convergence of these dynamical trajectories has been verified; as shown in Fig.~\ref{fig:twa_nx_4000}(b), the standard error of the mean for these ensembles is extremely narrow, indicating that the observed deviations are physically genuine. The conclusion is that, for strong fluctuations, the decoherent effects captured by the fuller TWA treatment are at least as important or likely more important than purely energetic effects, including those arising from a LHY energy functional in the equation of state.

This effect is robust to modifying the cutoff $k_c$ in the numerical model, provided $k_c$ is not too low, as seen in Fig.~\ref{fig:gew}(a). For example,  $k_c=7.5$ (yellow data in Fig.~\ref{fig:gew}(a)) and $k_c=17.5$ (green data in Fig.~\ref{fig:gew}(a)) agree well, whereas $k_c=3.75$ (grey data in Fig.~\ref{fig:gew}(a)) is too low and introduces quantitative shifts, although the qualitative difference compared to the coherent models remains dominant.

As the interaction strength decreases, a behaviour closer to that of the beyond-mean-field EGPE is recovered. Panels (c-d) in Fig.~\ref{fig:gew} show the deviation from the GPE for $g=0.01$. For the central density at $x=0$ shown in panel (c), the mean of the TWA trajectories lies closer to the EGPE result than to the GPE in terms of density deviations. For the kinetic energy per particle shown in panel (d), there is a global shift in the TWA that includes kinetic energy from small-scale fluctuations not captured by the kinetic-energy term of the EGPE. Beside this shift though, the same features of deviation from the GPE are observed in both models. However, the number of trajectories (realisations) required to extract these effects from the noise at such weak interactions is enormous.

One may therefore suspect that using an EGPE in regimes where the LHY effects are small, as in this case, may be not significantly better than using the mean-field GPE alone, since it appears to miss a substantial part of the beyond-mean-field effects that are dominated by fluctuations. This is especially so when the interaction is strong (large $\gamma_L=g/n$). Naturally, scenarios such as quantum droplets and supersolids for which the LHY physics is essential for a qualitative change of the state, are not included in this conjecture. These will be the focus of a follow-on paper.


\section{Conclusions}


We have presented a path and an algorithm to faithfully represent Lee-Huang-Yang (LHY) energy corrections and the corresponding physics in the Wigner representation of the quantum state, using bare interactions $g_0$. This is achieved by introducing appropriately matched Bogoliubov fluctuations in a Bogoliubov ground state (see Sec.~\ref{g0steps}). 
As might be expected in retrospect, the precise value of the bare delta-function interaction strength that generates a given mean-field interaction depends on the dimensionality and the details of the numerical lattice. 
Notably, achieving a correct energy match requires taking into account ``extra''  contributions that appear in the Wigner representation due to the amalgamation and indistinguishability of fluctuation and condensate fields. Higher-order contributions, such as $\delta N^2$, can no longer be separated out and removed in the TWA. 
This becomes increasing crucial as the Lieb 
parameter $\gamma_L=g/n$ becomes larger. 

Beyond the fundamental result of developing a path to implementing a TWA model that recovers the correct LHY energy, 
additional findings thanks to implementing this approach are as follow: 
\begin{itemize}
    \item For observables other than the total energy, some residual dependence on lattice parameters remains even after the bare interaction and condensate fraction are self-consistently matched. For example, quantum depletion requires a resolution several times finer than the healing length to approach its thermodynamic value, while kinetic and interaction energies can slowly diverge (with the divergences mutually cancelling) as the momentum cutoff becomes large.
    \item The large qualitative difference between the fuller TWA model and coherent models (EGPE and GPE) when $g$ is large. The coherent models produce numerous apparently spurious small-wavelength interference fringes that persist indefinitely, whereas the TWA does not, even in single trajectories. Neither two of the three models can be said to be clearly closer to each other in this regime. In particular, for quantities shown in Fig.~\ref{fig:gew}, such as the central density or energies, the TWA predictions unexpectedly do not lie closer to the EGPE than to the GPE in the single-component case.
    \item Convergence toward the coherent models improves as $g$ decreases. However, in this regime, single TWA trajectories do not distinguish between the presence or absence of the LHY energy, and very large numbers of realisations -- $\mc{O}(10^3)$ or more -- are required to resolve the LHY effects, which are completely swamped in single trajectories.
    \item The decoherence present in TWA evolution at large $g$ removes any superfluous interference fringes produced by the EGPE and GPE quite rapidly, and allows the system to settle into a stably fluctuating quantum state.
    \item Therefore, it appears that the decoherent effects captured by the fuller TWA treatment are at least as important as, or more important than, the energetic effects arising from including the LHY contribution in the equation of state -- at least while the LHY contributions to the ground state remain perturbative, as in the single component gas. This suggests that using an EGPE in cases with small LHY effects (such as the single component gas) may be not significantly better than using the mean-field GPE alone, as the EGPE misses a large part of the beyond-mean-field effects.
\end{itemize}
These differences between the EGPE and the TWA are much larger than previously assumed. Therefore, one of the main conclusions of this work is that such long-lifetime, small-wavelength wave features appearing in the EGPE should not be trusted unless $\gamma_L$ is very small.

We should mention that, if needed, beyond-LHY expressions derived in recent years \cite{Hu20,Ota20b,Cikojevic19,Cikojevic21,Zin22b,Valles-Muns24} can easily be used as input to the numerical dressing scheme of Sec.~\ref{g0steps}, replacing the base LHY expressions of \eqn{LHY}. The same applies to systems in a regime of dimensional crossover \cite{Zin18,Ilg18,Zin22}, for which appropriate LHY expressions not considered here can be used.

Finally, the mechanism developed in this work can be extended to more involved cases, particularly those involving phenomena seeded by quantum fluctuations but not necessarily perturbative in the long-time limit. Examples include non-uniform systems such as bright solitons, trapped gases, and multicomponent systems. Notably, this approach is potentially beneficial for the study of quantum droplets in which the LHY effect is essential for the formation of the self-bound quantum state. The quantum fluctuations represented by samples of the Wigner representation 
can presumably replace the effect of the LHY correction in the formation of quantum droplets, leading to a qualitatively different 
description of the phenomenon than previously obtained. Implementation of this extension is currently in progress.



\section*{Acknowledgements}


The Authors would like to thank our colleagues: Dmitri Petrov, 
Nick Proukakis, Stefano Giorgini, Karen Kheruntsyan, Blair Blakie, and Ashton Bradley, who have discussed this topic with us and given valuable inspiration.


\paragraph{Funding information}
This research was supported by the (Polish) National Science Center Grant No. 2018/31/B/ST2/01871.

\bibliography{artnew_k}

\begin{appendix}
\numberwithin{equation}{section}



\section{Use of bare versus dressed quantities} 
\label{BAREDRESS}


There are several places in the model and analysis that involve a choice of whether to use bare or dressed interactions, condensate or full density, and where the ``right'' choice is not obvious.
\begin{itemize}
    \item The choice of units or healing length, e.g., $gn=1$, $g_0n_0=1$, or similar.
    \item Whether to use $gn$ or $g_0n_0$ (etc\dots) in the LHY energy expressions \eqn{LHY} when matching with the Bogoliubov/TWA expressions in \eqn{LHSRHS}.
    \item A certain inconsistency in the Bogoliubov expansion (Sec.~\ref{BOGDES}), where the spectrum \eqn{epsilon}, coefficients $U_{\bo{k}}, V_{\bo{k}}$, and other quantities explicitly involve the bare quantities $g_{\bo{k}}n_0$, and the discarded $\mc{O}(\delta\op{\Psi})$ terms are zero when the condensate wavefunction $\phi(\bo{x})$ is a solution of the GPE involving $g_{\bo{k}}$ and $n_0$. However, the GPE and mean-field expressions used elsewhere typically involve the dressed interaction and density.
\end{itemize}
The resolution is that the difference between the various choices lies at a higher order in $\delta n$ and/or $\delta g = g-g_0$ than treated in the base Bogoliubov theory. Therefore, when $\delta n\ll n$ and $\delta g \ll g$, either choice is safe, provided it is followed consistently in the numerical work. 
Note, though, that the $g_0$ and $n_0$ appearing explicitly in the expressions for Bogoliubov observables $G_2$ \eqn{G2_1} and $E_{\rm int}$ \eqn{Eint} must always be the bare values $g_0$ and $n_0$, not the dressed values $g$ and $n$. 

In principle, a choice of units is arbitrary, but in practice one must take care. When running the matching algorithm and during the dynamics in Secs.~\ref{OBS} and~\ref{DYN}, we used units $g_{\rm set}n_{\rm set}=gn=1$. However, this means that $g_0n_0\neq1$, which affects the Bogoliubov spectrum \eqn{epsilon} and hence the coefficients $U_{\bo{k}},V_{\bo{k}}$ and all observable sums. Therefore, these must be recalculated with each trial choice of $g_0$ and each iteration of the algorithm in Sec.~\ref{g0steps}. 
If one wants to apply the estimates from Sec.~\ref{OBS} in that case, it is necessary to include the correct prefactors from dimensional analysis: $(g_0n_0)^{d/2}$ on direct energies such as \eqn{Hke}, and $(g_0n_0)^{d/2-1}$ on $\delta n$ and $\wb{m}$, respectively. Also, quantities such as $k_c,k_L$ appearing in the expressions in Sec.~\ref{OBS} should be replaced by $k_c/\sqrt{g_0n_0}$, etc.

Regarding the LHY energy value used in the $\mathrm{LHS}$ of \eqn{LHSRHS}, we decided to use $g_{\rm set}n_{\rm set}$ to keep the postulated and Bogoliubov calculations separate. However, $g_0n_0$ would also be valid if used consistently, leading only to corrections of relative order $\delta g/g$ and $\delta n/n$ in any calculated quantities. 

Regarding the apparent inconsistency in Bogoliubov quantities, consider the following. If we
replace $U(\bo{r})$ in the Hamiltonian \eqn{H1D} with the bare delta-function interaction $g_0$ as per \eqn{deltag}, make the Bogoliubov wavefunction replacement \eqn{BOGpsi}, and take $i\dot{\phi} = d\op{H}/d\phi^*$, then the GPE is found to be
\eq{BogGP}{
i\frac{d\phi}{dt} = \left(-\frac{\nabla^2}{2}+g_0|\phi|^2N_0\right)\phi,
}
which must vanish to be stationary. Conveniently, this removes the $\mc{O}(\delta\op{\Psi})$ terms in the Bogoliubov Hamiltonian \eqn{Hbogbog}. 
The evolution equations 
for the Bogoliubov fields $\delta\op{\Psi}$ are:
\eqa{BogEq}{
i\frac{d\delta\op{\Psi}}{dt} &=& \left[\op{H},\delta\op{\Psi}\right]\\
&=& \left(-\frac{\nabla^2}{2}+g_0|\phi|^2N_0\right)\phi\,\op{a}_0 + \left(2g_0|\phi|^2N_0+\frac{\nabla^2}{2}\right)\delta\op{\Psi} + g_0\phi^2\op{a}_0{}^2\delta\dagop{\Psi},\nonu
}
where higher-order terms in $\delta\dagop{\Psi}$ are neglected, 
and the first term 
vanishes if we assume \eqn{BogGP}. Suppose though, now, that instead of \eqn{BogGP}, it was the GPE equation containing full $n=|\phi|^2N_0+\delta\dagop{\Psi}\delta\op{\Psi}$ that is assumed stationary:
\eq{nGP}{
i\frac{d\phi}{dt} = \left(-\frac{\nabla^2}{2}+g_0|\phi|^2N_0+g_0\delta\dagop{\Psi}\delta\op{\Psi}\right)\phi = 0.
}
Then
\eq{nGPstat}{
\left(-\frac{\nabla^2}{2}+g_0|\phi|^2N_0\right)\phi\op{a}_0 = -g_0\delta\dagop{\Psi}\delta\op{\Psi}\phi\op{a}_0.
}
Substituting \eqn{nGPstat} into \eqn{BogEq} now gives 
\eqa{BogEq2}{
i\frac{d\delta\op{\Psi}}{dt} &=& \left(2g_0|\phi|^2N_0+\frac{\nabla^2}{2}\right)\delta\op{\Psi} + g_0\phi^2\op{a}_0{}^2\delta\dagop{\Psi} -g_0\delta\dagop{\Psi}\delta\op{\Psi}\phi\op{a}_0.\nonu\\
}
Thus, the extra term in the Bogoliubov equations is higher order in $\delta\op{\Psi}$, so it can be safely ignored within the Bogoliubov framework, as it lies outside the approximation's scope. If our GPE and condensate use $g$ and/or $n$, any resulting corrections to the Bogoliubov evolution (due to the $g_0$ and $n_0$ in GPE not being exactly matching the $g$ and $n$) are also higher order in the fluctuations than the leading contributions, and are therefore irrelevant and should be discarded. Similarly, the shift $|g-g_0|\ll|g_0|$ is assumed to be small. It is important, however, to remain internally consistent throughout.

\section{Square lattice}
\label{SQUARE}

In Sec.~\ref{OBS}, quantities were estimated via an approximation \eqn{continuum}, which assumed a kinetic energy cutoff $E_c=k_c^2/2$, so that only plane-wave modes with  momenta below $k_c$ were retained. Numerical implementations on a square lattice, however, often retain the entire plane-wave $k$-space, in which the wavevectors $k_j$ in each direction $j=x,y,z$ can take all values up to $|k_j|\le k_{\rm max}$, without an additional energy cutoff. In 2D and 3D systems, this includes additional ``corners'' in $k$-space whose effect must be taken into account when implementing LHY physics on such a lattice. Numerically, this is straightforward, but for estimates an integral form is helpful. 
Here we present some results related to integrating on a square lattice that includes the corners. 

\begin{figure}[h]
\centering
\includegraphics[width=0.4\columnwidth]{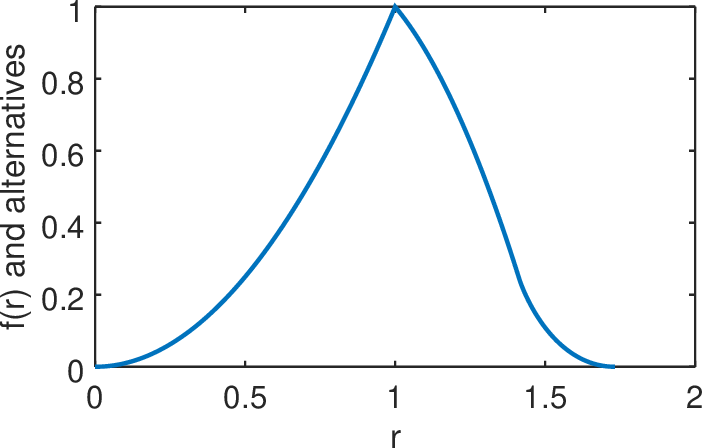}
\caption{Integrand from \eqn{corners} for polar integration over the corners in a 3D system, showing $f(r)$ for $r\ge\sqrt{2}$.}
\label{fig:corners}
\end{figure}

In three dimensions, the integral approximation \eqn{continuum}-\eqn{ddk} becomes
\eq{corners}{
\frac{1}{V}\sum_{\bo{k}}\to\frac{1}{2\pi^2}\int_0^{\sqrt{3}k_{\rm max}}\\dk\, 
\left\{
\begin{array}{c@{\ \text{if}\ }l}
 k^2,     &  r<=1\\
 k^2\left(\frac{3}{r}-2\right),     & 1<r<\sqrt{2}\\
 f(r),     & r>\sqrt{2}
\end{array}
\right.
}
where $r=k/k_{\rm max}$ and 
\eqa{fr}{
f(r) =kk_{\rm max}\bigg[ 1&+&\frac{4}{\pi}\cot^{-1}\sqrt{r^2-2}-\frac{4r}{\pi}\cot^{-1}\left(r\sqrt{r^2-2}\right)-\frac{4}{\pi}\sec^{-1}\sqrt{r^2-1}\nonu\\
&+&\frac{4r}{\pi}\tan^{-1}\sqrt{1-\frac{2}{r^2}}-\frac{4}{\pi}\tan^{-1}\sqrt{r^2-2}\bigg].
}
The above integrand is plotted in Fig.~\ref{fig:corners}.

For 2D, the analogue of \eqn{corners} is 
\eq{corners2d}{
\frac{1}{V}\sum_{\bo{k}}\to\frac{1}{2\pi}\int_0^{\sqrt{2}k_{\rm max}}\\dk\, 
\left\{
\begin{array}{c@{\ \text{if}\ }l}
 k,     &  r<=1\\
 k\left(1-\frac{4}{\pi}\cos^{-1}\frac{1}{r}\right),     & 1<r<\sqrt{2}
\end{array}
\right.
}

The corresponding estimates for $g_1$ are
\eqa{a1deltaint}{
g_1^{(3d)} &=& -\frac{g_0^2}{2\pi^2}\ k_{\rm max}C_{3d},\\
g_1^{(2d)} &=& -\frac{g_0^2}{2\pi}\left(\log\frac{k_{\rm max}}{k_L}+\log 2-\frac{2C}{\pi}\right),\\
g_1^{(1d)} &=& -\frac{g_0^2}{\pi}\left(\frac{1}{k_L}-\frac{1}{k_{\rm max}}\right),
}
where $C_{3d}=1.22137$, $C=0.915966$ is the Catalan constant, and $k_L$ is defined as before, with default $k_L=\pi/L$.


\section{Estimation of the dressed-to-bare interaction shift}
\label{A:DG}

Let us set units according to the chosen reference values: $g_{\rm set}n_{\rm set}=1$; label $n=n_{\rm set}$, $g=g_{\rm set}\to1/n$, and define the relative interaction shift as per \eqn{dgest1d} as
\eq{dgdef}{
g_0=g_{\rm set}(1+\delta_g).
}
Substituting the contributions into \eqn{LHSRHS}, we have 
\eq{a3-1}{
gn^2/2+\ve_{LHY}(g,n,k_c) =\ve_{\rm kin}+\frac{g(1+\delta_g)}{2}\left[n^2+2n_0(\delta n+\wb{m})+\delta n^2+\wb{m}^2\right],
}
including the practically important $E_{LHY}^{\rm extra}$ contributions. The $g_0$ in the interaction energy on the $\mathrm{RHS}$ has already been substituted via \eqn{dgdef}.
Here $\ve_{LHY} = E_{LHY}^{\infty}(g_{\rm set},n_{\rm set},k_c)/V$ is the targeted LHY energy density from \eqn{LHY} after substituting the target values, and $\ve_{\rm kin} = E_{\rm kin}/V$.
For example, $\ve_{LHY}^{(1d)}=-2/3\pi$.
Next, we replace $n_0\to n-\delta n$ in the explicit $n_0$ on the $\mathrm{RHS}$, yielding
\eq{a3-3}{
\ve_{LHY}(g,n,k_c) = 
\ve_{\rm kin}+(1+\delta_g)(\delta n+\wb{m}) +\frac{n\delta_g}{2}
+g(1+\delta_g)\left(\frac{\wb{m}^2-\delta n^2}{2}-\delta n\,\wb{m}\right).
}
To expand in small $\delta_g$, one needs to express all quantities in $g$ and $\delta_g$. Therefore, we identify the leading $g_0\to g$ contributions and the next $\mc{O}(\delta_g)$ contributions (after using units $gn=1$) as per
\eq{a3-2}{
\delta n=\delta n_0+\delta_g\delta n';\quad
\wb{m}=\wb{m}_0+\delta_g\wb{m}';\quad
\ve_{\rm kin}=\ve^{\rm kin}_0+\delta_g\ve_{\rm kin}'.
}
Keeping terms up to $\mc{O}(\delta_g)$ and $\mc(O)(g=1/n)$, one obtains an equation of the form $\ve_{LHY}=\mc{A}_0+\delta_g\mc{A}_1$, which is solved as $\delta_g=(\ve_{LHY}-\mc{A}_0)/\mc{A}_1$ to give \eqn{dgest1d}. Explicitly in orders of $n$, this reads
\eq{a3-4}{
\delta_g = \frac{\ve_{LHY}+A+gB}{\tfrac{n}{2}+\wt{A}+g\wt{B}},
}
where
\eqs{a3-AB}{
A&=&-\delta n_0-\wb{m}_0-\ve^{\rm kin}_0,\\
\wt{A}&=&\delta n_0+\wb{m}_0-\delta n'-\wb{m}'-\ve_{\rm kin}',\\
B&=& \frac{\delta n^2-\wb{m}^2}{2}+\delta n_0\wb{m}_0,\\
\wt{B}&=&\delta n_0(\delta n'+\wb{m}')+\wb{m}_0(\delta n'-\wb{m}')-B.
}
Using the large-box estimates from Sec.~\ref{OBS} for 1D, for example, one recovers the terms as seen in \eqn{dgest1d}. Notably, all the $gB$ and $g\wt{B}$ terms arise from contributions that are omitted at the standard Bogoliubov order -- cross terms such as $\delta^2+\wb{m}^2$ in $E_{LHY}^{\rm extra}$ and corrections due to $n_0$ being smaller than $n$.

In 2D and 3D systems with spherical cutoffs $k_c$ and corresponding $\mc{L}_c=\log(4/k_c)$, one finds, respectively: 
\eqa{dgest2d}{
\delta_g^{(2d)} &\approx& \frac{1-\frac{gk_c^2}{4\pi}\left(2\mc{L}_c^2+2\mc{L}_c-\frac{1}{2}+\frac{3+2\mc{L}_c}{k_c^2}\right)}{2n\pi k_c^2-5+k_c^2\left(\frac{7}{4}-3\mc{L}_c\right)+\frac{gk_c^2}{4\pi}\left(4\mc{L}_c^2+5\mc{L}_c-\frac{5}{2}+\frac{12+2\mc{L}_c}{k_c^2}\right)},\\
\delta_g^{(3d)} &\approx& \frac{k_c-\frac{g}{3\pi^2}\left(\frac{3k_c^2}{2}-4k_c-\frac{5}{3}\right)}{2n\pi^2 +\frac{649}{120}-5k_c+\frac{g}{3\pi^2}\left(\frac{15k_c^2}{2}-\frac{9k_c}{8}-\frac{29}{3}\right)}.
\label{dgest3d}
}

\end{appendix}


\end{document}

\subsection{Units -- not for publication}
\pin{My understanding of King's units:
$\wt{u}_x=\hbar/\sqrt{2mgn}$, $\wt{u}_m=m$, $\wt{u}_t=\hbar/gn\sqrt{2}$, leading to $\wt{u}_E=gn$, $\wt{u}_{\hbar}=\hbar/\sqrt{2}$, $\wt{u}_g=\hbar^d(gn)^{1-d/2}/(2m)^{d/2}$ Therefore, if we change SI quantities to King quantities (tilde), $\wt{m}=1$, $\wt{g}\wt{n}=1$, $\wt{\hbar}=\sqrt{2}$, $\wt{\xi}_0=\sqrt{2}$ where $\xi_0=\hbar/\sqrt{mgn}$.

otoh, Piotr's units are like this:
$\wb{u}_x=\hbar/\sqrt{mgn}$, $\wb{u}_m=m$, $\wb{u}_t=\hbar/gn$, leading to $\wb{u}_E=gn$, $\wb{u}_{\hbar}=\hbar$, $\wb{u}_g=\hbar^d(gn)^{1-d/2}/m^{d/2}$. Therefore, if we change SI quantities to Piotr quantities (bar), $\wb{m}=1$, $\wb{g}\wb{n}=1$, $\wb{\hbar}=1$, $\wb{\xi}_0=1$, but healing length $\hbar/\sqrt{2mgn}\to1/\sqrt{2}$. 


Converting, generally the rule is  $\wb{O} = (\wt{u}_O/\wb{u}_O) \wt{O}$. Such that $\wt{E}=\wb{E}, \wb{x}=\wt{x}/\sqrt{2},
\wb{k}=\sqrt{2}\wt{k},
\wb{t}=\wt{t}/\sqrt{2},
\wb{g}=\wt{g}/2^{d/2}, 
\wb{n}=2^{d/2}\wt{n}$. Note that sometimes there are hidden $\hbar$, such as in $E_{LHY}$ or $E_{\rm kin}$ which get extra factors in King's units compared to what is written previously here.

}